\documentclass[aps,pra,twocolumn,letterpaper,superscriptaddress]{revtex4-2}
\usepackage{amssymb,amsthm,amsmath,amsfonts}
\usepackage{graphicx,enumerate,bbm,bm}
\usepackage{graphicx}
\usepackage{dcolumn}
\usepackage{bm}
\usepackage{amsmath}
\usepackage{xcolor}
\usepackage{changes}
\usepackage{xfrac}
\usepackage{mathtools}

\usepackage{upgreek}
\usepackage{gensymb}
\usepackage{appendix}
\usepackage{subfiles}
\usepackage{nccmath}
\usepackage{amsfonts}
\usepackage{xcolor}
\usepackage[T1]{fontenc}
\usepackage[utf8]{inputenc}
\usepackage{braket} 
\usepackage{todonotes}
\usepackage{lmodern}
\usepackage[hidelinks]{hyperref}

\begin{document}

\preprint{APS/123-QED}

\title{Exceptional rings in nonlinear non-Hermitian planar optical microcavities: \\ implementation, signal enhancement, and topology}

\author{Jan Wingenbach}
 \affiliation{Department of Physics and Center for Optoelectronics and Photonics Paderborn (CeOPP), Paderborn University, Warburger Strasse 100, 33098 Paderborn, Germany}
 \affiliation{Institute for Photonic Quantum Systems (PhoQS),
Paderborn University, Warburger Straße 100, 33098 Paderborn, Germany}%

\author{Laura Ares}
 \affiliation{Department of Physics and Center for Optoelectronics and Photonics Paderborn (CeOPP), Paderborn University, Warburger Strasse 100, 33098 Paderborn, Germany}
 \affiliation{Institute for Photonic Quantum Systems (PhoQS),
Paderborn University, Warburger Straße 100, 33098 Paderborn, Germany}%

\author{Xuekai Ma}
 \affiliation{Department of Physics and Center for Optoelectronics and Photonics Paderborn (CeOPP), Paderborn University, Warburger Strasse 100, 33098 Paderborn, Germany}%
 
\author{Nai H. Kwong}
\affiliation{Wyant College of Optical Sciences, University of Arizona, Tucson, AZ 85721, USA}%

\author{Jan Sperling}
 \affiliation{Department of Physics and Center for Optoelectronics and Photonics Paderborn (CeOPP), Paderborn University, Warburger Strasse 100, 33098 Paderborn, Germany}
 \affiliation{Institute for Photonic Quantum Systems (PhoQS),
Paderborn University, Warburger Straße 100, 33098 Paderborn, Germany}%

\author{Rolf Binder}
\affiliation{Wyant College of Optical Sciences, University of Arizona, Tucson, AZ 85721, USA}%
\affiliation{Department of Physics, University of Arizona, Tucson, AZ 85721, USA}%
 
\author{Stefan Schumacher}
 \affiliation{Department of Physics and Center for Optoelectronics and Photonics Paderborn (CeOPP), Paderborn University, Warburger Strasse 100, 33098 Paderborn, Germany}%
 \affiliation{Institute for Photonic Quantum Systems (PhoQS),
Paderborn University, Warburger Straße 100, 33098 Paderborn, Germany}%
 \affiliation{Wyant College of Optical Sciences, University of Arizona, Tucson, AZ 85721, USA}%

\begin{abstract}

Non-Hermitian systems hosting exceptional points (EPs) exhibit signal enhancement and unconventional mode dynamics. Going beyond isolated EPs, here we report on the existence of exceptional rings (ERs) in planar optical resonators with specific form of circular dichroism and TE–TM splitting. Such exceptional rings possess intriguing topologies as discussed earlier for condensed matter systems, but they remain virtually unexplored in presence of nonlinearity, for which our photonic platform is ideal. We find that when Kerr-type nonlinearity (or saturable gain) is introduced, the linear ER splits into two concentric ERs, with the larger-radius ring being a ring of third-order EPs. Transitioning from linear to nonlinear regime, we present a rigorous analysis of (spectral and band) topologies and report enhanced and adjustable perturbation response in the nonlinear regime. Whereas certain features are specific to our system, the results on non-Hermitian topologies and nonlinearity-enhanced perturbation response are generic and equally relevant to a broad class of other nonlinear non-Hermitian systems, providing a universal framework for engineering ERs and EPs in nonlinear non-Hermitian systems. 
\end{abstract}

\maketitle

\section{Introduction} 
Non-Hermitian systems have attracted significant attention in recent years. In contrast to Hermitian systems they exhibit potentially complex eigenvalues and non-orthogonal eigenvectors enabling the investigation of striking and often counter-intuitive phenomena. Exceptional points are singularities at which two or more complex eigenvalues coalesce, and their respective eigenvectors become parallel~\cite{kato.1966, GARRISON1988177, Berry2004, heiss2004exceptional, Bender2007, Heiss2012,ashida-etal.2021}. The number of modes $n$ that coalesce at an exceptional point defines the order $n$ of the exceptional point. Exceptional points have been investigated in various platforms, including microwave resonators~\cite{PhysRevLett.86.787, PhysRevLett.90.034101, RevModPhys.87.61}, atomic systems~\cite{PhysRevLett.104.153601}, plasmonic nanostructures~\cite{PhysRevB.94.201103, park2019observation}, optical waveguides~\cite{PhysRevLett.103.093902}, microresonators~\cite{PhysRevLett.103.134101, chang2014parity, peng2014parity}, and non-reciprocal systems~\cite{Fruchart2021}. So-called Liouvillian exceptional points can be investigated in systems obeying a Lindblad master equation~\cite{PhysRevA.100.062131}. The coalescence of eigenvectors often leads to counterintuitive behavior close to the exceptional point such as loss-induced transparency~\cite{Zhang:18}, loss-induced suppression and revival of lasing~\cite{doi:10.1126/science.1258004,li2022switching}, unidirectional invisibility, and reflectivity exceeding unity~\cite{PhysRevLett.103.093902, PhysRevLett.106.213901}. Unlike diabolic points in Hermitian systems, the eigenvalue splitting scales with the $n^\mathrm{th}$-root of the perturbation strength close to an exceptional point of order $n$, rendering them highly sensitive~\cite{chen2017exceptional,Wiersig:20}. Although the discussion about noise enhancement remains ongoing, this sensitivity makes exceptional points promising candidates for new sensor designs~\cite{PhysRevA.93.033809,hodaei2017enhanced, Wiersig2020}, while some works demonstrate limitations to these prospects~\cite{langbein2018no,lau2018fundamental,mortensen2018fluctuations,wolff2019time,zhang2019quantum,chen2019sensitivity,naikoo2023multiparameter,loughlin2024exceptional}. Recent studies show that nonlinear systems provide excellent platforms to manipulate exceptional points and their characteristic features. Prominent examples include coupled optical and electrical resonators, and hybrid light-and-matter quasiparticles such as exciton-polaritons. Notable phenomena in these systems include mode-switching around exceptional points in bistable domains~\cite{wang2019dynamics,PhysRevA.103.043510}, abrupt global changes in the stability due to coinciding exceptional points~\cite{PhysRevResearch.7.013326}, phase-transitions~\cite{Opala:23,PhysRevB.109.085311}, robust wireless power transfer~\cite{assawaworrarit2017robust}, as well as saturable gain and Kerr-nonlinearity induced shifting~\cite{Ramezanpour:24,PhysRevResearch.6.013148} and rotation of exceptional points~\cite{PhysRevResearch.6.013148}. Remarkably, exceptional-like points can also be studied in nonlinear Hermitian systems~\cite{PhysRevB.111.L161102}.

\begin{figure*}[!tb]
  \centering
   \includegraphics[width=1.0\textwidth]{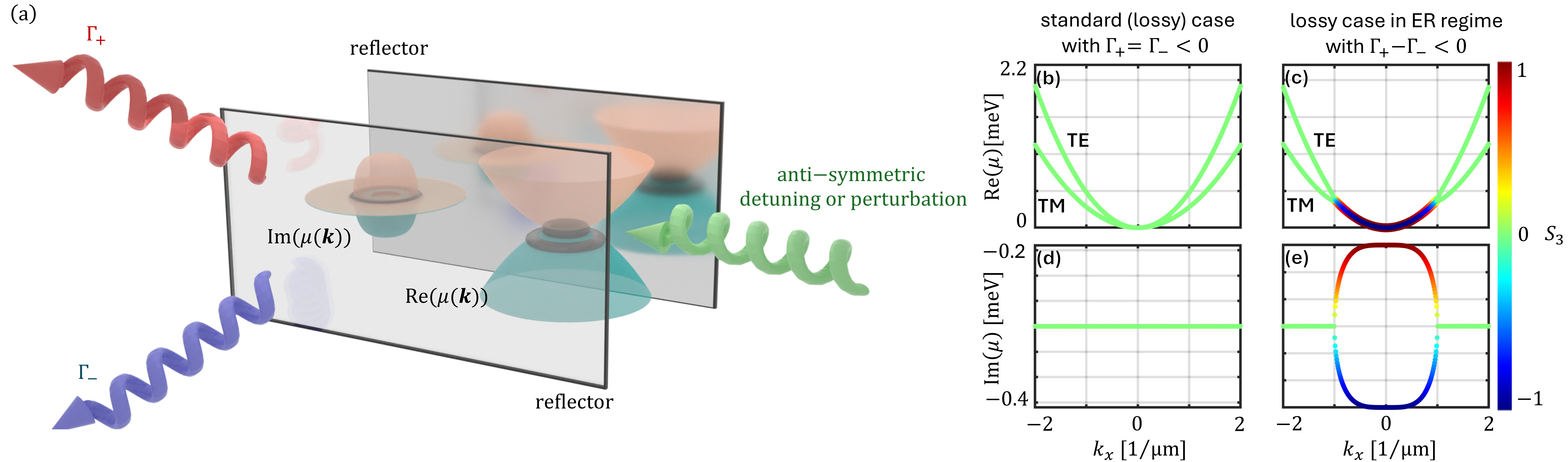}
  \noindent
  \caption{\textbf{Concept of exceptional ring in planar microcavities with circular dichroism and spin-orbit coupling.} (a) Sketch of planar resonator with circular dichroism in the form $\Gamma_+\neq\Gamma_-$ and finite TE-TM splitting. The complex-valued in-plane resonator dispersions, $\mu(\bm{k})$, host an ER centered at $k=0$ in reciprocal space. Nonlinearity splits that ring into mulitple rings (marked as black donuts). (b,d) Isotropic in-plane dispersions of the circularly polarized modes split into TE and TM modes by their coupling. (c,e) In the ER regime, when circular dichroism is introduced, $\Gamma_+\neq\Gamma_-$, TE-TM splitting is substracted inside an exceptional ring (ER) forming in $k$-space at $|\bm{k}_{\mathrm{ER}}|=\sqrt{|\gamma|/|\Delta_\mathrm{LT}|}=1/\mathrm{\upmu m}$; the dispersions become circularly/elliptically polarized as shown by their $S_3$ Stokes vector element. Dispersions are rotationally invariant in $k$-space and $y$-axes in (b,c) are relative to the dispersion minimum.}
  \label{fig:1}
\end{figure*}    

While the interplay of nonlinear and non-Hermitian physics is fascinating, in the nonlinear regime it is not always easy to tune the system to the vicinity of an isolated EP~\cite{PhysRevResearch.6.013148}. In contrast to distinct pairs of exceptional points, \textit{exceptional rings} (ERs) form a continuous, gapless loop of coalescing eigenvectors and eigenvalues and offer a promising platform to drive a system to an exceptional point. It was shown that ERs may emerge from diabolic points in reciprocal space of non-Hermitian spin-orbit coupled systems~\cite{zhen2015spawning}. To date, they have been studied in helical waveguide arrays~\cite{cerjan2019experimental}, high-spin ultracold atoms~\cite{10.1088/1674-1056/addcbe}, mechanical and thermal systems~\cite{yoshida2019exceptional,he2020floquet,xu2022observation,xu2022observation}, honeycomb~\cite{yoshida2019symmetry} and superconducting Lieb lattices~\cite{cao2020band}, photonic and sonic crystals~\cite{PhysRevB.100.165134, liu2022experimental,isobe2023symmetry,wang2023exceptional, isobe2025topological, zhao2025magnetically, clbx-mh6y}, as well as the excitation spectrum of light-matter systems~\cite{Opala:23}. In Ref.~\cite{kolkowski2021pseudochirality} a design for an ER-laser with pseudochiral optical response was presented in active plasmonic metasurfaces. By tweaking the gain and loss profiles, ERs can be tuned into different one-dimensional contours~\cite{cerjan2018effects}. In Ref.~\cite{liu2021higher} topological phase transitions between ERs with opposite topological charges were investigated. Recently, also the interplay of an ER and the nonlinear Hall effect was investigated~\cite{8g3q-qrpg}.

In the present paper we show that the in-plane dispersion relations in spin-orbit coupled planar optical resonators with a specific form of circular dichroism naturally host ERs (Section~\ref{sec:linear}). For these we discuss in detail their topology and perturbation response in both linear and nonlinear regimes. In Section~\ref{sec:umbilic} we introduce nonlinearity in the system and discuss the nature of nonlinear ERs. We report that introduction of a Kerr-type medium (or saturable gain) splits the ring into an inner ring of $2^\mathrm{nd}$- and an outer ring of $3^\mathrm{rd}$-order exceptional points. As derived in Ref.~\cite{kwong2025universal}, finite mode detuning reveals more ERs of similiar kind, whose radii form an elliptic umbilic singularity structure in nonlinear parameter space: for a given nonlinearity three fold lines of exceptional points (exceptional arcs) intersect at three exceptional points of $3^\mathrm{rd}$-order (exceptional nexus). In planar optical resonators with Kerr medium this leads to the emergence of rings formed by points on exceptional arcs (exceptional arc rings) and rings formed by exceptional nexuses (exceptional nexus rings), respectively. In Section~\ref{sec:sensing} we discuss perturbation features of nonlinear ERs and show ultra-enhanced and adjustable perturbation response. In Section~\ref{sec:topology} we show that the linear ER can (partially) be recovered even in anisotropic systems. With mode-switching applications in mind, we report that the ring forms a tilted quasi-toroidal structure in the nonlinear regime. We provide vorticity and Berry phase calculations both in the linear and nonlinear regime. While parts of our work focus on the specific system of planar optical resonators, and by that present a novel platform to study ERs, the ultra-enhanced and adjustable perturbation response and complex spectral topological features introduced are very generic and apply to a wide range of nonlinear non-Hermitian Hamiltonians \cite{kwong2025universal}.

\section{Exceptional ring in planar resonators with spin-orbit coupling}\label{sec:linear}

In spin-orbit coupled optical cavities and exciton-polariton systems, individual exceptional point pairs have been observed in $k$-space via linear dichroism, i.e. losses dependent on the linear polarization of the cavity mode~\cite{Krol2022, PhysRevResearch.6.013324}. Pairs of exceptional points have also been observed for  non-monotonic momentum dependence of the mean polariton linewidth~\cite{doi:10.1126/sciadv.abj8905, PhysRevB.108.115404}. In Ref.~\cite{Opala:23} a natural ER is observed in the excitation spectrum of an exciton-polariton system in the transition from weak to strong light-matter coupling (transition from exciton and photon to polariton modes). In the present paper we introduce ERs resulting from the interplay of coupled circular polarization components and circular dichroism and discuss in detail the impact of nonlinearities on these ERs. 

We start with the linear problem where we find that our microcavity shows the interesting feature of a natural ER. Only two basic ingredients are required for this observation: (i) TE-TM splitting and (ii) different losses for left- and right-circularly polarized field components ($\Gamma_+\neq\Gamma_-$), a specific form of circular dichroism. We note that TE-TM splitting is typically intrinsically present (the specific functional form is not important here). Circular dichroism may be introduced, for example, through gyroscopic materials, assuming degeneracy of the resonances in the two circular components for the scalar case at zero TE-TM splitting~\cite{Rajaei2019,PhysRevApplied.22.054016}. If non-degenerate, the modes would have to be tuned back into resonance for example by applying an external magnetic field~\cite{PhysRevB.84.165325, Polimeno2021} or external off-resonant pumping in active semiconductor materials~\cite{PhysRevLett.109.036404, waldherr2018observation}. 
As a specific realization, in much of the following, we study exciton polaritons in a semiconductor microcavity. The polariton amplitudes (or wave functions) of the lower polariton branch are denoted by $\psi$; they depend on the polariton wave vector $\bm{k}$. The two circular polarization states (right and left circularly polarized) are indicated by the subscricts, $\psi_+$ and  $\psi_-$. The physics of interest comes from the (in general different) decay rates in the two polarization channels and the coupling of the two channels.  

We model the linear system by an effective $2\times 2$ Hamiltonian and solve the corresponding eigenvalue problem for each $\bm{k}$ on a discretized 2D $k$-space grid
\begin{equation}
\begin{aligned}
\begin{bmatrix}
\frac{\hbar^2 \bm{k}^2}{2m}+i\Gamma_+
&
\Delta_\mathrm{LT}(ik_x+k_y)^2
\\
\Delta_\mathrm{LT}(ik_x-k_y)^2
&
\frac{\hbar^2 \bm{k}^2}{2m}+i\Gamma_-
\end{bmatrix}
\begin{bmatrix}
\psi_+ \\ \psi_-
\end{bmatrix}
=
\mu
\begin{bmatrix}
\psi_+ \\ \psi_-
\end{bmatrix}\,.
\label{eq:eigenvalue}
\end{aligned}
\end{equation}
Here, for sufficiently small $|\bm{k}|$, we assume parabolic in-plane dispersions for the left- and right-circularly polarized modes in absence of spin-orbit interaction ($\Delta_\mathrm{LT}=0$) and set the dispersion minimum to zero. The dispersion curvature is given by the effective mass $m = 10^{-4}m_\mathrm{e}$, with $m_\mathrm{e}$ being the free electron mass. $\psi_\pm$ describes the field of the right- and left-circularly polarized modes with losses $\Gamma_\pm$, respectively. As a measure for the circular dichroism we define $\gamma = \sfrac{|(\Gamma_+-\Gamma_-)|}{2}$. The mode coupling, TE-TM splitting, is introduced by the off-diagonal elements, with the factor $(ik_x+k_y)^2$ signifying a spin-orbit coupling with an exchange of $2 \hbar$ between the orbital motion and the polaritonic spin (circular polarization) states. The coupling  strength is given by $\Delta_\mathrm{LT}=0.1~\mathrm{meV\upmu m^2}$ and here assumed to be independent of the magnitude of the wavevector, $|\bm{k}|$. Such a dependence depends on the structural details of the system, and does not affect the occurance of an ER as long as the total TE-TM splitting is a monotonically increasing function of $|\bm{k}|$ at least up to the $|\bm{k}|$ at which the ER is observed~\cite{panzarini1999exciton, kavokin2004quantum,shelykh-etal.2005, PhysRevB.76.245324}. In other words, the specific functional form of the dispersion is not important for the observation of the ER and even in absence of $k$-space isotropy a deformed ER structure is found. The value of TE-TM splitting strength chosen here is in good agreement with measured values in both GaAs and $\mathrm{CsPbBr_3}$ perovskite microcavities~\cite{PhysRevB.59.5082,doi:10.1126/sciadv.abj7667}. Note that while here we focus on the lossy case of negative $\Gamma_\pm$ in the following, the eigenvalue problem also shows an ER for overall gain. 

In the linear regime, the eigenvalue problem~(\ref{eq:eigenvalue}) can be solved straightforwardly. In Figure~\ref{fig:1}(b,d) the real and imaginary parts of the solutions in the linear isotropic regime (and $\gamma = 0$) show the well-established TE and TM frequency splitting and equal losses along $k_x$. The displayed results are rotationally invariant in $k$-space. The coloring of the lines indicates the $S_3=\frac{|\psi_+|^2-|\psi_-|^2}{|\psi_+|^2+|\psi_-|^2}$ Stokes vector element for the respective eigenvector $\ket\psi=\begin{bmatrix}\psi_+ & \psi_-\end{bmatrix}^\mathrm{T}$ as a measure for the degree of circular polarization. Left and right circular, as well as linear polarizations are given by $S_3=-1$, $S_3=1$, and $S_3=0$, respectively. 

In Figure~\ref{fig:1}(c,e) the respective solutions are shown in the linear non-Hermitian regime with circular dichroism ($\gamma = 0.1~\mathrm{ps^{-1}}$) with $\{\Gamma_+,\Gamma_-\}=\{-0.2,-0.4\}~\mathrm{ps^{-1}}$. The corresponding lifetimes are in a realistic range as measured in GaAs and $\mathrm{CsPbBr_3}$ perovskite microcavities~\cite{PhysRevLett.118.016602,su2018room}. In this regime an ER with radius $k_\mathrm{ER} =|\bm{k}_\mathrm{ER}|= \sqrt{\sfrac{|\gamma|}{|\Delta_\mathrm{LT}|}} = 1~\mathrm{\upmu m^{-1}}$ is formed in the dispersion (the displayed results are rotationally symmetric in $k$-space). The mode splitting is canceled inside the ER where the two solutions become circularly polarized [cf. Fig.~1(c)]. The radius of the ER is determined by the ratio of TE-TM splitting $\Delta_\mathrm{LT}$ and degree of circular dichroism $\gamma$. Specific linewidths only matter for the visual appearance of the ring; the dispersion and spectral properties are best visible when the lines close to the ER radius do not overlap too much. 

\begin{figure}[!tb]
  \centering
    \includegraphics[width=1.0\columnwidth]{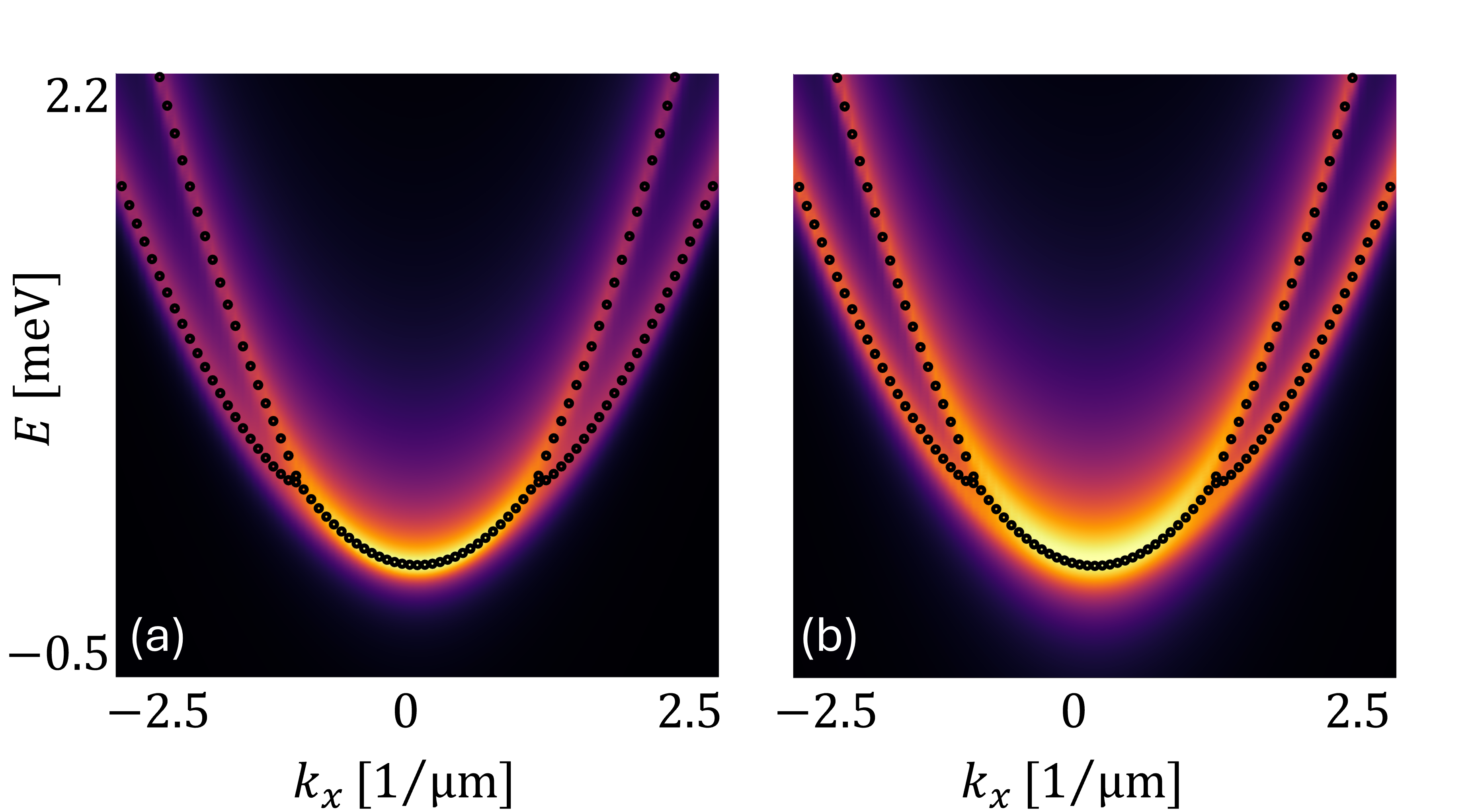}
  \caption{\textbf{Exceptional ring in in-plane resonator dispersion in linear regime.} Diagonal elements of response to left (a) and right (b) circularly polarized light, plotted along $k_x$. Comparing (a) and (b), the difference in linewidths as expected inside the exceptional ring are clearly visible [cf. Fig.~1(e)]. These full model calculation is in good agreement with the dispersions of the simplified 2x2 system in Eq.~(1) (black circles). The displayed results are rotationally symmetric in $k$-space and energies are measured relative to the dispersion minimum.}
  \label{fig:2}
\end{figure}

We verify these results and our model Hamiltonian by simulating the complex spatio-temporal dynamics of the respective cavity fields in time-dependent, two-dimensional real-space calculations. In line with our simplified $k$-space model above, Eq.~(\ref{eq:eigenvalue}), the equations governing the in-plane field propagation and dispersions in the circular polarization basis read:
\begin{align}
    i\hbar\partial_t\psi_\pm=\left(-\frac{\hbar^2\nabla^2}{2m}+ i\Gamma_\pm\right) \psi_\pm +J_\pm\psi_{\mp}+R_\pm\,.
\label{eq:PULSE_psi}
\end{align}
In contrast to the eigenvalue problem in Eq.~(\ref{eq:eigenvalue}) this equation describes the field dynamics in two-dimensional real-space with $\psi_\pm=\psi_\pm(\mathbf{r},t)$, which is driven by an external coherent field $R_\pm=R_\pm(\mathbf{r},t)$ with $\mathbf{r} = (x,y)$. Consistent with the $k$-space representation in Eq.~(\ref{eq:eigenvalue}), the TE-TM splitting operator is defined as $J_\pm= {\Delta_\mathrm{LT}}(\partial_x \mp i\partial_y)^2$. To extract the system response that shows the dispersions, we pulse the system with a short resonant pulse $R_\pm(\textbf{r},t) = R_{0}\mathrm{exp}\left(-\frac{\textbf{r}^2}{p_\mathrm{w}^2}\right)\mathrm{exp}\left(-i\omega t\right)\mathrm{exp}\left(\frac{-(t-t_0)^2}{\sigma^2}\right)$, separately in each polarization channel in two independent time evolution calculations. The calulations are performed with our CUDA compatible open-source solver for 2D nonlinear Schrödinger equations, "PHOENIX"~\cite{wingenbach2024phoenixpaderbornhighly}, and the code for the specific application shown here is included in the PHOENIX Github~\cite{phoenix_example}. In Fig.~\ref{fig:2} we plot $-\mathrm{Im}\left(\chi\right)$ for the diagonal elements of the response tensor, $\chi_{_{++}} = \frac{\mathcal{F}\left(\psi_+\right)}{\mathcal{F}\left(R_+\right)}$ and $\chi_{_{--}}= \frac{\mathcal{F}\left(\psi_-\right)}{\mathcal{F}\left(R_-\right)}$, which dominate the system response for small TE-TM splitting. Here, $\mathcal{F}(X)$ is the Fourier transform of the respective quantity $X$ into $k$- and $\omega$-space.

The resulting dispersions in Fig.~\ref{fig:2} show perfect agreement with the results of the eigenvalue problem, Eq.~(1). Both energy and linewidth behavior in the time-dependent calculations agree with the results of our reduced 2x2 model as displayed in Fig.~1(c,e). We note that, in principle, for suitably tailored $\gamma(\bm{k})$ or for non-$k$-isotropic systems, different shapes of the exceptional structure in 2D $k$-space can be realized, allowing for the formation of exceptional ellipses and more complex exceptional curves. In  Appendix~\ref{append:gp_extended}, for the specific realization in an exciton polariton system, we explicitly show that off-resonant pumping preserves the assumed $k$-independence of the polarization dependent gain difference, $\gamma$, inferred in the present study.

\section{Nonlinearity-induced third-order exceptional ring}\label{sec:umbilic}
So far the optical resonators considered here were assumed to be linear systems, however, their reponse can become significantly nonlinear if a nonlinear optical medium is inserted. The nonlinearity may result from a Kerr-type nonlinear material response or, in systems that form exciton polaritons, stem from polariton-polariton interactions, which can typically lead to strong coherent optical nonlinearities. Besides symmetries~\cite{PhysRevResearch.6.023205}, nonlinearities allow for the realization of higher-order exceptional points in lower-dimensional systems~\cite{bai2023nonlinearity,PhysRevLett.130.266901}. This led to the investigation of an exceptional nexus, i.e. a higher-order exceptional point forming at the intersection of multiple lines of second-order exceptional points, so-called exceptional arcs~\cite{tang2020exceptional, bai2023nonlinearity}.

Here, we take into account coherent Kerr-type nonlinearity and cross-mode coupling from TE-TM splitting. Note that in the specific case of exciton polaritons the cross-nonlinearity between circularly polarized spinor components, $\psi_+$ and $\psi_-$, is known to be weaker and attractive compared to the repulsive self-interaction nonlinearity \cite{PhysRevB.76.245324}. Also in the nonlinear regime we make the (in most cases valid) assumption of unidirectional coherent light-field propagation with negligible scattering into other $k$-modes (typically also justified for polariton systems if exciting below the magic-angle \cite{PhysRevLett.84.1547}; numerically explicitly shown in Appendix~E). The extended nonlinear eigenvalue problem then reads
\begin{equation}
\begin{aligned}
\begin{bmatrix}
M_+
&
\Delta_\mathrm{LT}(ik_x+k_y)^2
\\
\Delta_\mathrm{LT}(ik_x-k_y)^2
&
M_-
\end{bmatrix}
\begin{bmatrix}
\psi_+ \\ \psi_-
\end{bmatrix}
=
\mu
\begin{bmatrix}
\psi_+ \\ \psi_-
\end{bmatrix},
\end{aligned}
\label{eq:eigenvalue_nonlinear}
\end{equation}
with $M_\pm =\frac{\hbar^2 \bm{k}^2}{2m} + i\Gamma_\pm+g_\mathrm{c}|\psi_\pm|^2+g_\mathrm{x}|\psi_\mp|^2\pm\Delta$. 
Here, $g_\mathrm{c}$ and $g_\mathrm{x}$ define the Kerr-like nonlinearity and cross-mode interaction strength, respectively. We first focus on the degnerate case of $\Delta=0$ for which the fully analytic solutions of the nonlinear eigenvalue problem are of tractable length (see Appendix~\ref{append:math}). Later, through $\Delta\neq0$ we will include finite mode detuning of different origins; examples we discuss are through magnetic field induced Zeeman splitting and electric field tunable Rashba-Dresselhaus splitting.

To solve nonlinear eigenvalue problems, several approaches have been proposed, including the polar representation~\cite{PhysRevA.79.013812,PhysRevE.83.066608} or Stokes parameters and momentum operators incorporation~\cite{Graefe_2008,PhysRevA.82.013629,PhysRevA.82.043803,Graefe_2012}. However, these methods are often limited to specific Hamiltonian structures. More recently, iterative techniques based on self-consistent field approaches have been applied to non-Hermitian nonlinear eigenvalue problems~\cite{PhysRevA.103.043510, GU2024107736}. Despite their utility, iterative methods, such as the Newton-Raphson method show convergence issues in bistable regimes close to the exceptional point. In Ref.~\cite{kwong2025universal}, we provide the general solutions (with analytical expressions in the case of zero detuning) of the eigenvalue problem~(\ref{eq:eigenvalue_nonlinear}); the solution approach used was inspired by the separability eigenvalue equations~\cite{PhysRevLett.111.110503} as discussed in more detail in Appendices~\ref{append:math} and \ref{app:sperling}. The resulting solutions apply to any real normalization $|\psi_+|^2+|\psi_-|^2 = N^2$. The given solutions are valid also for the left eigenvalue problem under complex conjugation (Appendix~\ref{append:math}). Our analytical solutions are validated using them as initial conditions for the Newton-Raphson method. Here, we scale the solutions to norm $N=1~\mathrm{\upmu m^{-2}}$, and introduce $\alpha = \sfrac{(g_\mathrm{c}-g_\mathrm{x})}{2}$ as a measure of nonlinearity.
\begin{figure*}[!tb]
  \centering
   \includegraphics[width=1.0\textwidth]{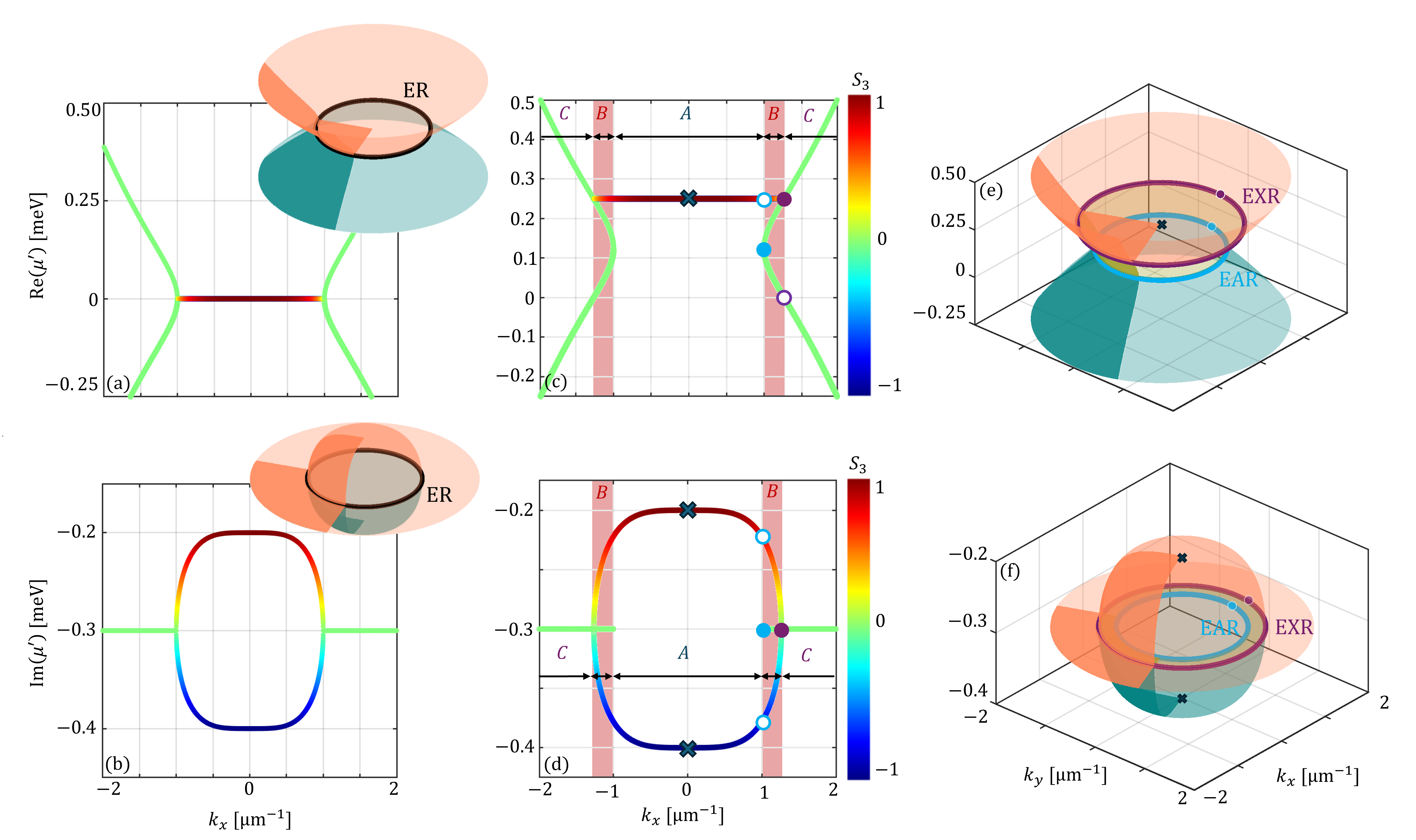}
    \noindent
  \caption{\textbf{Exceptional rings in the complex-valued energy spectra in nonlinear regime at zero detuning ($\Delta = 0 $).} Central cuts of the (a) real and (b) imaginary parts of energy spectra of the linear eigenvalue problem (neglecting the $k^2$ dispersion). The respective surfaces are depicted in the insets with the exceptional ring (ER) marked in black. Central cuts of the (c) real and (d) imaginary parts of energy spectra of the nonlinear eigenvalue problem. The respective surfaces are depicted in (e-f). In contrast to the linear regime, the energy spectrum in the nonlinear regime hosts two exceptional rings, one with radius $|\bm{k}|^2=\sfrac{|\gamma|}{|\Delta_\mathrm{LT}|}=1~\mathrm{\upmu m}^{-1}$ (solid blue dot and labeled by EAR) and a second ring with radius $|\bm{k}|^2=\sfrac{\sqrt{\gamma^2+\alpha^2}}{|\Delta_\mathrm{LT}|}$ (solid violet dot and labeled by EXR). Between these two rings [marked in red in section B in panel (c-d)] four solutions are found. Inside the inner (outside the outer) ring two circularly (linearly) polarized solutions are found [section A and C in panel (c,d)]. The circles mark points where a solution crosses the boundary between two segments without tracing through an exceptional ring and the cross marks $k=0$. These markers are used in Fig~\ref{fig:4}(b) to analyze the bifurcation behavior near the two rings. The results displayed are rotationally symmetric in $k$-space. Simulation parameters: $\{\alpha,g_\mathrm{c},g_\mathrm{x}\} = \{0.125,0.25,0\}~\mathrm{meV\upmu m}^2$.} 
  \label{fig:3}
\end{figure*}

We begin the discussion of the nonlinear eigenvalue problem for the case of zero detuning, $\Delta=0$, showing eigenvalues and further analysis in Figs. \ref{fig:3} and \ref{fig:4}. As shown in Appendix~\ref{append:math}, we have two solutions in region A and C, and four solutions in region B. In the following figures, for clarity, we show the eigenvalues $\mu^\prime = \mu-\frac{\hbar^2 \bm{k}^2}{2m}$, where the parabolic dispersion is substracted. For a better comparison with the results in Section~\ref{sec:linear}, we set $\{\Gamma_+,\Gamma_-\}=\{-0.2, -0.4\}~\mathrm{ps^{-1}}$ and $\Delta_\mathrm{LT}=0.1~\mathrm{meV\upmu m^2}$ for the remainder of this work. Figure~\ref{fig:3} shows the resulting real and imaginary parts of the complex-valued energy spectrum and central cuts for $\{\alpha,g_\mathrm{c},g_\mathrm{x}\} = \{0.125,0.25,0\}~\mathrm{meV\upmu m}^2$ [see Fig.~\ref{fig:3}(c-f)] in comparison to the linear results [see Fig.~\ref{fig:3}(a-b)]. These parameter values are in good agreement with measured blueshifts in both GaAs and $\mathrm{CsPbBr_3}$ perovskite microcavities ~\cite{PhysRevB.100.035306,polimeno2024room}. In the nonlinear regime the complex-valued energy spectrum hosts two concentric rings marked by solid blue and violet dots in the central cuts in Fig.~\ref{fig:3}(c-d), and labled by EAR and EXR in Fig.~\ref{fig:3}(e-f), separating three radial segments (A-C). The inner segment A hosts two circularly polarized solutions, the segment B between the rings, marked in red, features four solutions (two circularly and two linearly polarized), and the outer segment C shows two linearly polarized solutions. Note, that nonlinearity shifts the non-linearly polarized real parts of the solutions in segment A and B by $2\alpha$, while the linearly polarized solutions in segment B and C are shifted by $\alpha$. In Figure~\ref{fig:4}(a-b) these segments are sketched for an arbitrary nonlinearity $\alpha\neq0$ to visualize the bifurcation behavior at the two rings. Analyzing this behavior, we conclude that in the nonlinear regime the linear ER splits into an inner ring of fold points (blue dot) along $|\bm{k}|^2=\sfrac{|\gamma|}{|\Delta_\mathrm{LT}|}$ and an outer ring of cusp points (violet dot) along $ \sqrt{\sfrac{(\alpha^2+\gamma^2)}{\Delta_\mathrm{LT}^2}}$, which grows with increasing Kerr-like nonlinearity $g_\mathrm{c}$ or cross-interaction strength $g_\mathrm{x}$. The fold and cusp points can be identified with exceptional points of $2^\mathrm{nd}$ and $3^\mathrm{rd}$-order, respectively. Hence, the outer ring is a $3^\mathrm{rd}$-order exceptional ring. The circles mark points where a solution crosses the boundary between two segments without tracing through an ER (e.g. the two green colored solutions form an ER at the blue solid dot, at the border of segment A and B, while the blue and red colored solutions cross the border of segment A and B at the blue circle without forming an ER). For $\alpha<0$ the cusp ring is redshifted in relation to the fold ring [this scenario is displayed in Figure~\ref{fig:1}(a)]. While previous work realized individual higher-order exceptional points in two-dimensional nonlinear parameter space and thus reduced the parameter tuning requirements~\cite{PhysRevLett.130.266901, bai2023nonlinearity}, the ring structure observed here reduces these parameter tuning requirements even further, making higher-order exceptional point structures in low dimensions more accessible. 

At this point, we need to discuss the nature of exceptional points of order $n>2$ in nonlinear $2\times2$ systems. In previous work it was exemplarily shown that mapping a nonlinear and non-Hermitian Hamiltonian to a $n\times n$ linear and $\mathcal{PT}$-symmetric Hamiltonian features the same exceptional point of order $n$~\cite{PhysRevLett.130.266901, bai2023nonlinearity}. Accordingly, the singularity in the nonlinear system is called a “nonlinear exceptional point”, containing at least one dynamically unstable solution. It is straightforward to show that the four solutions in section B in Figure~\ref{fig:3}(c-d) form an overcomplete basis. This is reflected in the Petermann factor $\mathrm{PF}=\left| \braket{\psi_L |\psi_R} \right|^{-2}$, which serves as a measure for non-orthognality of left $\bra{\psi_L}$ and right $\ket{\psi_R}$ eigenvectors~\cite{wiersig2023petermann}. $\mathrm{PF} = 1$ indicates orthogonal modes, whereas $\mathrm{PF} > 1$ signals mode non-orthogonality, resulting in a divergent Petermann factor at linear exceptional points. At the inner ring the $\mathrm{PF}$ diverges, highlighting the presence of an ER with coalescing solutions in the nonlinear regime. At low nonlinearity the nonlinear $3^\mathrm{rd}$-order ER exhibits diverging $\mathrm{PF}$. With increasing nonlinearity the Petermann factor at the nonlinear $3^\mathrm{rd}$-order ER decreases until it converges to $1$. This means that the eigenvectors at the ER become increasingly orthogonal and do not coincide in the nonlinear regime. However, it has been shown that nonlinear exceptional points still feature an intriguing spectral topology and the $n^\mathrm{th}$-root perturbation response of their linear counterparts \cite{Wiersig:20}. This suggests that the unstable mode still contributes to the order of the nonlinear exceptional point~\cite{PhysRevLett.130.266901, bai2023nonlinearity}, which we will prove for our $3^\mathrm{rd}$-order ER in Section~\ref{sec:sensing}. In that sense we believe that the nonlinear ER appears to show sufficient similarities with its linear counterpart to justify its name and order classification.

\begin{figure}[!tb]
  \centering
    \includegraphics[width=1.0\columnwidth]{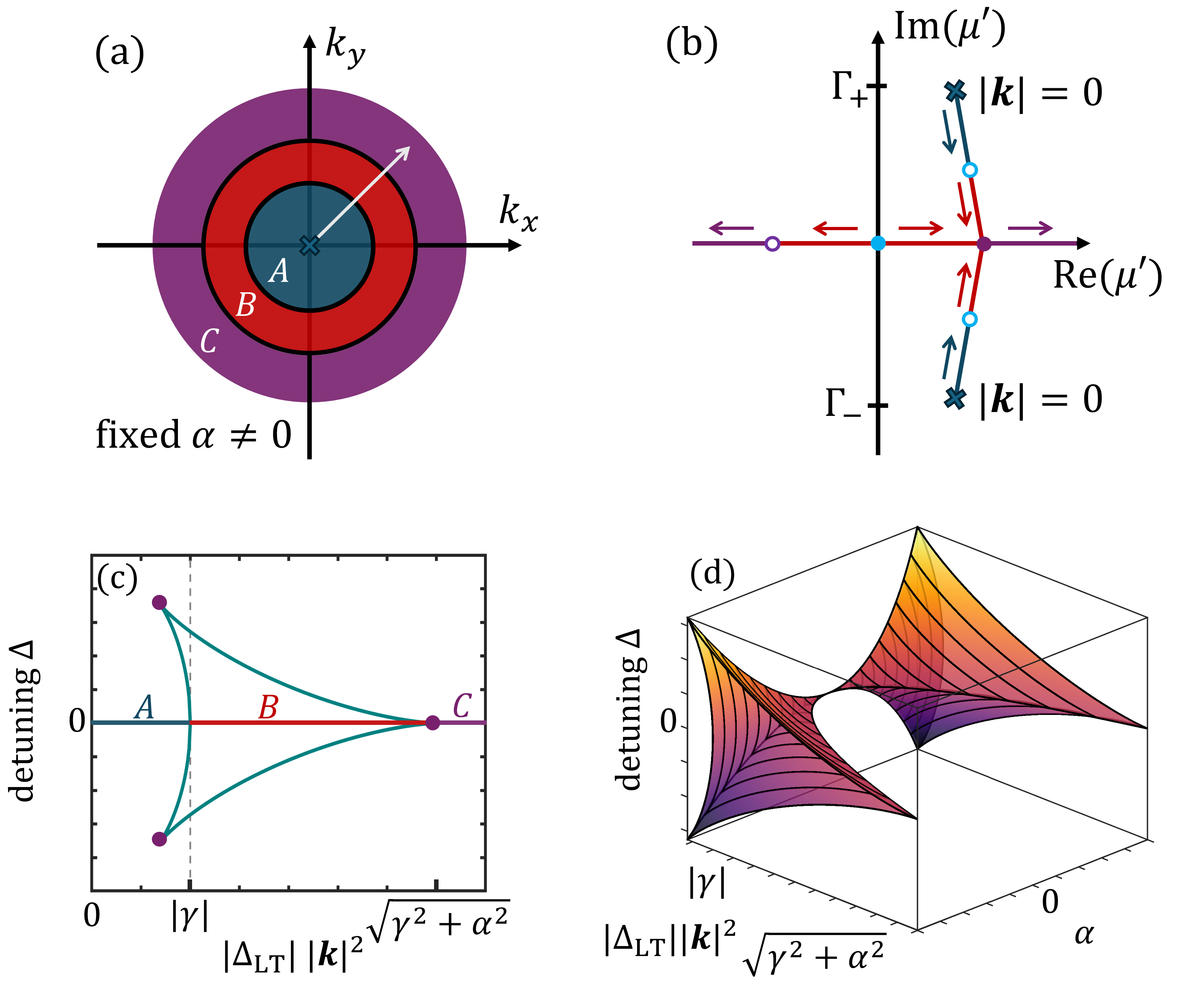}
  \caption{\textbf{Elliptic umbilic structure of exceptional rings and exceptional points in nonlinear regime.} (a) Sketch of the three sections A-C [as displayed in Fig.~\ref{fig:3}(c)] with two, four, and two solutions, respectively. (b) The bifurcation behavior on the edges of ring B characterizes the inner (outer) ring as a ring of fold (cusp) points. (c) Radii of fold rings (turquoise lines) and cusp-rings (violet dots) for degnerate modes (segments marked as above) and for finite mode detuning $\Delta$. The umbilic structure shows that fold-rings are part of exceptional arcs and cusp-rings are located at exceptional nexuses. (d) Cross section in (c) for variable nonlinearity $\alpha$. The depicted elliptic umbilic predicts exceptional ring radii and type in the full parameter space. The central organization point represents a point on the linear exceptional ring discussed in Section~\ref{sec:linear}.}
  \label{fig:4}
\end{figure}

To understand the nature of the two ERs in the nonlinear regime in a more general setting, we have to solve the nonlinear eigenvalue problem~(\ref{eq:eigenvalue_nonlinear}) for finite detuning $\Delta\neq0$, which, while not rendering it impossible, greatly complicates the fully analytic solution. However, in Ref.~\cite{kwong2025universal}, we have derived the $4^\mathrm{th}$-order polynomial that determines the solutions for this eigenvalue problem. Solving it numerically allows us to analyze the neighborhood of the two ERs. In Ref.~\cite{kwong2025universal} we have shown that for finite $\alpha$ a single linear exceptional point splits into three exceptional nexuses ($3^\mathrm{rd}$-order exceptional points) connected by three exceptional arcs (lines of $2^\mathrm{nd}$-order exceptional points) in the plane of mode coupling (here: TE-TM splitting strength $\Delta_\mathrm{LT}$) and mode detuning $\Delta$. In 2D $k$-space each exceptional point on this structure represents a radius of an ER. In Figure~\ref{fig:4}(c) the possible ER radii are sketched for finite $\alpha$ together with the segments along $\Delta=0$ discussed above, turquoise lines depict exceptional arcs and violet dots the exceptional nexuses. Thus, the inner and outer ER shown in Figure~\ref{fig:3} are formed by the fold point of the exceptional arc and by the exceptional nexus along $\Delta = 0$, respectively. 

Extending this picture for variable nonlinearity the cross section in Figure~\ref{fig:4}(c) forms an elliptic umbilic in parameter space along the nonlinearity strength $\alpha$ [Fig.~\ref{fig:4}(d)]. This shows that in contrast to the linear ER, the two rings in the nonlinear regime are not individual singularities in parameter space, but surrounded by a surface of other fold rings on the respective exceptional arcs and more cusp rings at the other exceptional nexuses on the umbilic even for finite detuning. Following this convention, we will refer to the rings formed by the fold rings on the umbilic as \textit{exceptional arc rings} (EARs) and rings formed by cusp points of the umbilic as \textit{exceptional nexus rings} (EXRs). Note that these exceptional rings do not only reduce the parameter number to tune the system towards an exceptional point but can even be found for finite detuning. 
\begin{figure}[!tb]
  \centering
    \includegraphics[width=1.0\columnwidth]{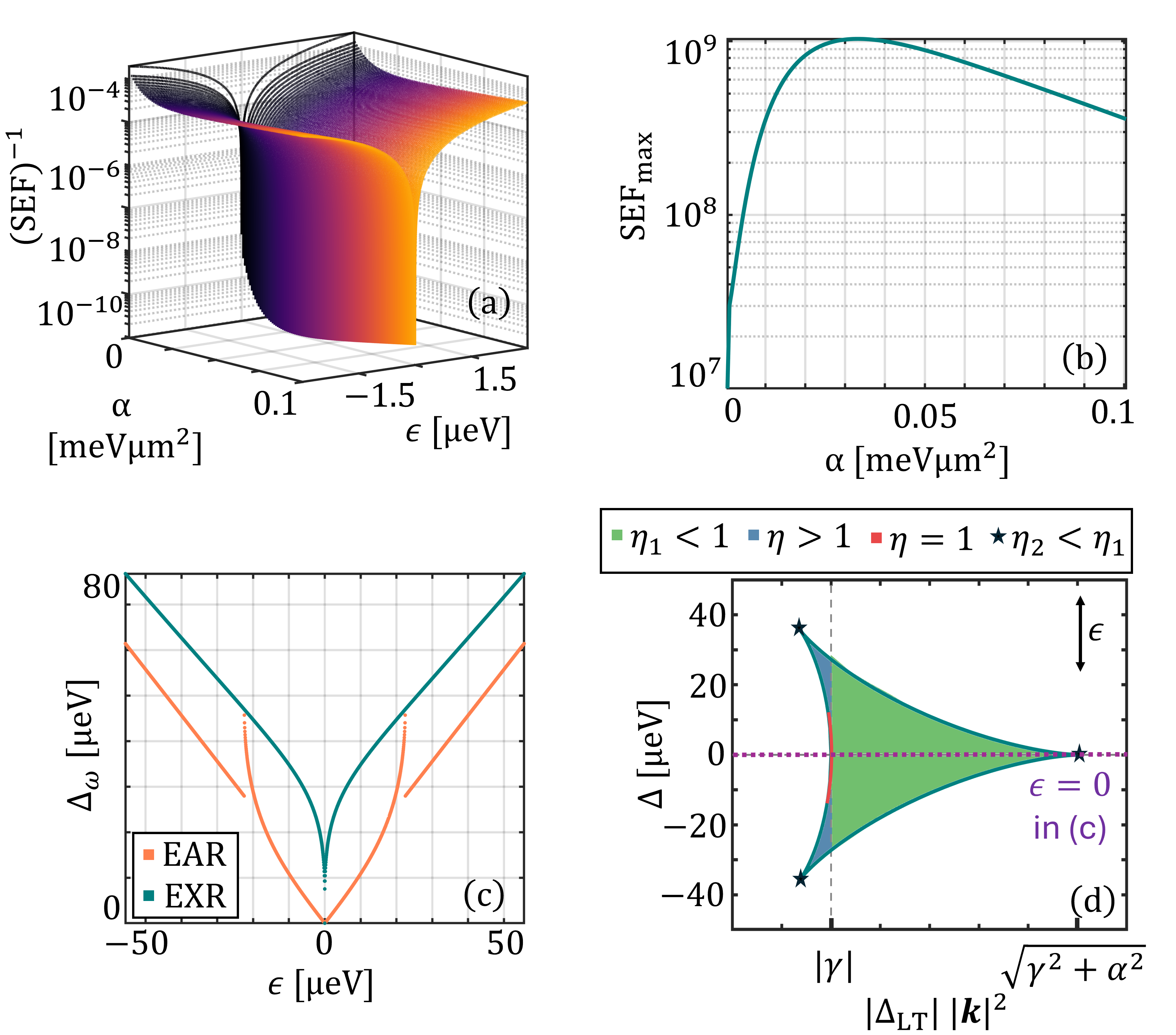}
   \caption{\textbf{Enhanced perturbation response at nonlinear exceptional rings and points}. In (a-b) a $3^\mathrm{rd}$-order exceptional point on the exceptional nexus ring [star in (d)] is traced for increasing $\alpha$. (a) The inverse signal-enhancement factor $(\mathrm{SEF})^{-1}$ decreases in the nonlinear regime. The coloring visualizes $\alpha$ in each curve (growing from black to yellow). (b) The maximum $\mathrm{SEF}_\mathrm{max}$ at the exceptional point as a function of $\alpha$. (c) Response $\Delta_\omega$ of the exceptional points on the exceptional arc ring (EAR) and exceptional nexus ring (EXR) on the elliptic umbilic along $\Delta=0$ [horizontal dashed line in (d) corresponds to $\epsilon=0$ in (c)] for a fixed nonlinearity $\alpha$. The response of the EAR significantly deviates from the square-root law of linear exceptional points. Here, small perturbations tune the system along the left exceptional arc of the umbilic [red line in (d)]. Larger perturbations tune it away from the arc due to its curvature [blue area in (d)] leading to growing response. (d) Colored cut through elliptic umbilic at finite nonlinearity $\alpha$. The coloring denotes the response function scaling factor $\eta$ when perturbing an exceptional point on the umbilic singularity structure. Detuning allows to realize different rings on the umbilic, which feature distinct perturbation responses. Perturbation from the lower to the upper arc (or vice versa) leads to characteristic square-root response $\eta_1$. Nexus perurbation leads to stronger response $\eta_2$. The perturbation $\epsilon$ is defined in relation to the respective point on the umbilic. Simulation parameters: $\{g_\mathrm{c},g_\mathrm{x}\} = \{2\alpha,0\}~\mathrm{meV\upmu m}^2$.} 
  \label{fig:sensing}
\end{figure}

While in the linear regime the relation between eigenstates and stationary solutions is straightforward, their relation becomes nontrivial when moving to the nonlinear regime. However, in Appendix~\ref{append:time} we show analytically, that the ER structures discussed above can also be found in the coherently driven nonlinear Schrödinger equations with resonant excitation. There, our solutions hold true up to a shift induced by the resonant driving force. Moreover, in Appendix~\ref{append:NSE} we provide one possible example to study the exceptional ring structure in the coherently driven nonlinear system. Specifically, we demonstrate that the resonance behavior in the spatio-temporal dynamics of the driven system exhibits the same features as the nonlinear eigenvalue problem discussed above. The resonance spectrum with the two co-centrical ERs in case of zero detuning is included in Fig.~\ref{fig:10} as a guide to the eye. We emphasize that the systematic study of the nonlinear system dynamics with off-resonant excitation requires more work and thus needs to be adressed in future works. However, this outlook lays the foundation for such more detailed investigations into nonlinear dynamics and nonlinear response bahavior.

In previous work~\cite{cerjan2019experimental} it was mentioned that the linear ER does not exist in the solutions of the eigenvectors for finite mode detuning. Here we have shown that in the nonlinear regime new ERs arise even for finite mode detuning. We use this knowledge to analyze the perturbation response of nonlinear ERs in Section~\ref{sec:sensing} and their topological features under mode detuning in Section~\ref{sec:topology}. In Appendix~\ref{append:hamilton} we detail that the elliptic umbilic structure applies to a wide range of non-Hermitian Hamiltonians when introducing nonlinearity. Therefore, the topolgoical features and perturbation response of nonlinear ERs on the umbilic we discuss below, are generic and can be applied to exceptional points in many other systems and constellations.

\section{Enhanced Perturbation Response at Nonlinear Exceptional Rings and Points}\label{sec:sensing}
Optical resonators are promising platforms to study the sensing capabilites of exceptional points~\cite{chen2017exceptional,jin2018high,niu2025enhancing}. In the linear regime, the characteristic root-law eigenvalue response to perturbations is compensated by noise enhancement due to linewidth broadening, which results from the coalescence of the eigenvectors~\cite{PhysRevA.98.023805,wang2020petermann}. This leads to limitations of the overall signal-to-noise ratio close to linear exceptional points. Recently, there is a growing interest in studying the effect of nonlinearity on the perturbation response of exceptional points. It was shown theoretically in Ref.~\cite{PhysRevResearch.6.013148} and later experimentally in Ref.~\cite{Li2024}, that nonlinearity can be used to tune the eigenvalue splitting close to an exceptional point. 

In Ref.~\cite{PhysRevLett.130.266901,PhysRevLett.134.133801} it was proposed that nonlinear exceptional points can exhibit a complete basis of eigenstates. In laser systems, the Petermann factor can be investigated as a measure of noise enhancement. Based on this, it was proposed in Ref.~\cite{bai2023nonlinearity} that nonlinear exceptional points can exhibit a divergent signal-to-noise ratio. More recently, however, it was shown that the conventional Petermann factor does not cover the interplay of noise and nonlinearity and that nonlinear exceptional points exhibit finite signal-to-noise ratio~\cite{PhysRevLett.134.133801}, with a generalized form of the Petermann factor at exceptional points derived in Ref.~\cite{kullig2025}. In the following, we discuss that nonlinear exceptional points possess adjustable and in some cases ultra-enhanced perturbation response. We will not make claims about the resulting signal-to-noise ratio in our system, as further work would be needed in that direction. However, we note that our results demonstrate how the signal enhancement can not only be increased but also tuned in the nonlinear regime. Our analysis covers the group of both real- and imaginary-valued smooth nonlinearities that lead to the elliptic umbilic exceptional point structure described above; here exemplarily evaluated for Kerr-type coherent nonlinearities. The elliptic umbilic structure significantly increases the complexity of an exceptional points neighborhood permiting sensor design for exceptional points at different positions on the singularity structure, which changes the response to perturbations not limited to $n^\mathrm{th}$-root laws.

For now we will focus on the EXR along $\Delta=0$ (without detuning) and show that the nonlinearity leads to drastically increased perturbation response at the EXR [marked by stars in Fig.~\ref{fig:sensing}(d)]. We specifically analyze the eigenvalue splitting resulting from perturbations $\epsilon$ of the form
\begin{equation}
    \hat{H}(\epsilon,\alpha)=\hat{H}_\mathrm{EP}(\alpha)+\epsilon\hat{H}_1,~\mathrm{with}~\hat{H}_1=\mathrm{diag}(1,-1).
\end{equation}
Here, $\hat{H}_\mathrm{EP}(\alpha)$ denotes the Hamiltonian of the nonlinear eigenvalue problem~(\ref{eq:eigenvalue_nonlinear}) at the investigated exceptional point. The perturbation $\epsilon$ is equivalent to a small additional detuning $\Delta \rightarrow \Delta + \epsilon $ in the nonlinear eigenvalue problem. We focus on perturbations of form $\hat{H}_1$ to investigate the straightforward interplay of perturbation and the elliptic umbilic singularity structure. We note that this interplay may become more complex for other forms of perturbation as discussed in Ref.~\cite{Wiersig:20}. In the latter case, the exceptional points on the umbilic can be perturbed into new dimensions in parameter space, making visualization difficult. 

For now, we investigate the signal enhancement at the EXR. Aside from the obvious signal enhancement resulting from the increased order of the exceptional points on the EXR compared to the linear ER, we calculate the signal-enhancement factor defined by the change in eigenvalue with perturbation~\cite{bai2023nonlinearity}:
\begin{equation}
     \mathrm{SEF(\epsilon,\alpha)} = \left|\frac{\partial\mu(\epsilon,\alpha)}{\partial\epsilon}\right|^2
    \label{eq:sef}
\end{equation}
with $\mu(\epsilon,\alpha)$ being the eigenvalue at the exceptional point described by $\hat{H}(\epsilon,\alpha)$. In Fig.~\ref{fig:sensing}(a) the inverse signal-enhancement factor $(\mathrm{SEF})^{-1}$ is displayed. Note that in the linear regime the noise enhancement covered by the non-orthogonality of eigenstates compensates the signal enhancement. In the nonlinear regime, the interplay between noise and nonlinearity does not allow conclusions to be drawn about noise amplification based solely on non-orthogonality~\cite{PhysRevLett.134.133801}.

\begin{figure*}[!t]
  \centering
    \includegraphics[width=1.0\textwidth]{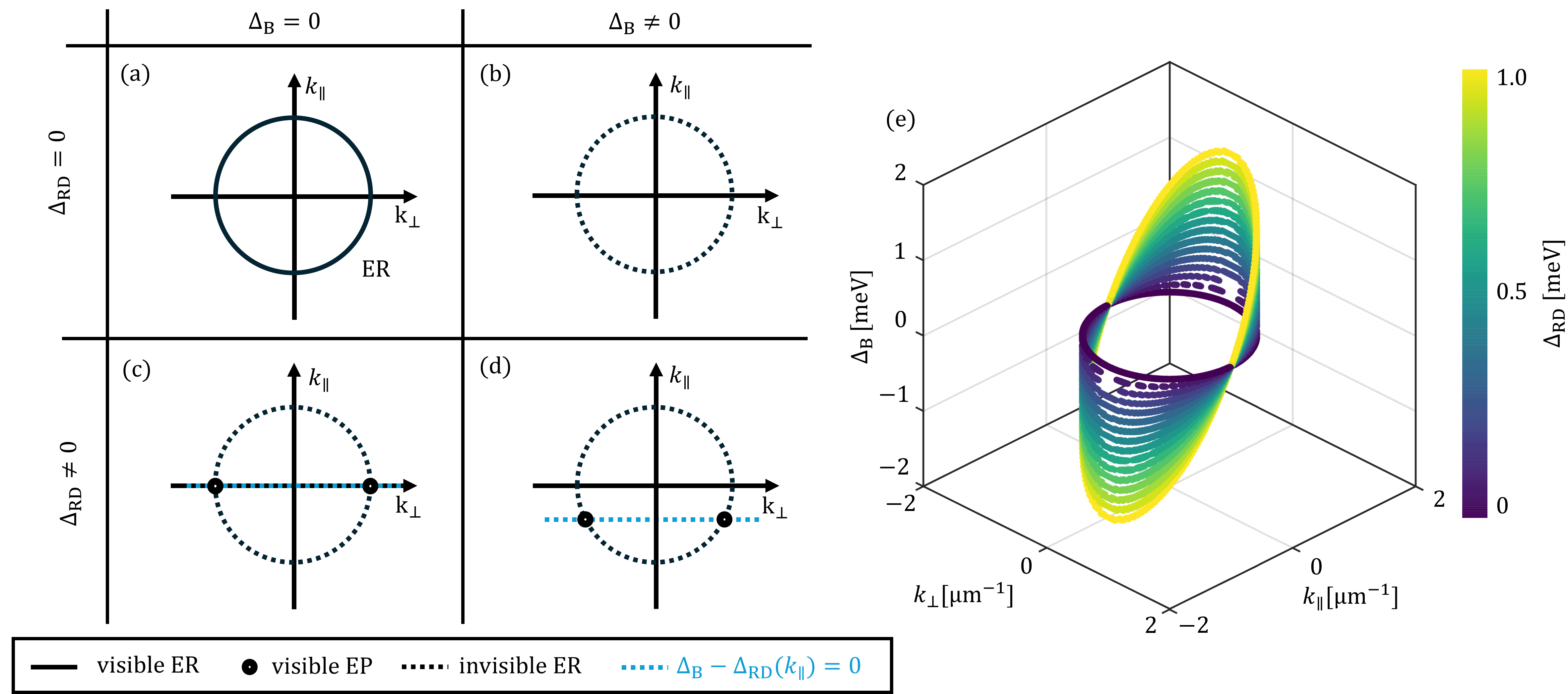}
    \noindent
  \caption{\textbf{Topological neighborhood of linear exceptional rings in 4D parameter space.} (a) For degenerate modes ($\Delta_\mathrm{B}=\Delta_\mathrm{RD}=0$) the complete ER is visible in the original reduced parameter space (2D $k$-space). (b) In case of finite Zeeman splitting ($\Delta_\mathrm{B}\neq0$), the ER vanishes from the reduced parameter space as the degeneracy is lifted for all $k$-values. (c) In case of Rashba-Dresselhaus splitting ($\Delta_\mathrm{RD}\neq0$), the ER vanishes from the reduced parameter space but a pair of exceptional points remains along $k_\mathrm{\perp}$. (d) If Zeeman and Rashba-Dresselhaus splitting are combined ($\Delta_\mathrm{B}\neq0$, $\Delta_\mathrm{RD}\neq0$), the ER vanishes but the pair of exceptional points remains along $k_\mathrm{\perp}$ at $k_\mathrm{\parallel}$ for which Zeeman and Rashba-Dresselhaus splitting compensate each other in the extended parameter space. (e) Exceptional-point surface consisting of tilted ERs (gaps in the surface result from limited discretization). The tilting in $k_\mathrm{\parallel}$-direction is proportional to the Rashba-Dresselhaus splitting, $\Delta_\mathrm{RD}$. The exceptional points are located at those points in parameter space at which the effect of Zeeman and Rashba-Dresselhaus splitting compensates each other.}
  \label{fig:tilted_ring}
\end{figure*}

In Figure~\ref{fig:sensing}(b) the maximum of the signal-enhancement factor $\mathrm{SEF}_\mathrm{max}$ is plotted against $\alpha$. The maximum signal enhancement of $\mathrm{SEF} = 1.9\times10^{9}$ is found for a nonlinearity strength $\sfrac{(g_\mathrm{c}-g_\mathrm{x})}{2}=68~\mathrm{\upmu eV\upmu m^2}$. Here, the signal enhancement scales with the perturbation as $\mathrm{SEF}\propto|\epsilon|^{-1.29}$. In contrast to an isolated linear exceptional point or ER, the elliptic umbilic allows for different exceptional points to be perturbed depending on the initial detuning $\Delta$. Since the perturbation $\epsilon$ is parallel to $\Delta$ this means that depending on which exceptional point we chose to perturb, the response depends on its position on the umbilic. In planar nonlinear resonators external electric and magnetic fields can be used to tune the mode detuning via Zeeman and Rashba-Dresselhaus splitting for instance. The response is defined by the eigenvalue splitting $\Delta_\omega(\epsilon,\alpha)$ from the exceptional point eigenvalue $\mu(\epsilon=0,\alpha)$ in dependence of the perturbation strength:
\begin{equation}
    \Delta_\omega(\epsilon,\alpha)=|\mu(\epsilon,\alpha)-\mu(0,\alpha)|.
\end{equation}
In Figure~\ref{fig:sensing}(c) the response of the EXR (turquoise) is shown together with the response function of EAR (orange) for $\alpha = 0.1~\mathrm{meV\upmu m^{2}}$ and $\Delta = 0$. Note, that the response function of the EAR clearly deviates from $n^\mathrm{th}$-root law of a linear exceptional point. This is due to the fact that the exceptional point is initially perturbed parallel to the exceptional arc of the umbilic [red line in Fig.~\ref{fig:sensing}(c)], resulting in linear response ($\Delta_\omega\propto|\epsilon|^\eta$ with $\eta = 1$) to perturbation. Only when the curved shape of the arc starts to deviate from the perturbation direction, the eigenvalues split faster ($\eta > 1$) [blue in Fig.~\ref{fig:sensing}(c)]. If the perturbation pushes the system outside the elliptic umbilic, the system jumps to another solution, which then has a linear response in relation to the perturbed exceptional point [white area outside the umbilic in Fig.~\ref{fig:sensing}(c)]. If finite detuning is used, the system can also be perturbed from one exceptional arc to another (top to bottom or vice versa). In this case, the exceptional point then shows a classic root-shaped response ($\eta_1 < 1$) [green in Fig.~\ref{fig:sensing}(c)]. Perturbation of the EXR leads to stronger response $\eta_2<\eta_1$. Importantly, these findings are valid for all exceptional points located along the ERs, owing to the rotational symmetry of the structure in $k$-space. This underlines the versatility of exceptional points in nonlinear systems, opening the door to highly tunable sensor designs. These findings further illustrate that in the nonlinear regime the response of an exceptional point depends on its position in parameter space (relative to the other arcs on the umbilic) and the specific form of perturbation.

\section{Topology of Exceptional Rings}\label{sec:topology}
In the literature, ERs are intensively studied for their non-trivial band topologies~\cite{xu2017weyl}. In the following, we discuss several related aspects for our system. We have evaluated the vorticity for bands $n=u$ (upper band) and $n=l$ (lower band) in the case without nonlinearity, defined in Ref.~\cite{shen-etal.2018}
\begin{equation}
   \nu_{ul} = - \frac{1}{2\pi}\oint_{\cal{L}}  \nabla_k {\rm arg} ( \mu'_{u}(\bm{k}) - \mu'_{l}(\bm{k})) \cdot d\bm{k}
   \label{eq:vorticity}
\end{equation}
over a closed loop $\cal{L}$ encircling one point of the ER once, without piercing the ER. In particular, we have used a loop in the $\Delta - k_y$ plane encircling the ER on the positive $k_y$ axis, i.e. at
$(k_y, \Delta) = (k_\mathrm{ER}, 0)$.
We call this configuration a Hopf--link configuration, because the ER and the loop $\cal{L}$ form a Hopf link. We obtain a vorticity of $\nu_{ul} = 0.5$, consistent with the square-root singularity of the exceptional point \cite{mailybaev-etal.2005} and the results in \cite{xu2017weyl,shen-etal.2018}.

We have numerically verified that the result is independent of the details of the loop \cite{mailybaev-etal.2005}, for example using circular loops with various radii ($ k_{\cal{L}} <  2 k_\mathrm{ER}$) and elliptical loops. Without nonlinearity, the vorticity can be obtained for a loop in the $k_x - k_y$ plane with $\Delta = 0$ without numerical evaluation. In this case, both eigenvalues are real (imaginary) for $k_{\cal{L}} > k_\mathrm{ER}$ ($k_{\cal{L}} <   k_\mathrm{ER}$) and hence the argument of the eigenvalue difference is $k$-independent on $\cal{L}$, resulting in the vorticity being zero.

In Appendix~\ref{append:topol} we provide more details regarding the topological properties of the ER in the $k_x-k_y$ plane, i.e. using $\Delta=0$. We find that the closed-loop Berry phase and the integral over the $z$-componennt of the Berry curvature depend on the choice of eigenvectors (left or right) used to calculate the Berry connection. Next, we calculate an integral over the Berry curvature that we call a quasi-Chern integral

\begin{equation}
    C^{ab}_\mathrm{n} = \frac{1}{2\pi}\int\Omega_n^{ab}(\bm{k}) d\textbf{S}\,.
   \label{eq:chern_ana}
\end{equation}

If we use only right eigenvectors ($ab = RR$), we find that the integral over the $z$-componennt of the Berry curvature exists and, without nonlinearity, is equal to $C^{RR}_\mathrm{n} =\pm 1$ if the integration area includes the ER. Note that for combinations of left and right eigenvectors (i.e. $ab = LR$ or $ab = RL$), self-orthognality leads to divergence of quasi-Chern integral at the ER. A detailed discussion of the scaling of the integral value is given in Appendix~\ref{append:topol}.

With the objective of analyzing the nonlinear solutions, the analytical results for the quasi-Chern integral are first validated numerically. For this purpose, we discretize the $k$-space grid and use the $U(1)$ link variable to calculate the Berry curvature for each lattice site $\bm{k}$ following Ref.~\cite{Wang2020}:

\begin{equation}
    C^{ab}_\mathrm{n} = \frac{1}{2\pi}\sum_k \Omega_n^{ab}(\bm{k})\delta k^2.
   \label{eq:chern_num}
\end{equation}

In the nonlinear regime, we compute the quasi-Chern integral on three types of round disks: (i) inside the inner ring, (ii) enclosing only the inner ring, and (iii) enclosing both the inner and outer rings. To calculate the quasi-Chern integral of the EAR and the EXR the integral is evaluated over the eigenvalue surfaces forming the respective ring. Here, both rings exhibit quasi-Chern integral of $C^{RR}_\mathrm{n} = \pm1$. For weak nonlinearity, the proximity of the two rings leads to a significant overlap of their Berry curvature distributions. Consequently, integrating over the disk enclosing only the inner ring yields finite quasi-Chern integral values on the order of $\approx 10^{-1}$ for the outer ring, despite the integration disk lying entirely inside the ring.

\begin{figure}[!t]
  \centering
    \includegraphics[width=1.0\columnwidth]{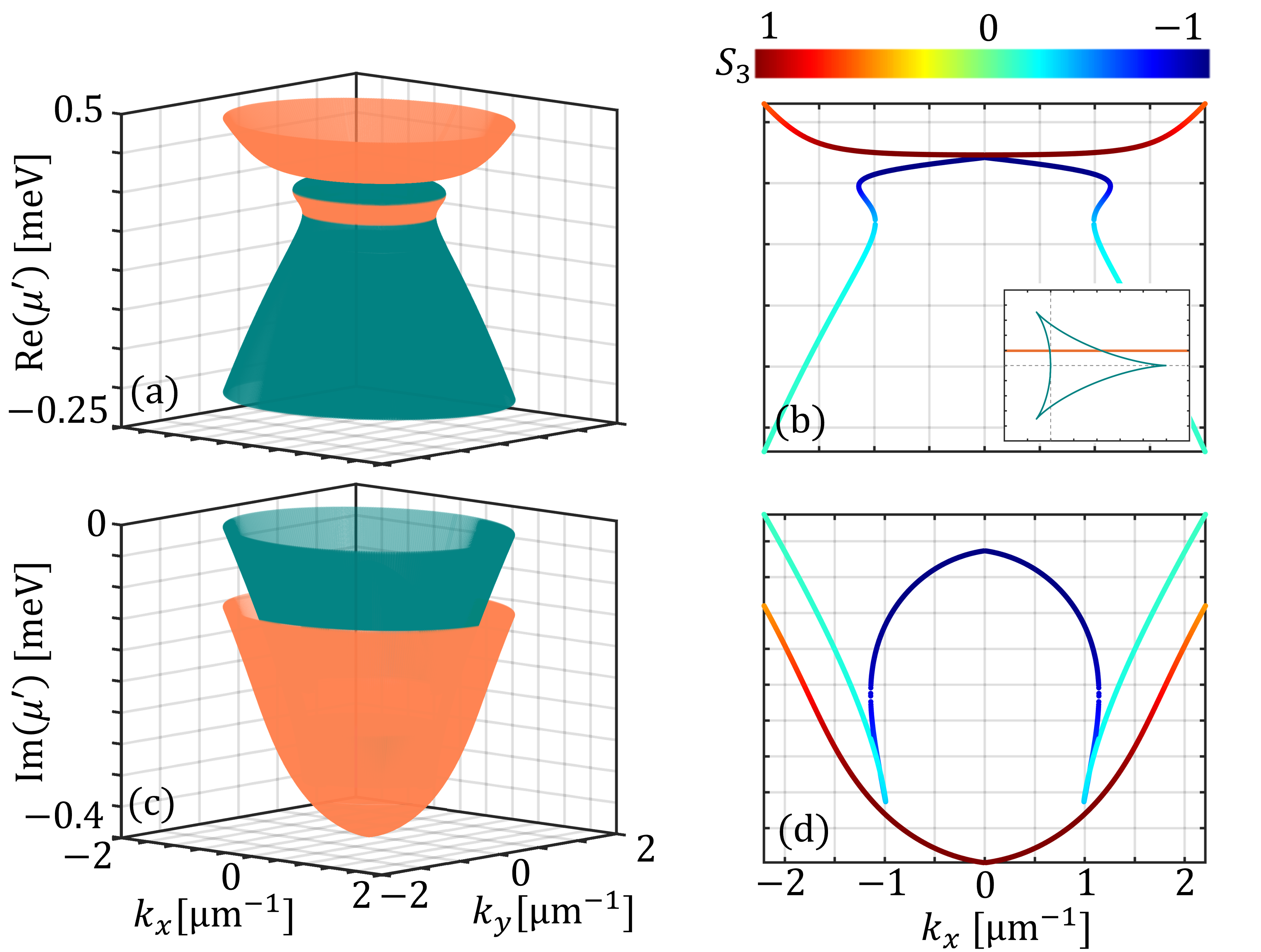}
  \caption{\textbf{Complex-valued energy spectrum with exceptional rings for finite Zeeman splitting in nonlinear regime.} (a,c) Real- and imaginary part of eigenvalues $\mu$. (b,d) Cross sections of $k$-isotropic complex-valued energy spectrum in (a,c). The main panels show the real and imaginary parts of the energy spectrum and central cuts that host the two concentric exceptional arc rings. The turquoise line in the inset in (b) shows the cross-section of the umbilic for finite $\alpha=0.1~\mathrm{meV\upmu m^2}$. The orange line displays the Zeeman splitting-induced detuning (increasing with increasing $\Delta_\mathrm{B}$). The displayed exceptional rings form at radii defined by the intersections with the two exceptional arcs. The coloring in (b) and (d) illustrates the mode polarization in form of the $S_3$ Stokes vector element for positive $\Delta_\mathrm{B}$. Simulation parameters: $\{\alpha,g_\mathrm{c},g_\mathrm{x}\} = \{0.1,0.2,0\}~\mathrm{meV\upmu m}^2$}
  \label{fig:umbilic}
\end{figure}

As also discussed in Ref.~\cite{xu2017weyl} the ER only exists in the eigenvalues of the matrix in Eq.~(\ref{eq:eigenvalue}) if the two modes are degenerate in energy ($\Delta=0$) over all $\bm{k}$ in the vicinity of the ER [Fig.~\ref{fig:tilted_ring}(a)]. Here, the eigenvalue problem~(\ref{eq:eigenvalue}) exibits passive $\mathcal{PT}$-symmetry (the system is $\mathcal{PT}$-symmetric up to a shift in eigenvalues), without detuning resulting in codimension of one for our exceptional points (which therefore form a gapless ring)~\cite{PhysRevB.107.224306}. Introducing detuning breaks the $\mathcal{PT}$-symmetry and increases the exceptional points' codimension to two~\cite{PhysRevResearch.6.023205, privateKunst}, resulting in isolated exceptional points as reported in Ref.~\cite{xu2017weyl}. In microcavities, however, different physical effects (e.g., Zeeman and Rashba-Dresselhaus splitting) can induce different forms of, even $k$-dependent, detuning, which adds dimensions to the parameter space that can be accessed and explored. 

To show that the ER is not destroyed with finite detuning but instead can be recovered in a higher-dimensional parameter space, we extend the linear eigenvalue problem~(\ref{eq:eigenvalue}) by finite ($k$-dependent) detuning (with Zeeman and Rashba-Dresselhaus splittings as our specific examples of physical causes of this detuning). To this end $\Delta_\pm=\pm\Delta_\mathrm{B}\pm\Delta_\mathrm{RB}(k_\parallel)$ is added to the first (+) and second (-) diagonal elements in Eq.~(\ref{eq:eigenvalue}). Here $\Delta_\mathrm{B}$ describes the magnetic field induced Zeeman splitting and $\Delta_\mathrm{RB} = 2\alpha_\mathrm{RD}k_\parallel$ the Rashba-Dresselhaus splitting~\cite{rechcinska2019engineering, PhysRevApplied.17.014041, Li2022} with splitting strength $\alpha_\mathrm{RD}$. In specifically designed semiconductor microcavities $\alpha_\mathrm{RD}$ is adjustable by an external electric field acting on an optically anisotropic liquid crystal inside the cavity \cite{li2022switching,rechcinska2019engineering,PhysRevApplied.17.014041,Sedov2024}. Here we denote the Rashba-Dresselhaus splitting direction in $k$-space as $k_\parallel$ and the direction perpendicular to the splitting as $k_\perp$. We emphasize that the exact mathematical form and physical origin of detuning is not important for our results, those transfer to any form of detuning breaking $k$-isotropy in a similar fashion. 

By introducing the Zeeman shift $\Delta_\mathrm{B}$ to the circularly polarized modes via a magnetic field (which may be enabled with an optically active matter component inside the resonator), the two modes become non-degenerate and the ER vanishes [Fig.~\ref{fig:tilted_ring}(b)]. As discussed, in certain structures and materials, an electric field induced Rashba-Dresselhaus splitting $\Delta_\mathrm{RD}$ along $k_\mathrm{\parallel}$ may be used to lift the degenercy of TE and TM modes in $k_\mathrm{\parallel}$-direction \cite{Li2022}. Thus, in the $k_x$-$k_y$ plane only the exceptional point pair along $k_\mathrm{\perp}=0$ remains visible [Fig.~\ref{fig:tilted_ring}(c)]. The combination of these two effects lets us see the two exceptional points on the ring positioned where $k_\mathrm{\perp}\neq0$ with $\Delta_\mathrm{RD}(k_\parallel)-\Delta_\mathrm{B} = 0$ [Fig.~\ref{fig:tilted_ring}(d)]. With this approach, we can validate that the ER structure stays intact in higher-dimensional parameter space ($k_x,k_y,\Delta_\mathrm{B},\alpha_\mathrm{RD}$) even for finite detuning. In Fig.~\ref{fig:tilted_ring}(e) the ER is shown in this 4D parameter space, the dots mark the location of exceptional points and thus the points for which the Zeeman shift compensates the Rashba-Dresselhaus splitting (note that the ERs form a continuous structure; for discretization reasons this is not fully resolved in the plot). Here, the tilt of the individual representations of the ER is induced by the respective Rashba-Dresselhaus splitting marked in color. While we are at present unable to calculate the quasi-Chern intergrals for the tilted rings, we note that we expect each colored ring to possess a $1^\mathrm{st}$ quasi-Chern intergral of $C^{RR} = \pm1$. Note that the radius of the ER projected onto the $k_x-k_y$ plane remains unchanged by these extensions. The detuning strengths chosen here and below are in agreement with possible Zeeman and Rashba-Dresselhaus splittings measured in microcavity systems with the perovskite semicondcutor $\mathrm{CsPbBr_3}$ as active medium~\cite{Li2022,yakovlev2024exciton}. This observation shows that in higher dimensional parameter space the ER is part of an exceptional surface and the ring is resilient against detuning. Increasing both parameter space dimensionality and codimension of the exceptional points explains the occurance of the exceptional surface, as discussed in~\cite{privateKunst}. While detuning values might not be fully tunable in experiments, this result illustrates that the linear ER can be (partially) recovered even in $k$-anisotropic samples. Note that this analysis extends to any similar form of detuning or anisotropy in this and other physical systems. Moreover, the tilt of the ER permits new ways for tracing trajectories in higher-order dimensional space, such as Solomon's knots around the ER. This is promising for both mode switching applications as well as investigation of topological characteristics. In the following we will extend this investigation into the nonlinear regime, to show that the elliptic umbilic leads to even richer (spectral and band) topologies and exceptional point structures in the nonlinear regime.

\begin{figure}[!tb]
  \centering
   \includegraphics[width=1.0\columnwidth]{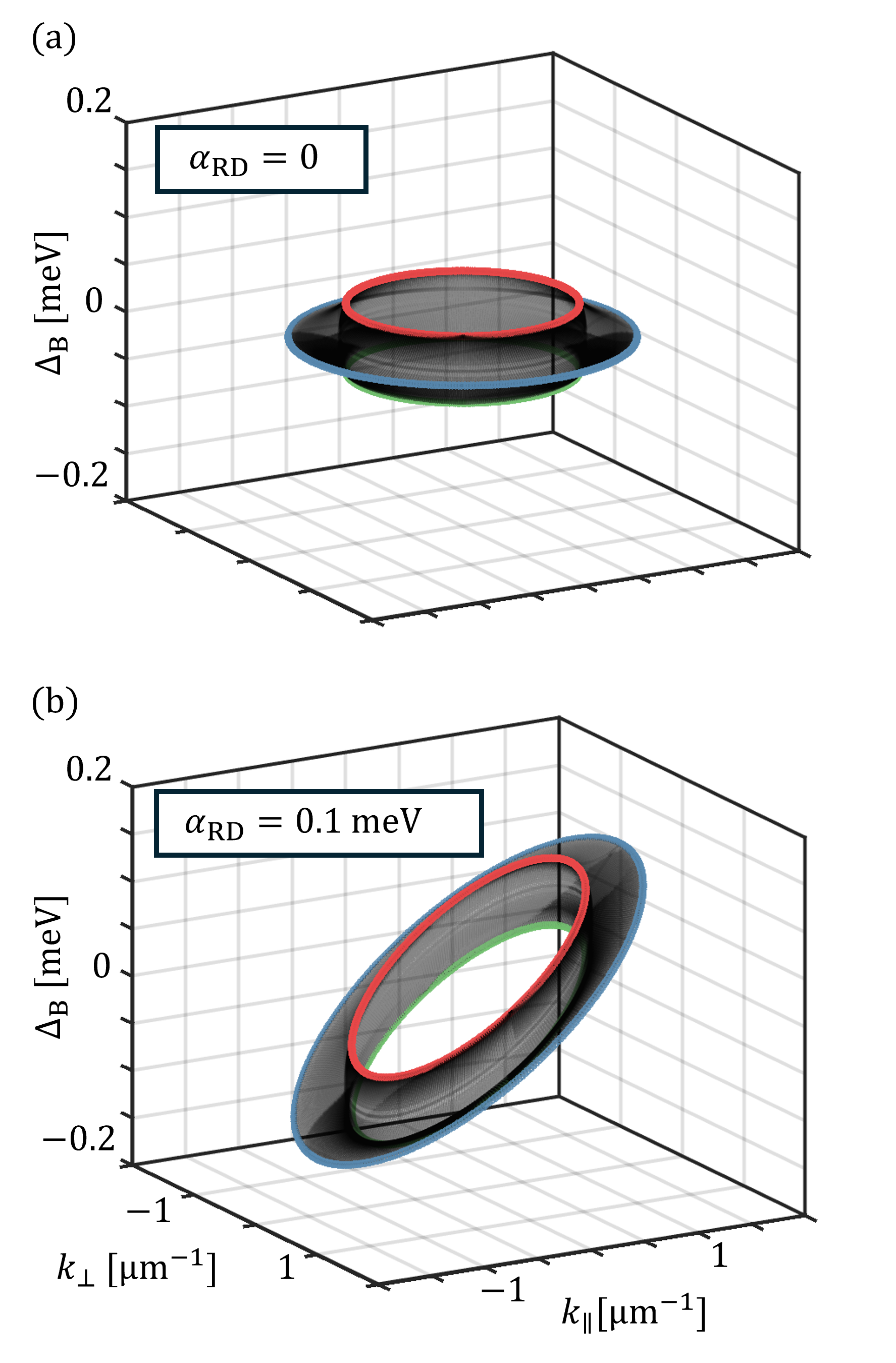}
  \caption{\textbf{Topological structure of nonlinear exceptional rings in 4D parameter space.} (a) With Zeeman splitting, a single linear exceptional ring at $\Delta_\mathrm{B}=0$, in the nonlinear regime expands into a quasi-torus of exceptional nexuses (colored) connected by surfaces of exceptional arcs (black). (b) Rashba-Dresselhaus splitting in $k_\parallel$-direction leads to a tilt of the quasi-torus along $k_\parallel$ (shown for $\alpha_\mathrm{RD}=0.1~\mathrm{meV}$). Simulation parameters: $\{\alpha,g_\mathrm{c},g_\mathrm{x}\} = \{0.1,0.2,0\}~\mathrm{meV\upmu m}^2$}
  \label{fig:torus}
\end{figure}

We use the solutions of the nonlinear eigenvalue problem~(\ref{eq:eigenvalue_nonlinear}) with finite detuning to investigate topological features by applying Rashba-Dresselhaus and Zeeman splitting to the system. In contrast to the linear ER, the elliptic umbilic allows the formation of new ER constellations for finite detuning $\Delta$. In Fig.~\ref{fig:umbilic} the complex-valued energy spectrum and central cuts are displayed for a finite Zeeman splitting of $\Delta_\mathrm{B} = 12.5~\mathrm{\upmu eV}$. The inset in Fig.~\ref{fig:umbilic}(b) shows the umbilic cut at $\alpha = 0.1~\mathrm{meV\upmu m}^2$ in turquoise together with the chosen Zeeman splitting in orange. The intersections mark the visible ERs for the given setup. Here, two rings are formed by points on the exceptional arcs. Notably, reversing the direction of the Zeeman-induced energy shift inverts the polarization in the complex-valued energy spectrum. We note that similar energy spectra can be achieved for stronger Zeeman splitting if the nonlinearity is increased simultanously (to increase the size of the umbilic in the regime where its structure is valid). A relatively weak nonlinearity was selected to limit the blueshift, ensuring the resulting structures remain within an easily visualizable energy range. These findings highlight, that while they both feature identical 
quasi-Chern integrals, the neighborhood topology of linear and nonlinear ERs is drastically different, as detuning is introduced.

Building on the understanding that the linear ER is tilted in parameter space for $k$-anisotropy (in this case explored through Rashba-Dresselhaus and Zeeman splitting [see Fig.~\ref{fig:tilted_ring}(e)]), we now examine how this affects the arrangement of umbilic cross-sections in the nonlinear regime. While the ER vanishes in the linear regime for finite Zeeman and vanishing Rashba-Dresselhaus splitting ($\Delta_\mathrm{B}\neq0$, $\Delta_\mathrm{RD}=0$) [see Fig.~\ref{fig:tilted_ring}(b)], in the nonlinear regime the elliptic umbilc profile forms a donut-like quasi-torus in parameter space, as displayed in Fig.~\ref{fig:torus}(a). Here, the colored rings and the connecting black surfaces form EXRs and surfaces of exceptional arcs, respectively. To our knowledge, this represents the first theoretical observation of an exceptional nexus ring, emerging from the intersection of two exceptional arc surfaces. This structure extends the concept of an exceptional nexus (previously point-like) into higher dimensions. The size of the quasi-torus scales with nonlinearity. These results demonstrate that in contrast to linear ERs, finite detuning induces a change in topology in the nonlinear regime. In case of finite Zeeman and Rashba-Dresselhaus splitting ($\Delta_\mathrm{B}\neq0$, $\Delta_\mathrm{RD}\neq0$), the quasi-torus is tilted in parameter space in $k_\mathrm{\parallel}$-direction. In Fig.~\ref{fig:torus}(b) such a tilted quasi-torus is displayed for Rashba-Dresselhaus splitting of $\alpha_\mathrm{RD}=0.1~\mathrm{meV}$. While the tilt of the linear ER allows for new loops such as the Solomon's knot, the topology of nonlinear ERs would allow for even more complex loops. Analogously to the linear ER, stronger Rashba-Dresselhaus splitting leads to a stronger tilt of the quasi-torus. These findings demonstrate the robustness and versatility of exceptional points and ER singularity with elliptic umbilic structure, even with variations of the simplest model. The highly tunable nature of energy spectra associated with exceptional points on the umbilic presents exciting opportunities for mode-switching applications, where different trajectories on the quasi-torus can be strategically exploited and which represents an exciting avenue for future investigations.

\begin{figure}[!tb]
  \centering
   \includegraphics[width=1.0\columnwidth]{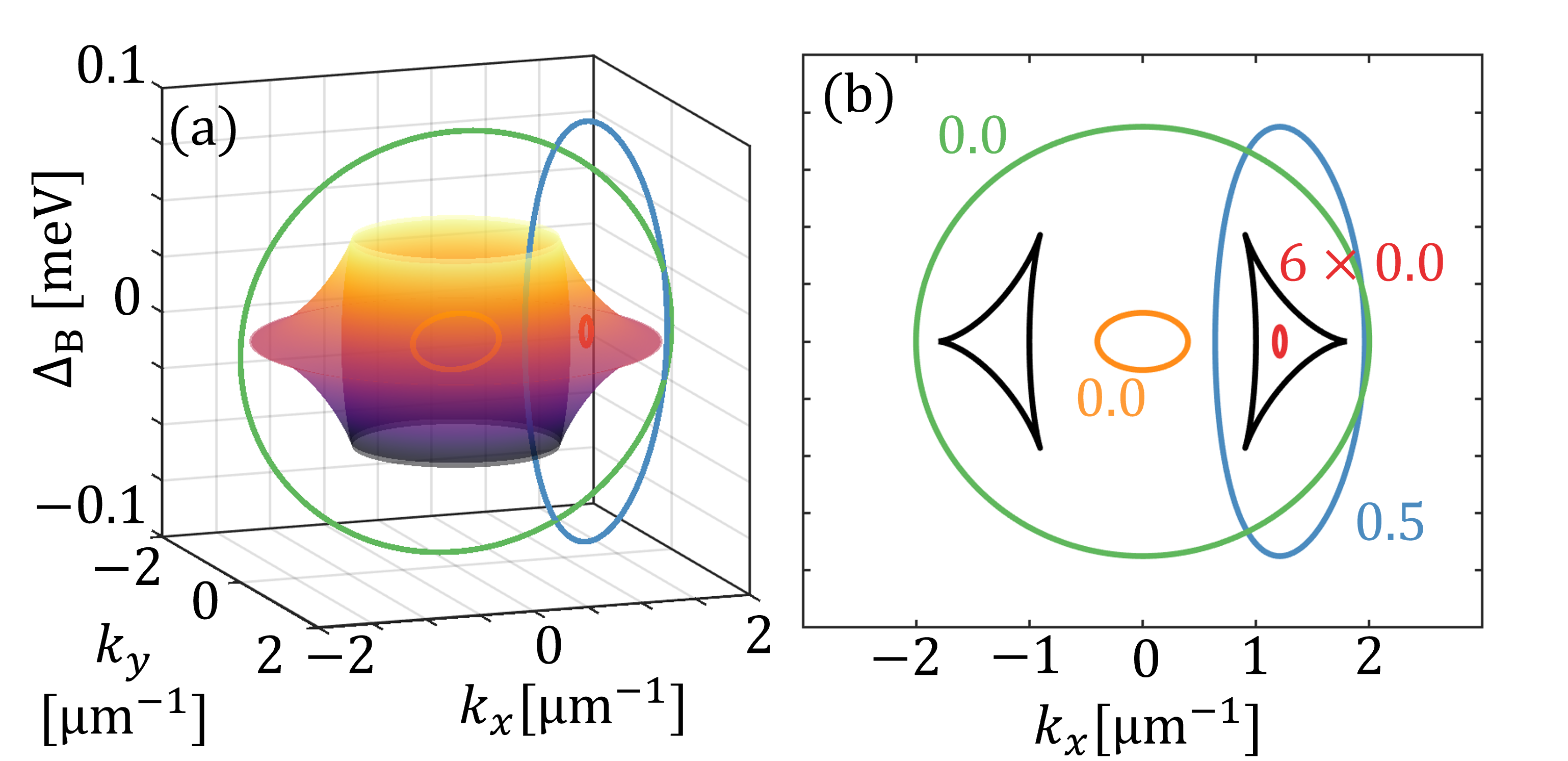}
  \caption{\textbf{Vorticity of nonlinear exceptional quasi torus.} (a) Full and (b) central cut of exceptional quasi torus arising from an exceptional ring in the nonlinear regime. Closed loops as  highlighted are used to calculate the vorticity of the torus. The quasi-torus carries a vorticity of $\nu_{ul} = 0.5$, while the bands inside the torus carry no vorticity. Simulation parameters $\{\alpha,g_\mathrm{c},g_\mathrm{x}\} = \{0.15,0.3,0\}~\mathrm{meV\upmu m}^2$ were used.}
  \label{fig:vorticity}
\end{figure}

Next we extend our analysis of vortictiy $\nu$ and Berry phase $\gamma^{a b}_\mathrm{Berry, n}$ to the exceptional quasi torus emerging from the ER when entering the nonlinear regime. To this end, we evaluate Eq.~(\ref{eq:vorticity}) and Eq.~(\ref{eq:berry}) numerically using the solutions of the nonlinear eigenvalue problem. In Fig.~\ref{fig:vorticity}(a) the exceptional quasi torus is shown together with multiple closed loops in the $\Delta-k_x$ plane used to calculate the vorticity. Note that the blue ring represents a Hopf-link configuration for the linear ER, while the red ring would not form a Hopf link with the linear ER, since it is fully enclosed by the umbilic and thus would run outside the linear ER. If the Hopf link is large enough to enclose all exceptional points on the umbilic the vorticity amounts to $\nu_{ul} = 0.5$ for the nonlinear exceptional quasi torus. Any loop that does not encircle the torus (orange and green in Fig.~\ref{fig:vorticity}) amounts to vorticities of $\nu_{ul} =0$. Lastly, all unique permutations of bands inside the quasi-torus (6 in total) amount to vorticities of $\nu_{ul} =0$ as indicated in the figure. We note that depending on the radius chosen, numerical inaccuracies of about $1\%$ may occur.

These results show that certain topological invariants (here the vorticity) can be preserved even if nonlinearity is introduced into the system. In the present case, the second-order ER in the linear case is a point-like singularity (e.g., in the $k_x -\Delta_B$ plane) and
obstruction 
to Stokes theorem and therefore yields the expected vorticity  of 0.5. In contrast, in the nonlinear regime the cross section of the torus (again, e.g., in the $k_x -\Delta_B$ plane) is an extended object with singularities on the surface of the triangular cross section. Nevertheless, we find from our numerical studies that this extended object in the nonlinear regime has the same vorticity as the isolated point-like singularity in the linear regime.

Numerical evaluation of the Berry phase $\gamma^{a b}_\mathrm{Berry, n}$ in the $k_x-k_y$ plane for EAR and EXR in the nonlinear regime leads to $\gamma^{R R}_\mathrm{Berry, \mathrm{EAR}} = 0$ and $\gamma^{R R}_\mathrm{Berry, \mathrm{EXR}} \in \{0,2 \pi\}$. Here the indices "EAR" and "EXR" indicate that we are using bands which are forming the corresponding exceptional ring. Since the EXR features reduced self-orthogonality we can calculate the Berry phase for mixed left and right eigenvectors. Then the Berry phase amounts to $\gamma^{R L}_\mathrm{Berry, EXR} \in \{0,2\pi\}$.

\section{Conclusion}

We have demonstrated that exceptional rings (ERs) naturally emerge inside the in-plane dispersion of planar optical resonators with TE-TM splitting and circular dichroism (leading to different losses in the oppositely circularly polarized resonator modes). These ERs can even be recovered in $k$-space-anisotropic systems, where adjusting mode detuning can be used to recover them in higher-dimensional parameter space. Introducing a Kerr-type medium (or saturable gain) splits the linear ER into mutliple ERs forming an elliptic umbilic in nonlinear parameter space. Here, $3^\mathrm{rd}$-order exceptional nexus rings and $2^\mathrm{nd}$-order exceptional arc rings form concentrically in $k$-space. We performed a rigorous analysis of the nonlinear ERs, showing  diverse perturbation-response enhancement at different points on the umbilic singularity structure, and intriguing topological features. We show that a linear ER in higher-dimensional parameter space splits into a complex quasi-toroidal structure of ERs, making nonlinear ERs a promising candidate for mode-switching applications. Since a wide range of nonlinear non-Hermitian Hamiltonians feature exceptional points or contours that are organization point of an elliptic umbilic \cite{kwong2025universal}, our results on topology and perturbation response can be applied to a large group of physical systems hosting exceptional points that are subject to real (or imaginary) valued smooth nonlinearities. Examples are Kerr-type nonlinearity or saturable gain, but also higher-order nonlinearities and nonlinearities of different functional form. Our work lays the foundation for a range of further research questions: the experimental realization of ERs in planar optical resonators and exciton polariton systems, the comprehensive investigation of the sensitivity properties (such as signal-to-noise ratio) of exceptional points on the elliptic umbilic, taking into account noise-nonlinearity interactions, and the analysis of topological invariants in other nonlinear and non-Hermitian systems. Moreover, we also lay the foundation for the detailed investigation of the relation of the nonlinear eigenvalue problem and the system dynamics for different possible excitation setups.

\begin{acknowledgments}
    The authors gratefully acknowledge financial support for the Paderborn groups, by the Deutsche Forschungsgemeinschaft (DFG, German Research Foundation) through Grant No.~467358803 and the transregional collaborative research center TRR 142 (Projects A04 and C10, Grant No. 231447078), and for the Arizona group from the US National Science Foundation (NSF) under Grant No. DMR-1839570. The authors gratefully acknowledge the computing time provided to them on the high-performance computer Noctua 2 at the NHR Center PC$^2$ under project hpc-prf-hermep. LA and JS acknowledge support through the QuantERA project QuCABOoSE. We thank Jan Wiersig and Julius Kullig for fruitful discussions, specifically on the aspects of nonlinearity and signal enhancement. We further thank Anton Montag and Flore Kunst for a fruitful exchange on the role that symmetries of our system play in the linear regime.
\end{acknowledgments}

\appendix
\section{Exceptional ring generation and control by off-resonant pumping}\label{append:gp_extended}
In this section, as a specific example, we discuss the option of non-resonant pumping in a system with active (semiconductor) medium, as a source for tunable non-Hermiticity. Given that pump polarization is preserved to some degree, circular-dichroism can be tuned optically in this case \cite{waldherr2018observation}. Thus, the system could be tuned in and out of the non-Hermitian regime, meaning switching the exceptional ring on and off, as well as tuning its size. In this case the spinor Gross-Pitaevskii equations read \cite{PhysRevLett.109.036404}:
\begin{align}
    i\hbar\partial_t\psi_\pm=\left(H_\pm+g_\mathrm{r}n_\pm + i\Gamma_\pm\right) \psi_\pm +J_\pm\psi_{\mp}+R_\pm,
\label{eq:PULSE_psi_n}
\end{align}
\begin{align}
    \partial_t n = P_{\pm} - (\gamma_\mathrm{r}+r|\psi_\pm|^2)n_\pm.
\label{eq:PULSE_res}
\end{align}
Here, the field $\psi_\pm=\psi_\pm(\mathbf{r},t)$ is coupled to an incoherent excitation reservoir $n_\pm=n_\pm(\mathbf{r},t)$ driven by continous-wave pump $P_\pm=P_\pm(\mathbf{r},t)$ in real-space $\mathbf{r}=(x,y)$. The excitation reservoir exhibits a loss rate $\gamma_\mathrm{r}$ and its scattering into the field is determined by the stimulated scattering rate $r$. The loss term now holds $\Gamma_\pm = \frac{\hbar}{2}[rn_\pm-\gamma_\pm]$. The pump is below threshold to avoid condensation, however, in prinicple the eigenvalue problem~(\ref{eq:eigenvalue}) shows an exceptional ring in the gain regime as well. For the calculations with the extended model in equations~(\ref{eq:PULSE_psi_n},\ref{eq:PULSE_res}) we find the results (not shown) to be in good agreement with the results where the difference in loss rate was introduced by hand in Fig.~2(a) above, with minor differences in line profiles. Note that to realize the exceptional ring in this case the reservoir-induced blueshift $g_r n_\pm$ resulting from polariton-exciton interaction has to be compensated (for example by Zeeman splitting) to preserve mode degeneracy. The results displayed in Figure~\ref{fig:2} can be reproduced with the polarized reservoir as source for circular dichroism.

\section{Analytical solution of the nonlinear eigenvalue problem at zero detuning}\label{append:math}

In this section we solve the nonlinear eigenvalue problem~(\ref{eq:eigenvalue_nonlinear}) without detuning.
First we make the valid assumption $\Gamma_1,\Gamma_2,g_\mathrm{c},g_\mathrm{x},k_x,k_y,\Delta_\mathrm{LT}\in\mathbb R$. For a better visualization of the eigenvalues, we substract the parabolic term, $\mu^\prime=\mu-\frac{\hbar^2 \bm{k}^2}{2m}$, from the dispersion in the following. We write the eigenvalue problem as:
\begin{equation}
        \begin{aligned}
        	\begin{bmatrix}
                H
                &
                \beta
                \\
                \beta
                &
                -H
        	\end{bmatrix}
        	\begin{bmatrix}
        		\psi_+ \\ \psi_-^\prime
        	\end{bmatrix}
            =
            (\mu^\prime-\mu_0)
        	\begin{bmatrix}
        		\psi_+ \\ \psi_-^\prime
        	\end{bmatrix}
        \end{aligned}
\label{eq:eigenvalue_rewritten}
\end{equation}
with $H = i\frac{\Gamma_1-\Gamma_2}{2}+\frac{g_c-g_x}{2}(|\psi_+|^2-|\psi_-^\prime|^2)$, 
$\mu_0 =  \frac{i(\Gamma_1+\Gamma_2)+g_c+g_x}{2}$, $\psi_-^\prime=\psi_- e^{-2\mathrm{i}\varphi}$, $\varphi =  \mathrm{arg}(k_x+\mathrm{i}k_y)$, and $\beta=-\Delta_\mathrm{LT}|\bm{k}|^2$. After some rewriting this leads to the following nonlinear eigenvalue equation: 
\begin{equation}
\begin{aligned}
    {}&
    \begin{bmatrix}
        \alpha(|a_0|^2-|a_1|^2)+i\gamma
        &
        \beta
        \\
        \beta
        &
        -\alpha(|a_0|^2-|a_1|^2)-i\gamma
    \end{bmatrix}
    \begin{bmatrix}
        a_0 \\ a_1
    \end{bmatrix}
    \\
     {}&=
   (\mu^\prime-\mu_0)
    \begin{bmatrix}
        a_0 \\ a_1
    \end{bmatrix}
    \label{eq:separability}
\end{aligned}
\end{equation}
with $\gamma=\sfrac{|\Gamma_1-\Gamma_2|}{2}$ and $\alpha=\sfrac{(g_\mathrm{c}-g_\mathrm{x})}{2}$ as also introduced in the main part of the text and $a_0=\psi_+$ and $a_1=\psi_-^\prime$. We note that Eq~(\ref{eq:separability}) has the same form as the separability eigenvalue equation known in a more general context in entanglement theory~\cite{PhysRevA.79.022318,PhysRevLett.111.110503} as elaborated in Appendix~\ref{app:sperling}. Using the solution ansatz
\begin{equation}
    |a\rangle=\frac{1}{\sqrt{1+|w|^2}}
    \begin{bmatrix}
        1 \\ w
    \end{bmatrix},
    \text{ with }
    w\in\mathbb C\setminus\{0\}
\end{equation}
we find the equation
\begin{equation}
    0=\frac{1}{2}\left(w-\frac{1}{w}\right)
    +\left(
        \frac{\alpha}{\beta}\frac{\frac{1}{|w|}-|w|}{\frac{1}{|w|}+|w|}+i\frac{\gamma}{\beta}
    \right),
\end{equation}
for $\beta\neq 0$. The ansatz $w=\exp(\eta+i\vartheta)$ with $\eta,\vartheta\in\mathbb R$ leads to
\begin{equation}
\begin{aligned}
    0
    &=\left(
        \sinh(\eta)\cos(\vartheta)
        -\frac{\alpha}{\beta}\frac{\sinh(\eta)}{\cosh(\eta)}
    \right)\nonumber\\
    &+i\left(
    \cosh(\eta)\sin(\vartheta)
    +\frac{\gamma}{\beta}
    \right)\,.
    \end{aligned}
\end{equation}
Solving this equation yields the solutions, determining the eigenvectors $|a\rangle$ with
\[
w =
\begin{cases}
\pm\sqrt{1-\frac{\gamma^2}{\beta^2}}-i\frac{\gamma}{\beta} & \text{for } |\gamma|\leq |\beta| \\ \frac{\sqrt{\alpha^2+\gamma^2}\pm\sqrt{\alpha^2+\gamma^2-\beta^2}}{\beta}\frac{\alpha-i\gamma}{\sqrt{\alpha^2+\gamma^2}} & \text{for } |\beta|\leq\sqrt{\alpha^2+\gamma^2}
\end{cases}
\]
In each of the two regions indicated in this equation we have two solutions, as can be seen from the $\pm$ in front of the square roots. However, there is a region where the two regions overlap, and in this overlap region we therefore have a total of four solutions. In his case, the eigenvalues can also be given in closed form. For non-zero $\beta$, we have
\begin{equation}
\mu' = \mu_0\pm \sqrt{\beta^2 - \gamma^2}
\end{equation}
for $|\beta| \ge |\gamma|$   (region B and C in Fig. \ref{fig:4}(a)), and
\begin{equation}
\mu' = \mu_0+\alpha \pm i \gamma \sqrt{1 - \frac{\beta^2}{\alpha^2 +  \gamma^2} }
\end{equation}
for $|\beta| \le \sqrt{\alpha^2 + \gamma^2}$ (region A and B in Fig. \ref{fig:4}(a)). 
This again shows that we have two solutions in region A and C, and four solutions in region B. Note, that the solution is not well-defined for $\beta = 0$, i.e. at $k=0$ and $\gamma \neq 0$. The resulting eigenstates have arbitrary phases. In the nonlinear regime, $\alpha\neq 0$, all four solutions of $w$ apply, leading to multistability. Note that the solutions are normalized to $|\psi_+|^2+|\psi_-|^2=N^2$, where $N>0$. For $N\neq1$ the rewritten eigenvalue equation~(\ref{eq:eigenvalue_rewritten}) can be rescaled by substituting $\tilde a_0=a_0/N$ and $\tilde a_1=a_1/N$, now obeying $|\tilde a_0|^2+|\tilde a_1|^2=1$. 

\section{Separability eigenvalue equations}\label{app:sperling}
Taking a more general view, the eigenvalue problem~(\ref{eq:eigenvalue_rewritten}) can be cast into a form known from quantum information theory in the context of separability eigenvalue equations \cite{PhysRevLett.111.110503}. This is due to the fact that nonlinearities of vector-valued quantities can be expressed in terms of tensor products, such as $|a\rangle^{\otimes 2}=|a\rangle\otimes|a\rangle$ representing a square of the vector $|a\rangle$ in the tensor algebra~\cite{lang2012algebra}. 

In this section we derive the separability eigenvalue equation~(\ref{eq:separability}). We start deriving the operator for the separability eigenvalue equation~\cite{PhysRevA.79.022318,PhysRevLett.111.110503}
\begin{equation}
    \hat L_a |a\rangle=\lambda |a\rangle.
    \label{eq:sepa_ansatz}
\end{equation}
with the partially reduced operator
\begin{equation}
        \hat L_a=(\langle a|\otimes\hat 1)\hat L(|a\rangle\otimes\hat 1)
\end{equation}
where $\hat 1$ denotes the identity and $\otimes$ the respective tensor product.
Note that $\hat L_a$ can be also understood in terms of the computational basis $\{|i\rangle:i=0,1\}$. That is, the matrix elements of the partially reduced operator are $\langle i|\hat L_a|j\rangle=\langle a,i|\hat L|a,j\rangle$, with the common tensor-product notation $|a\rangle\otimes|b\rangle=|a,b\rangle$. The general form of exchange-symmetric $\hat L$ expanded into Pauli matrices reads 
\begin{equation}
    \hat L=\sum_{w,w'\in\{0,x,y,z\}} L_{w,w'}\hat \sigma_w\otimes\hat \sigma_{w'},
\end{equation}
where $\hat L_{w,w'}=\hat L_{w',w}$. This leads to the specific expression
\begin{equation}
    \hat L=
    \alpha\hat\sigma_z\otimes\hat\sigma_z
    +(\beta\hat\sigma_x+i\gamma\hat\sigma_z)\otimes\hat\sigma_0
    +\hat\sigma_0\otimes(\beta\hat\sigma_x+i\gamma\hat\sigma_z),
\end{equation}
where $\alpha,\beta,\gamma\in\mathbb R$. The resulting partially reduced operator takes the form
\begin{equation}
    \hat L_a
    =
    \alpha\langle a|\hat\sigma_z|a\rangle\hat\sigma_z
    +\beta\hat\sigma_x+i\gamma\hat\sigma_z,
\end{equation}
with $|a\rangle=[a_0~a_1]^\mathrm{T}$. When substituted into Eq.~(\ref{eq:sepa_ansatz}), this results in an eigenvalue problem of the form of the rewritten eigenvalue problem~(\ref{eq:separability}).

\section{Stationary solutions of the nonlinear system for resonant driving}\label{append:time}
For resonant excitation of the  the solutions of our nonlinear eigenvalue problem those are also stationary solutions of the time-dependent system. For Eq.~(\ref{eq:sepa_ansatz}) the respective time-dependent nonlinear Schrödinger equation with excitation $\ket{c(t)}$ is given by
\begin{equation}
    \begin{aligned}
    \partial_t\ket{a} = \hat{L}_a\ket{a}+\ket{c(t)}.
    \end{aligned}
\end{equation}
For continuous-wave resonant driving of an eigenmode with $\ket{c(t)} = \ket{c} = \nu\ket{a}$ and $\nu\in\mathbb{R}$, the resulting steady state obeys:
\begin{equation}
    \begin{aligned}
    \hat{L}_a\ket{a}+\nu\ket{a} = \lambda\ket{a}.
    \end{aligned}
\end{equation}
This can be recast into
\begin{equation}
    \begin{aligned}
    \hat{L}_a\ket{a}= \tilde{\lambda}\ket{a}.
    \end{aligned}
\end{equation}
with $\tilde{\lambda} = (\lambda-\nu)$, which is of the same form as Eq.~(\ref{eq:sepa_ansatz}). Hence, for resonant continuous-wave driving the eigenmodes found are also steady states of the time-dependent system with shifted eigenenergies.

\section{ER structures in nonlinear spatio-temporal system dynamics}\label{append:NSE}
Here we explicitly implement a method to analyze the EAR and EXR in an exciton-polariton system. We simulate the complex spatio-temporal dynamics of the respective cavity fields by solving the coherent driven spinor nonlinear Schrödinger equations with circular dichroism:

\begin{eqnarray}
\label{eq:PULSE_psi_nl}
    i\hbar\partial_t\psi_\pm
    = \biggl(& - &\frac{\hbar^2\nabla^2}{2m}+g_\mathrm{c}|\psi_\pm|^2+g_\mathrm{x}|\psi_{\mp}|^2 +  i\Gamma_\pm \biggr) \psi_\pm\nonumber\\&+&J_\pm\psi_{\mp}+ R_{\pm,\ket{a}}\,.
\end{eqnarray}

In contrast to the eigenvalue problem in Eq.~(\ref{eq:eigenvalue_nonlinear}) this equation describes the field dynamics in two-dimensional real-space with $\psi_\pm=\psi_\pm(\mathbf{r} = (x,y),t)$. For visualization purposes we choose $\Gamma_\pm=\{0.1;0.05\}~\mathrm{meV}$. $g_\mathrm{c}=6~\mathrm{\upmu eV\upmu m^2}$ and $g_\mathrm{x}=-0.1g_\mathrm{c}$ are polariton-polariton interaction strengths within and between circular polarization components. These parameter values are in agreement with lifetime and blueshift measurements in GaAs microcavities~\cite{PhysRevLett.118.016602, PhysRevB.100.035306}. We note that in the nonlinear regime, a wide range of observables and excitation setups can be used to further investigate or utilize the ER structure: among others are pump-probe setups or noise-induced nonlinear response.

Here we chose a specific nonlinear optical setup to further illustrate the results found in the nonlinear eigenvalue problem~(\ref{eq:eigenvalue_nonlinear}). For simplicity, we use a fixed pump amplitude and polarization state and scan the relevant parameter space ($k_0,\omega_0$ of the continuous and plane-wave coherent pump) by evolving Eq.~(\ref{eq:PULSE_psi_nl}) in time and computing the stationary solutions (details below). We demonstrate that this approach shows the ER features in the nonlinear regime discussed in the main part (note that blueshifts are induced by the pump that in principle may be different for modes with different imaginary parts). We note that an alternative approach would be to fix the nonlinearity by keeping the steady-state density at a constant value for every parameter pair pump frequency $\omega_0$ and pump momentum $k_0$. This, however, would result in quite the  procedure as it would require optimization of pump parameters at each frequency value to map out the resonance features. We use the following pump profile:
\begin{align}
    R_{\pm,\ket{a}} = R_0\cdot\mathrm{exp}\left(\frac{-\mathbf{r}^2}{2\sigma^2}\right)\mathrm{exp}\left(ik_0x+i\omega_0 t\right)\ket{a}.
\end{align}
Here, $R_0 = 0.5~\mathrm{\upmu eV\upmu m^{-1}}$ is the pump amplitude used. $\sigma = 25~\mathrm{\upmu m}$ denotes the width of the Gaussian real-space profile of the pump, which is large enough such that the pump drives a narrow range in $(k,\omega)$-space around $k_0$ and $\omega_0$. Since the system is rotationally symmetric in $k$-space, we focus on the $k_\mathrm{y}=0$ plane. $\ket{a}$ denote the eigenvectors of the nonlinear eigenvalue problem, Eq.~(\ref{eq:eigenvalue_nonlinear}), at a given $k_0$.

\begin{figure}[!h]
  \centering
    \includegraphics[width=1.0\columnwidth]{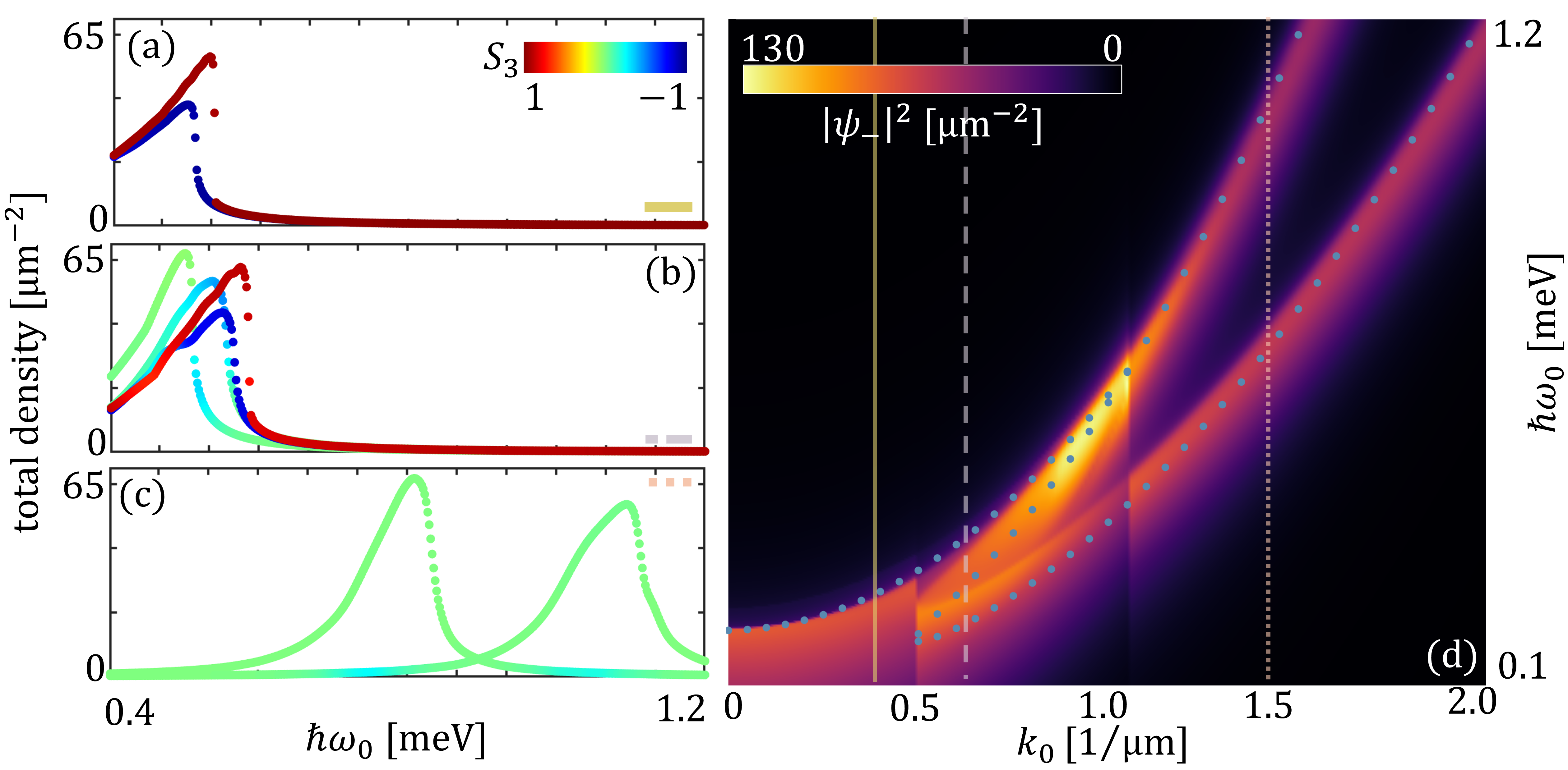}
  \caption{\textbf{Exceptional rings in nonlinear system response.} The plots on the left show resonance peaks over pump energy, scanned along the vertical lines marked in (d). (a) For excitation inside the inner ring two circularly polarized modes are observed [also indicated by the solid line in (d)], (b) between the concentrical rings we observe a total of four peaks (two circularly polarized and two almost linearly polarized ones) [dashed line in (d)]. Finite peak width leads to modified polarization character of the upper linearly polarized peak. (c) Outside the outer ring two linearly polarized peaks are found [dotted line in (d)]. (d) Momentum-frequency scan of the resonance structure. The real parts of nonlinear eigenvalues (without the spectral shift induced by driving) are included as a guide to the eye as dahsed lines. The discontinuities visible in the profile mark the two exceptional rings at $k_0 = k_x=0.5$ and $k_0 = k_x\approx1.05$. The observed resonance features (including number and polarization) in the driven nonlinear dynamics agree with the nonlinear eigenvalue problem.}
  \label{fig:10}
\end{figure}

Keeping $k_0$ fixed we sweep $\omega_0$ for all $\ket{a}$ found at $k_0$. This leaves us with two (four) resonance peaks in segment A and C (segment B) of Fig.~\ref{fig:3}(c,d). In Fig.~\ref{fig:10}(a-c) the resulting stationary total density is shown over the pump energy $\hbar\omega_0$, color coded according to the resulting polarization state of the stationary solutions at any given frequency. The polarization state is calculated at the global real-space peak density,  $\mathrm{max}\left(|\psi_\pm|^2\right)$.

The obtained resonance structures in segments A and C are in agreement with the solutions of our nonlinear eigenvalue problem: in Fig.~\ref{fig:10}(a) the stationary solutions possess circular polarization, with one resonance peak being significantly smaller than the other due to the different imaginary parts inside the EXR [cf. Fig.~\ref{fig:3}(c)]. In Fig.~\ref{fig:10}(c) the two peaks are linearly polarized as expected. The difference in height results from the different density of states, which is slightly lower on the steeper branch of the $k$-space dispersions. In Fig.~\ref{fig:10}(b) we observe the two circularly polarized solutions as expected. For the two linearly polarized solutions in the eigenvalue problem we observe that the upper one is slightly elliptically polarized in the time-domain calculations. This behavior is explained by the finite peak width and hence finite spectral overlap in the response of the dynamical system. Here, the linearly polarized solution overlaps with the circularly polarized solution, resulting in elliptical polarization as both are simultaneosuly driven to a certain extent. However, the results still clearly demonstrates the existance of the multistability and nonlinear ER structure discussed in the present manuscript. Eq.~(E1) includes a contact potential in configuration space, i.e. a scalar $g_c$ ($g_x$) describing the strength of the scattering processes at position $\mathbf{r}$. In this case, all four-wave mixing (FWM) or higher-order scattering events that fulfill wavevector conservation of the four wavevectors involved in the scattering will be allowed. However, if initially we have only one wavector ($k_0$) excited, i.e. the three waves on the rhs of Eq.~(E1) are plane waves at $k_0$, then the condition of wavevector conservation restricts the wavevector on the lhs of Eq.~(E1) to $k_0$. In other words, no new k-components with significant amplitudes will be created under these conditions (as confirmed by our calculations presented in Fig.~\ref{fig:10}). Hence, this case corresponds to the assumption used in Eq.~(3) of negligible coupling between different k-states, whose validity is explicitly demonstrated for the conditions used in our full numerical simulations in the present Appendix.

Next, we scan the nonlinear response with a large number of independent calculations in $k$-space. Resolving the resulting spectral features requires the calculation of about $3\times10^4$ time-evolutions on a large two-dimensional spatial grid. This numerically expensive exercise is made possible by using our PHOENIX code \cite{wingenbach2024phoenixpaderbornhighly} together with parallel for-loops. In Fig.~\ref{fig:10}(d) the total density in the $\psi_-$-component is depicted as an example (each data point in this plot is extracted from four [two] individual separate time evolutions for segements B [segment A and C]), showing good agreement with the energy spectra resulting from the nonlinear eigenvalue problem [eigenvalues are included as dots in Fig.~\ref{fig:10}(d)]. We emphasize that the shift induced by the resonant driving is visible (for fixed pump intensity this shift may be different for each individual calculation, which is most visible in Fig.~\ref{fig:10}(d) where the phase boundaries are crossed). However, the singularity structure from the nonlinear eigenvalue problem is clearly also recovered in the nonlinear driven system. We further note that the asymmetric shape of the resonance peaks that are clearly visible in  Fig.~\ref{fig:10} reflect the nonlinear nature of the system response and are the result of the specific excitation scenario with fixed intensity, here chosen for simplicity. Other excitation scenarios including off-resonant excitation  will be adressed in future work. 

\section{Generalization of signal enhancement and topological features to a large group of non-Hermitian Hamiltonians}\label{append:hamilton}
We would like to highlight that the elliptic umbilic neighborhood topology holds true for any exceptional point (group) in a non-Hermitian Hamiltonian of the form 
\begin{equation}
\begin{aligned}
\hat{H} =
\begin{bmatrix}
\Delta+i\gamma+\alpha F(\psi)
&
\beta
\\
\beta
&
-\Delta-i\gamma-\alpha F(\psi)
\end{bmatrix}
\label{eq:eigenvalue_general}
\end{aligned}
\end{equation}
where $F(\psi)$ is a smooth and real (or imaginary) valued function of the eigenvector with non-zero first derivative at the EP. A linear exceptional point described by the above Hamiltonian $\hat{H}$ for $\alpha = 0$ is an organizing point of the elliptic umbilic described above (exceptions and a more detailed discussion of generality are given in \cite{kwong2025universal}). Therefore, the signal enhancement and topological features of nonlinear exceptional points on the elliptic umbilic can be adapted to any of those linear exceptional points, making our results highly generic and applicable to many nonlinear and non-Hermitian systems.

\section{Topological properties for zero detuning}\label{append:topol}

In this appendix, we evaluate topological properties of the ER in the $k_x - k_y$ plane, i.e. for the case of zero detuning. We do this first for the case without nonlinearity, and then extend the discussion to non-zero nonlinearity. 

We define the closed-loop (sometimes called quantized) Berry phase of band $n$ calculated with left ($L$) or right ($R$) eigenvectors as
\begin{equation}
\gamma^{a b}_\mathrm{Berry, n} =   \oint_{\cal{L}} \bm{A}^{a b}_{n}(\bm{k}) \cdot d\bm{k}
\label{eq:berry}
\end{equation}
with superscripts $a$ and $b$ indicating the use of $L$ or $R$ eigenvectors, $\bm{A}^{a b}_{n}(\bm{k}) = i \langle \Psi^a_{n}( \bm{k}) | \nabla_k |   \Psi^{b}_{n}( \bm{k}) \rangle $ being the Berry connection, $|   \Psi^{R}_{n}( \bm{k}) \rangle   = [ \psi^n_+(\bm{k}), \psi^n_-(\bm{k}) ]^T $ the solution of the right eigenvalue problem (and similarly $|   \Psi^{L}_{n}( \bm{k}) \rangle$ being the solution of the left eigenvalue problem), and, again, the loop $\cal{L}$ is over one cycle.

Taking, for example, the upper band solution ($n=u$), we find the Berry phase for a closed loop in the $k_x -k_y$ plane with $\Delta = 0$, after analytically diagonalizing Eq.~\eqref{eq:eigenvalue_rewritten},
\begin{equation}
\gamma^{R R}_\mathrm{Berry, u} =   2 \pi \, {\rm mod}  \, 2 \pi
\end{equation}
for $k_{\cal{L}} >  k_\mathrm{ER}$, and 
\begin{equation}
\gamma^{R R}_\mathrm{Berry, u} =   -2 \pi 
 \left(   1-\sqrt{   1-  (k_{\cal{L}}  / k_\mathrm{ER} )^4  }  \right)  \, {\rm mod}  \, 2 \pi
\end{equation}
for $k_{\cal{L}} <  k_\mathrm{ER}$.
Similarly, we find
\begin{equation}
\gamma^{L R}_\mathrm{Berry, u} =   2 \pi  \left(
1 +  \frac{i}{ \sqrt{  ( k_{\cal{L}} / k_\mathrm{ER}  )^4  - 1    }}
\right)
\;  {\rm mod}  \; 2 \pi
\end{equation}
for $k_{\cal{L}} >  k_\mathrm{ER}$, and 
\begin{equation}
\gamma^{L R}_\mathrm{Berry, u} =  - 2 \pi 
 \left(   1  - \frac{1}{\sqrt{   1-  (k_{\cal{L}} / k_\mathrm{ER} )^4  }}  \right)   \; {\rm mod}  \; 2 \pi
\end{equation}
for $k_{\cal{L}} <  k_\mathrm{ER}$. First, we note that in none of these cases is the closed-loop Berry phase equal to $\pi$. This is not in contradiction to the results for the vorticity in the Hopf-link configuration discussed above, or the findings in Ref. \cite{mailybaev-etal.2005}, because Ref. \cite{mailybaev-etal.2005} is limited to point singularities in the 2-dimensional parameter plane. In the present case of the ER and the loop $\cal{L}$ in the same plane $k_x - k_y$, the singularity is a 1-dimensional object (the ER) that breaks up the plane into two separate areas (the ring and its complement in the $k_x - k_y$ plane). The point singularity, in contrast, cannot break up the plane into two separate areas. Futhermore, for the case of the loop inside the ER, $k_{\cal{L}} < k_\mathrm{ER}$, we have no obstruction to Stoke's theorem, and the Berry phase is the circulation of the Berry connection around the loop $\cal{L}$, whose value must be equal to the flux through the disk with radius $k_{\cal{L}}$ calculated from Stoke's theorem (which we have verified analytically, see below). We note, however, that the Berry phase is different if calculated solely with right eivenvectors ($RR$) or with the combination of left and right eivenvectors ($LR$). In the latter case, we know that approaching the ER means that we are approaching self-orthogonality. This is reflected in the fact that the $LR$ Berry phase diverges when $k_{\cal{L}}$ appoaches $k_\mathrm{ER}$ from below. For the loop outside the ER, $k_{\cal{L}} > k_\mathrm{ER}$, we find that the $RR$ Berry phase is independent of $\cal{L}$, which renders it similar to a winding number. In contrast, the $LR$ Berry phase is complex and depends on $\cal{L}$. However, in the limit of $k_{\cal{L}} \rightarrow\infty$, the $LR$ Berry phase becomes real and equal to $-2 \pi \; {\rm mod}  \; 2 \pi$, in other words, equivalent to the $RR$ Berry phase.

Next, we calculate an integral over the Berry curvature that we call a quasi-Chern integral,
\begin{equation}
   C^{ab}_\mathrm{n} =  \frac{1}{2\pi}\int  \Omega^{ab}_\mathrm{n}(\bm{k}) 
   d\textbf{S}, 
   \label{eq:quasi-chern}
\end{equation}
with the Berry curvature
\begin{equation}
    \Omega^{ab}_\mathrm{n}(\bm{k})  =   \frac{\partial A^{a b}_{y,n}(\bm{k}) }{ \partial k_x } - 
   \frac{\partial A^{a b}_{x,n}(\bm{k}) }{ \partial k_y }. 
   \label{eq:quasi-Berry-curvature}
\end{equation}
Here, the area $S$ is a two-dimensional disk in the $k_x - k_y$ plane centered at $\bm{k}=0$ with radius $k_\mathrm{D}$, either smaller or larger than the ER. In the case without nonlinearity and with zero detuning, we have evaluated the Chern integral analytically and numerically, the latter in order to develop reliable numerics for the case with nonlinearity (discussed below). Since we take the domain of the Berry connection to be the entire $k_x - k_y$ plane (i.e. $\mathbb{R}^2$), rather than a compact and closed Brillouin zone (e.g. a torus), the usual proof of the Chern number being an integer (e.g. \cite{shen-etal.2018}) does not hold. While a compact $k$-surface can be divided into two surfaces with one common boundary (and no other boundaries), dividing $\mathbb{R}^2$ leads to an additional (non-shared) boundary at infinity \cite{araujo-etal.2019}. We have seen that in the $RR$ case, the Berry connection is independent of $k_{\cal{L}}$ for $k_{\cal{L}} > k_\mathrm{ER}$. This implies that the flux through any annulus bounded by radii $k_1$ and $k_2$ both larger than $k_\mathrm{ER}$ must vanish (because the circulation of the two boundaries of the annulus is equal in magnitude and opposite in sign). Indeed, for the $RR$ case, the Berry curvature is identically zero for all $|\bm{k}| > k_\mathrm{ER}$. Therefore the $RR$ Berry curvature becomes independent of the disk radius for $k_\mathrm{D} > k_\mathrm{ER}$. For $k_\mathrm{D} < k_\mathrm{ER}$, we find
\begin{equation}
   C^{RR}_\mathrm{u} =    - 1 + \sqrt{   1-  (k_\mathrm{D}  / k_\mathrm{ER} )^4  }  
\end{equation}
which is as expected from Stoke's theorem. Hence, outside the ER the quasi-Chern integral is
\begin{equation}
   C^{RR}_u = -1
\end{equation}
independent of $k_\mathrm{D}$. This is our motivation for calling this a quasi-Chern integral. However, we note that the $LR$ result is different. 
In the $LR$ case, the result for $k_\mathrm{D} < k_\mathrm{ER}$ is, in agreement with the circulation based on the Berry phase
\begin{equation}
   C^{LR}_\mathrm{u} =   - 1  + \frac{1}{\sqrt{   1-  (k_\mathrm{D} / k_\mathrm{ER} )^4  }}. 
\end{equation}
This diverges as $k_\mathrm{D}$ approaches $k_\mathrm{ER}$, and remains infinite for all $k_\mathrm{D} > k_\mathrm{ER}$. If the Hamiltonian is written in the standard form of $H = \bm{d(k)} \cdot \bm{\sigma}$ (with $\sigma$ being the Pauli matrices), we see that it maps the 2D $\bm{k}$ plane to a 2D $d_x$-$d_y$ plane with real-valued $d_x$ and $d_y$, and a $k$-independent complex z-component $d_z=\Delta + i \gamma$. In the $RR$ case, the quasi-Chern integral can be thought of describing the topological properties of $\bm{k}$ in that plane, with the caveat that it is not an integer for disks inside the ER and that it depends on the choice of eigenvectors ($RR$ vs $LR$). 


\begin{thebibliography}{122}%
\makeatletter
\providecommand \@ifxundefined [1]{%
 \@ifx{#1\undefined}
}%
\providecommand \@ifnum [1]{%
 \ifnum #1\expandafter \@firstoftwo
 \else \expandafter \@secondoftwo
 \fi
}%
\providecommand \@ifx [1]{%
 \ifx #1\expandafter \@firstoftwo
 \else \expandafter \@secondoftwo
 \fi
}%
\providecommand \natexlab [1]{#1}%
\providecommand \enquote  [1]{``#1''}%
\providecommand \bibnamefont  [1]{#1}%
\providecommand \bibfnamefont [1]{#1}%
\providecommand \citenamefont [1]{#1}%
\providecommand \href@noop [0]{\@secondoftwo}%
\providecommand \href [0]{\begingroup \@sanitize@url \@href}%
\providecommand \@href[1]{\@@startlink{#1}\@@href}%
\providecommand \@@href[1]{\endgroup#1\@@endlink}%
\providecommand \@sanitize@url [0]{\catcode `\\12\catcode `\$12\catcode `\&12\catcode `\#12\catcode `\^12\catcode `\_12\catcode `\%12\relax}%
\providecommand \@@startlink[1]{}%
\providecommand \@@endlink[0]{}%
\providecommand \url  [0]{\begingroup\@sanitize@url \@url }%
\providecommand \@url [1]{\endgroup\@href {#1}{\urlprefix }}%
\providecommand \urlprefix  [0]{URL }%
\providecommand \Eprint [0]{\href }%
\providecommand \doibase [0]{https://doi.org/}%
\providecommand \selectlanguage [0]{\@gobble}%
\providecommand \bibinfo  [0]{\@secondoftwo}%
\providecommand \bibfield  [0]{\@secondoftwo}%
\providecommand \translation [1]{[#1]}%
\providecommand \BibitemOpen [0]{}%
\providecommand \bibitemStop [0]{}%
\providecommand \bibitemNoStop [0]{.\EOS\space}%
\providecommand \EOS [0]{\spacefactor3000\relax}%
\providecommand \BibitemShut  [1]{\csname bibitem#1\endcsname}%
\let\auto@bib@innerbib\@empty
\bibitem [{\citenamefont {Kato}(1966)}]{kato.1966}%
  \BibitemOpen
  \bibfield  {author} {\bibinfo {author} {\bibfnamefont {T.}~\bibnamefont {Kato}},\ }\href@noop {} {\emph {\bibinfo {title} {Perturbation Theory for Linear Operators}}}\ (\bibinfo  {publisher} {Springer},\ \bibinfo {address} {New York, United States},\ \bibinfo {year} {1966})\BibitemShut {NoStop}%
\bibitem [{\citenamefont {Garrison}\ and\ \citenamefont {Wright}(1988)}]{GARRISON1988177}%
  \BibitemOpen
  \bibfield  {author} {\bibinfo {author} {\bibfnamefont {J.}~\bibnamefont {Garrison}}\ and\ \bibinfo {author} {\bibfnamefont {E.}~\bibnamefont {Wright}},\ }\bibfield  {title} {\bibinfo {title} {Complex geometrical phases for dissipative systems},\ }\href {https://doi.org/https://doi.org/10.1016/0375-9601(88)90905-X} {\bibfield  {journal} {\bibinfo  {journal} {Physics Letters A}\ }\textbf {\bibinfo {volume} {128}},\ \bibinfo {pages} {177} (\bibinfo {year} {1988})}\BibitemShut {NoStop}%
\bibitem [{\citenamefont {Berry}(2004)}]{Berry2004}%
  \BibitemOpen
  \bibfield  {author} {\bibinfo {author} {\bibfnamefont {M.~V.}\ \bibnamefont {Berry}},\ }\bibfield  {title} {\bibinfo {title} {Physics of non-{H}ermitian degeneracies},\ }\href {https://doi.org/10.1023/B:CJOP.0000044002.05657.04} {\bibfield  {journal} {\bibinfo  {journal} {Czechoslov. J. Phys.}\ }\textbf {\bibinfo {volume} {54}},\ \bibinfo {pages} {1039} (\bibinfo {year} {2004})}\BibitemShut {NoStop}%
\bibitem [{\citenamefont {Heiss}(2004)}]{heiss2004exceptional}%
  \BibitemOpen
  \bibfield  {author} {\bibinfo {author} {\bibfnamefont {W.}~\bibnamefont {Heiss}},\ }\bibfield  {title} {\bibinfo {title} {Exceptional points of non-{H}ermitian operators},\ }\href {https://doi.org/10.1088/0305-4470/37/6/034} {\bibfield  {journal} {\bibinfo  {journal} {J. Phys. A Math. Gen.}\ }\textbf {\bibinfo {volume} {37}},\ \bibinfo {pages} {2455} (\bibinfo {year} {2004})}\BibitemShut {NoStop}%
\bibitem [{\citenamefont {Bender}(2007)}]{Bender2007}%
  \BibitemOpen
  \bibfield  {author} {\bibinfo {author} {\bibfnamefont {C.~M.}\ \bibnamefont {Bender}},\ }\bibfield  {title} {\bibinfo {title} {Making sense of non-{H}ermitian {H}amiltonians},\ }\href {https://doi.org/10.1088/0034-4885/70/6/R03} {\bibfield  {journal} {\bibinfo  {journal} {Rep. Prog. Phys.}\ }\textbf {\bibinfo {volume} {70}},\ \bibinfo {pages} {947} (\bibinfo {year} {2007})}\BibitemShut {NoStop}%
\bibitem [{\citenamefont {Heiss}(2012)}]{Heiss2012}%
  \BibitemOpen
  \bibfield  {author} {\bibinfo {author} {\bibfnamefont {W.~D.}\ \bibnamefont {Heiss}},\ }\bibfield  {title} {\bibinfo {title} {The physics of exceptional points},\ }\href {https://doi.org/10.1088/1751-8113/45/44/444016} {\bibfield  {journal} {\bibinfo  {journal} {J. Phys. A Math. Theor.}\ }\textbf {\bibinfo {volume} {45}},\ \bibinfo {pages} {444016} (\bibinfo {year} {2012})}\BibitemShut {NoStop}%
\bibitem [{\citenamefont {Ashida}\ \emph {et~al.}(2021)\citenamefont {Ashida}, \citenamefont {Gong},\ and\ \citenamefont {Ueda}}]{ashida-etal.2021}%
  \BibitemOpen
  \bibfield  {author} {\bibinfo {author} {\bibfnamefont {Y.}~\bibnamefont {Ashida}}, \bibinfo {author} {\bibfnamefont {Z.}~\bibnamefont {Gong}},\ and\ \bibinfo {author} {\bibfnamefont {M.}~\bibnamefont {Ueda}},\ }\bibfield  {title} {\bibinfo {title} {{Non-Hermitian physics}},\ }\href@noop {} {\bibfield  {journal} {\bibinfo  {journal} {Advances in Physics}\ }\textbf {\bibinfo {volume} {69}},\ \bibinfo {pages} {249 } (\bibinfo {year} {2021})}\BibitemShut {NoStop}%
\bibitem [{\citenamefont {Dembowski}\ \emph {et~al.}(2001)\citenamefont {Dembowski}, \citenamefont {Gr\"af}, \citenamefont {Harney}, \citenamefont {Heine}, \citenamefont {Heiss}, \citenamefont {Rehfeld},\ and\ \citenamefont {Richter}}]{PhysRevLett.86.787}%
  \BibitemOpen
  \bibfield  {author} {\bibinfo {author} {\bibfnamefont {C.}~\bibnamefont {Dembowski}}, \bibinfo {author} {\bibfnamefont {H.-D.}\ \bibnamefont {Gr\"af}}, \bibinfo {author} {\bibfnamefont {H.~L.}\ \bibnamefont {Harney}}, \bibinfo {author} {\bibfnamefont {A.}~\bibnamefont {Heine}}, \bibinfo {author} {\bibfnamefont {W.~D.}\ \bibnamefont {Heiss}}, \bibinfo {author} {\bibfnamefont {H.}~\bibnamefont {Rehfeld}},\ and\ \bibinfo {author} {\bibfnamefont {A.}~\bibnamefont {Richter}},\ }\bibfield  {title} {\bibinfo {title} {Experimental observation of the topological structure of exceptional points},\ }\href {https://doi.org/10.1103/PhysRevLett.86.787} {\bibfield  {journal} {\bibinfo  {journal} {Phys. Rev. Lett.}\ }\textbf {\bibinfo {volume} {86}},\ \bibinfo {pages} {787} (\bibinfo {year} {2001})}\BibitemShut {NoStop}%
\bibitem [{\citenamefont {Dembowski}\ \emph {et~al.}(2003)\citenamefont {Dembowski}, \citenamefont {Dietz}, \citenamefont {Gr\"af}, \citenamefont {Harney}, \citenamefont {Heine}, \citenamefont {Heiss},\ and\ \citenamefont {Richter}}]{PhysRevLett.90.034101}%
  \BibitemOpen
  \bibfield  {author} {\bibinfo {author} {\bibfnamefont {C.}~\bibnamefont {Dembowski}}, \bibinfo {author} {\bibfnamefont {B.}~\bibnamefont {Dietz}}, \bibinfo {author} {\bibfnamefont {H.-D.}\ \bibnamefont {Gr\"af}}, \bibinfo {author} {\bibfnamefont {H.~L.}\ \bibnamefont {Harney}}, \bibinfo {author} {\bibfnamefont {A.}~\bibnamefont {Heine}}, \bibinfo {author} {\bibfnamefont {W.~D.}\ \bibnamefont {Heiss}},\ and\ \bibinfo {author} {\bibfnamefont {A.}~\bibnamefont {Richter}},\ }\bibfield  {title} {\bibinfo {title} {Observation of a chiral state in a microwave cavity},\ }\href {https://doi.org/10.1103/PhysRevLett.90.034101} {\bibfield  {journal} {\bibinfo  {journal} {Phys. Rev. Lett.}\ }\textbf {\bibinfo {volume} {90}},\ \bibinfo {pages} {034101} (\bibinfo {year} {2003})}\BibitemShut {NoStop}%
\bibitem [{\citenamefont {Cao}\ and\ \citenamefont {Wiersig}(2015)}]{RevModPhys.87.61}%
  \BibitemOpen
  \bibfield  {author} {\bibinfo {author} {\bibfnamefont {H.}~\bibnamefont {Cao}}\ and\ \bibinfo {author} {\bibfnamefont {J.}~\bibnamefont {Wiersig}},\ }\bibfield  {title} {\bibinfo {title} {Dielectric microcavities: Model systems for wave chaos and non-{H}ermitian physics},\ }\href {https://doi.org/10.1103/RevModPhys.87.61} {\bibfield  {journal} {\bibinfo  {journal} {Rev. Mod. Phys.}\ }\textbf {\bibinfo {volume} {87}},\ \bibinfo {pages} {61} (\bibinfo {year} {2015})}\BibitemShut {NoStop}%
\bibitem [{\citenamefont {Choi}\ \emph {et~al.}(2010)\citenamefont {Choi}, \citenamefont {Kang}, \citenamefont {Lim}, \citenamefont {Kim}, \citenamefont {Kim}, \citenamefont {Lee},\ and\ \citenamefont {An}}]{PhysRevLett.104.153601}%
  \BibitemOpen
  \bibfield  {author} {\bibinfo {author} {\bibfnamefont {Y.}~\bibnamefont {Choi}}, \bibinfo {author} {\bibfnamefont {S.}~\bibnamefont {Kang}}, \bibinfo {author} {\bibfnamefont {S.}~\bibnamefont {Lim}}, \bibinfo {author} {\bibfnamefont {W.}~\bibnamefont {Kim}}, \bibinfo {author} {\bibfnamefont {J.-R.}\ \bibnamefont {Kim}}, \bibinfo {author} {\bibfnamefont {J.-H.}\ \bibnamefont {Lee}},\ and\ \bibinfo {author} {\bibfnamefont {K.}~\bibnamefont {An}},\ }\bibfield  {title} {\bibinfo {title} {Quasieigenstate coalescence in an atom-cavity quantum composite},\ }\href {https://doi.org/10.1103/PhysRevLett.104.153601} {\bibfield  {journal} {\bibinfo  {journal} {Phys. Rev. Lett.}\ }\textbf {\bibinfo {volume} {104}},\ \bibinfo {pages} {153601} (\bibinfo {year} {2010})}\BibitemShut {NoStop}%
\bibitem [{\citenamefont {Kodigala}\ \emph {et~al.}(2016)\citenamefont {Kodigala}, \citenamefont {Lepetit},\ and\ \citenamefont {Kant\'e}}]{PhysRevB.94.201103}%
  \BibitemOpen
  \bibfield  {author} {\bibinfo {author} {\bibfnamefont {A.}~\bibnamefont {Kodigala}}, \bibinfo {author} {\bibfnamefont {T.}~\bibnamefont {Lepetit}},\ and\ \bibinfo {author} {\bibfnamefont {B.}~\bibnamefont {Kant\'e}},\ }\bibfield  {title} {\bibinfo {title} {Exceptional points in three-dimensional plasmonic nanostructures},\ }\href {https://doi.org/10.1103/PhysRevB.94.201103} {\bibfield  {journal} {\bibinfo  {journal} {Phys. Rev. B}\ }\textbf {\bibinfo {volume} {94}},\ \bibinfo {pages} {201103} (\bibinfo {year} {2016})}\BibitemShut {NoStop}%
\bibitem [{\citenamefont {Park}\ \emph {et~al.}(2019)\citenamefont {Park}, \citenamefont {Ndao},\ and\ \citenamefont {Kant{\'e}}}]{park2019observation}%
  \BibitemOpen
  \bibfield  {author} {\bibinfo {author} {\bibfnamefont {J.-H.}\ \bibnamefont {Park}}, \bibinfo {author} {\bibfnamefont {A.}~\bibnamefont {Ndao}},\ and\ \bibinfo {author} {\bibfnamefont {B.}~\bibnamefont {Kant{\'e}}},\ }\bibfield  {title} {\bibinfo {title} {Observation of exceptional points in passive plasmonic nanostructure},\ }in\ \href {https://doi.org/10.1364/CLEO_AT.2018.JTh2A.47} {\emph {\bibinfo {booktitle} {Frontiers in Optics}}}\ (\bibinfo {organization} {Optical Society of America},\ \bibinfo {year} {2019})\ pp.\ \bibinfo {pages} {JTu4A--79}\BibitemShut {NoStop}%
\bibitem [{\citenamefont {Guo}\ \emph {et~al.}(2009)\citenamefont {Guo}, \citenamefont {Salamo}, \citenamefont {Duchesne}, \citenamefont {Morandotti}, \citenamefont {Volatier-Ravat}, \citenamefont {Aimez}, \citenamefont {Siviloglou},\ and\ \citenamefont {Christodoulides}}]{PhysRevLett.103.093902}%
  \BibitemOpen
  \bibfield  {author} {\bibinfo {author} {\bibfnamefont {A.}~\bibnamefont {Guo}}, \bibinfo {author} {\bibfnamefont {G.~J.}\ \bibnamefont {Salamo}}, \bibinfo {author} {\bibfnamefont {D.}~\bibnamefont {Duchesne}}, \bibinfo {author} {\bibfnamefont {R.}~\bibnamefont {Morandotti}}, \bibinfo {author} {\bibfnamefont {M.}~\bibnamefont {Volatier-Ravat}}, \bibinfo {author} {\bibfnamefont {V.}~\bibnamefont {Aimez}}, \bibinfo {author} {\bibfnamefont {G.~A.}\ \bibnamefont {Siviloglou}},\ and\ \bibinfo {author} {\bibfnamefont {D.~N.}\ \bibnamefont {Christodoulides}},\ }\bibfield  {title} {\bibinfo {title} {Observation of $\mathcal{P}\mathcal{T}$-symmetry breaking in complex optical potentials},\ }\href {https://doi.org/10.1103/PhysRevLett.103.093902} {\bibfield  {journal} {\bibinfo  {journal} {Phys. Rev. Lett.}\ }\textbf {\bibinfo {volume} {103}},\ \bibinfo {pages} {093902} (\bibinfo {year} {2009})}\BibitemShut {NoStop}%
\bibitem [{\citenamefont {Lee}\ \emph {et~al.}(2009)\citenamefont {Lee}, \citenamefont {Yang}, \citenamefont {Moon}, \citenamefont {Lee}, \citenamefont {Shim}, \citenamefont {Kim}, \citenamefont {Lee},\ and\ \citenamefont {An}}]{PhysRevLett.103.134101}%
  \BibitemOpen
  \bibfield  {author} {\bibinfo {author} {\bibfnamefont {S.-B.}\ \bibnamefont {Lee}}, \bibinfo {author} {\bibfnamefont {J.}~\bibnamefont {Yang}}, \bibinfo {author} {\bibfnamefont {S.}~\bibnamefont {Moon}}, \bibinfo {author} {\bibfnamefont {S.-Y.}\ \bibnamefont {Lee}}, \bibinfo {author} {\bibfnamefont {J.-B.}\ \bibnamefont {Shim}}, \bibinfo {author} {\bibfnamefont {S.~W.}\ \bibnamefont {Kim}}, \bibinfo {author} {\bibfnamefont {J.-H.}\ \bibnamefont {Lee}},\ and\ \bibinfo {author} {\bibfnamefont {K.}~\bibnamefont {An}},\ }\bibfield  {title} {\bibinfo {title} {Observation of an exceptional point in a chaotic optical microcavity},\ }\href {https://doi.org/10.1103/PhysRevLett.103.134101} {\bibfield  {journal} {\bibinfo  {journal} {Phys. Rev. Lett.}\ }\textbf {\bibinfo {volume} {103}},\ \bibinfo {pages} {134101} (\bibinfo {year} {2009})}\BibitemShut {NoStop}%
\bibitem [{\citenamefont {Chang}\ \emph {et~al.}(2014)\citenamefont {Chang}, \citenamefont {Jiang}, \citenamefont {Hua}, \citenamefont {Yang}, \citenamefont {Wen}, \citenamefont {Jiang}, \citenamefont {Li}, \citenamefont {Wang},\ and\ \citenamefont {Xiao}}]{chang2014parity}%
  \BibitemOpen
  \bibfield  {author} {\bibinfo {author} {\bibfnamefont {L.}~\bibnamefont {Chang}}, \bibinfo {author} {\bibfnamefont {X.}~\bibnamefont {Jiang}}, \bibinfo {author} {\bibfnamefont {S.}~\bibnamefont {Hua}}, \bibinfo {author} {\bibfnamefont {C.}~\bibnamefont {Yang}}, \bibinfo {author} {\bibfnamefont {J.}~\bibnamefont {Wen}}, \bibinfo {author} {\bibfnamefont {L.}~\bibnamefont {Jiang}}, \bibinfo {author} {\bibfnamefont {G.}~\bibnamefont {Li}}, \bibinfo {author} {\bibfnamefont {G.}~\bibnamefont {Wang}},\ and\ \bibinfo {author} {\bibfnamefont {M.}~\bibnamefont {Xiao}},\ }\bibfield  {title} {\bibinfo {title} {Parity--time symmetry and variable optical isolation in active--passive-coupled microresonators},\ }\href {https://doi.org/10.1038/nphoton.2014.133} {\bibfield  {journal} {\bibinfo  {journal} {Nat. Photon.}\ }\textbf {\bibinfo {volume} {8}},\ \bibinfo {pages} {524} (\bibinfo {year} {2014})}\BibitemShut {NoStop}%
\bibitem [{\citenamefont {Peng}\ \emph {et~al.}(2014{\natexlab{a}})\citenamefont {Peng}, \citenamefont {{\"O}zdemir}, \citenamefont {Lei}, \citenamefont {Monifi}, \citenamefont {Gianfreda}, \citenamefont {Long}, \citenamefont {Fan}, \citenamefont {Nori}, \citenamefont {Bender},\ and\ \citenamefont {Yang}}]{peng2014parity}%
  \BibitemOpen
  \bibfield  {author} {\bibinfo {author} {\bibfnamefont {B.}~\bibnamefont {Peng}}, \bibinfo {author} {\bibfnamefont {{\c{S}}.~K.}\ \bibnamefont {{\"O}zdemir}}, \bibinfo {author} {\bibfnamefont {F.}~\bibnamefont {Lei}}, \bibinfo {author} {\bibfnamefont {F.}~\bibnamefont {Monifi}}, \bibinfo {author} {\bibfnamefont {M.}~\bibnamefont {Gianfreda}}, \bibinfo {author} {\bibfnamefont {G.~L.}\ \bibnamefont {Long}}, \bibinfo {author} {\bibfnamefont {S.}~\bibnamefont {Fan}}, \bibinfo {author} {\bibfnamefont {F.}~\bibnamefont {Nori}}, \bibinfo {author} {\bibfnamefont {C.~M.}\ \bibnamefont {Bender}},\ and\ \bibinfo {author} {\bibfnamefont {L.}~\bibnamefont {Yang}},\ }\bibfield  {title} {\bibinfo {title} {Parity--time-symmetric whispering-gallery microcavities},\ }\href {https://doi.org/10.1038/nphys2927} {\bibfield  {journal} {\bibinfo  {journal} {Nat. Phys.}\ }\textbf {\bibinfo {volume} {10}},\ \bibinfo {pages} {394} (\bibinfo {year} {2014}{\natexlab{a}})}\BibitemShut {NoStop}%
\bibitem [{\citenamefont {Fruchart}\ \emph {et~al.}(2021)\citenamefont {Fruchart}, \citenamefont {Hanai}, \citenamefont {Littlewood},\ and\ \citenamefont {Vitelli}}]{Fruchart2021}%
  \BibitemOpen
  \bibfield  {author} {\bibinfo {author} {\bibfnamefont {M.}~\bibnamefont {Fruchart}}, \bibinfo {author} {\bibfnamefont {R.}~\bibnamefont {Hanai}}, \bibinfo {author} {\bibfnamefont {P.~B.}\ \bibnamefont {Littlewood}},\ and\ \bibinfo {author} {\bibfnamefont {V.}~\bibnamefont {Vitelli}},\ }\bibfield  {title} {\bibinfo {title} {Non-reciprocal phase transitions},\ }\href {https://doi.org/10.1038/s41586-021-03375-9} {\bibfield  {journal} {\bibinfo  {journal} {Nature}\ }\textbf {\bibinfo {volume} {592}},\ \bibinfo {pages} {363} (\bibinfo {year} {2021})}\BibitemShut {NoStop}%
\bibitem [{\citenamefont {Minganti}\ \emph {et~al.}(2019)\citenamefont {Minganti}, \citenamefont {Miranowicz}, \citenamefont {Chhajlany},\ and\ \citenamefont {Nori}}]{PhysRevA.100.062131}%
  \BibitemOpen
  \bibfield  {author} {\bibinfo {author} {\bibfnamefont {F.}~\bibnamefont {Minganti}}, \bibinfo {author} {\bibfnamefont {A.}~\bibnamefont {Miranowicz}}, \bibinfo {author} {\bibfnamefont {R.~W.}\ \bibnamefont {Chhajlany}},\ and\ \bibinfo {author} {\bibfnamefont {F.}~\bibnamefont {Nori}},\ }\bibfield  {title} {\bibinfo {title} {Quantum exceptional points of non-{H}ermitian {H}amiltonians and {L}iouvillians: The effects of quantum jumps},\ }\href {https://doi.org/10.1103/PhysRevA.100.062131} {\bibfield  {journal} {\bibinfo  {journal} {Phys. Rev. A}\ }\textbf {\bibinfo {volume} {100}},\ \bibinfo {pages} {062131} (\bibinfo {year} {2019})}\BibitemShut {NoStop}%
\bibitem [{\citenamefont {Zhang}\ \emph {et~al.}(2018)\citenamefont {Zhang}, \citenamefont {Saif}, \citenamefont {Jiao},\ and\ \citenamefont {Jing}}]{Zhang:18}%
  \BibitemOpen
  \bibfield  {author} {\bibinfo {author} {\bibfnamefont {H.}~\bibnamefont {Zhang}}, \bibinfo {author} {\bibfnamefont {F.}~\bibnamefont {Saif}}, \bibinfo {author} {\bibfnamefont {Y.}~\bibnamefont {Jiao}},\ and\ \bibinfo {author} {\bibfnamefont {H.}~\bibnamefont {Jing}},\ }\bibfield  {title} {\bibinfo {title} {Loss-induced transparency in optomechanics},\ }\href {https://doi.org/10.1364/OE.26.025199} {\bibfield  {journal} {\bibinfo  {journal} {Opt. Express}\ }\textbf {\bibinfo {volume} {26}},\ \bibinfo {pages} {25199} (\bibinfo {year} {2018})}\BibitemShut {NoStop}%
\bibitem [{\citenamefont {Peng}\ \emph {et~al.}(2014{\natexlab{b}})\citenamefont {Peng}, \citenamefont {\c{S}. K.~Özdemir}, \citenamefont {Rotter}, \citenamefont {Yilmaz}, \citenamefont {Liertzer}, \citenamefont {Monifi}, \citenamefont {Bender}, \citenamefont {Nori},\ and\ \citenamefont {Yang}}]{doi:10.1126/science.1258004}%
  \BibitemOpen
  \bibfield  {author} {\bibinfo {author} {\bibfnamefont {B.}~\bibnamefont {Peng}}, \bibinfo {author} {\bibnamefont {\c{S}. K.~Özdemir}}, \bibinfo {author} {\bibfnamefont {S.}~\bibnamefont {Rotter}}, \bibinfo {author} {\bibfnamefont {H.}~\bibnamefont {Yilmaz}}, \bibinfo {author} {\bibfnamefont {M.}~\bibnamefont {Liertzer}}, \bibinfo {author} {\bibfnamefont {F.}~\bibnamefont {Monifi}}, \bibinfo {author} {\bibfnamefont {C.~M.}\ \bibnamefont {Bender}}, \bibinfo {author} {\bibfnamefont {F.}~\bibnamefont {Nori}},\ and\ \bibinfo {author} {\bibfnamefont {L.}~\bibnamefont {Yang}},\ }\bibfield  {title} {\bibinfo {title} {Loss-induced suppression and revival of lasing},\ }\href {https://doi.org/10.1126/science.1258004} {\bibfield  {journal} {\bibinfo  {journal} {Science}\ }\textbf {\bibinfo {volume} {346}},\ \bibinfo {pages} {328} (\bibinfo {year} {2014}{\natexlab{b}})}\BibitemShut {NoStop}%
\bibitem [{\citenamefont {Li}\ \emph {et~al.}(2022{\natexlab{a}})\citenamefont {Li}, \citenamefont {Ma}, \citenamefont {Hatzopoulos}, \citenamefont {Savvidis}, \citenamefont {Schumacher},\ and\ \citenamefont {Gao}}]{li2022switching}%
  \BibitemOpen
  \bibfield  {author} {\bibinfo {author} {\bibfnamefont {Y.}~\bibnamefont {Li}}, \bibinfo {author} {\bibfnamefont {X.}~\bibnamefont {Ma}}, \bibinfo {author} {\bibfnamefont {Z.}~\bibnamefont {Hatzopoulos}}, \bibinfo {author} {\bibfnamefont {P.~G.}\ \bibnamefont {Savvidis}}, \bibinfo {author} {\bibfnamefont {S.}~\bibnamefont {Schumacher}},\ and\ \bibinfo {author} {\bibfnamefont {T.}~\bibnamefont {Gao}},\ }\bibfield  {title} {\bibinfo {title} {Switching off a microcavity polariton condensate near the exceptional point},\ }\href {https://doi.org/10.1021/acsphotonics.2c00288} {\bibfield  {journal} {\bibinfo  {journal} {ACS Photonics}\ }\textbf {\bibinfo {volume} {9}},\ \bibinfo {pages} {2079} (\bibinfo {year} {2022}{\natexlab{a}})}\BibitemShut {NoStop}%
\bibitem [{\citenamefont {Lin}\ \emph {et~al.}(2011)\citenamefont {Lin}, \citenamefont {Ramezani}, \citenamefont {Eichelkraut}, \citenamefont {Kottos}, \citenamefont {Cao},\ and\ \citenamefont {Christodoulides}}]{PhysRevLett.106.213901}%
  \BibitemOpen
  \bibfield  {author} {\bibinfo {author} {\bibfnamefont {Z.}~\bibnamefont {Lin}}, \bibinfo {author} {\bibfnamefont {H.}~\bibnamefont {Ramezani}}, \bibinfo {author} {\bibfnamefont {T.}~\bibnamefont {Eichelkraut}}, \bibinfo {author} {\bibfnamefont {T.}~\bibnamefont {Kottos}}, \bibinfo {author} {\bibfnamefont {H.}~\bibnamefont {Cao}},\ and\ \bibinfo {author} {\bibfnamefont {D.~N.}\ \bibnamefont {Christodoulides}},\ }\bibfield  {title} {\bibinfo {title} {Unidirectional invisibility induced by $\mathcal{P}\mathcal{T}$-symmetric periodic structures},\ }\href {https://doi.org/10.1103/PhysRevLett.106.213901} {\bibfield  {journal} {\bibinfo  {journal} {Phys. Rev. Lett.}\ }\textbf {\bibinfo {volume} {106}},\ \bibinfo {pages} {213901} (\bibinfo {year} {2011})}\BibitemShut {NoStop}%
\bibitem [{\citenamefont {Chen}\ \emph {et~al.}(2017)\citenamefont {Chen}, \citenamefont {Kaya~{\"O}zdemir}, \citenamefont {Zhao}, \citenamefont {Wiersig},\ and\ \citenamefont {Yang}}]{chen2017exceptional}%
  \BibitemOpen
  \bibfield  {author} {\bibinfo {author} {\bibfnamefont {W.}~\bibnamefont {Chen}}, \bibinfo {author} {\bibfnamefont {{\c{S}}.}~\bibnamefont {Kaya~{\"O}zdemir}}, \bibinfo {author} {\bibfnamefont {G.}~\bibnamefont {Zhao}}, \bibinfo {author} {\bibfnamefont {J.}~\bibnamefont {Wiersig}},\ and\ \bibinfo {author} {\bibfnamefont {L.}~\bibnamefont {Yang}},\ }\bibfield  {title} {\bibinfo {title} {Exceptional points enhance sensing in an optical microcavity},\ }\href {https://doi.org/10.1038/nature23281} {\bibfield  {journal} {\bibinfo  {journal} {Nature}\ }\textbf {\bibinfo {volume} {548}},\ \bibinfo {pages} {192} (\bibinfo {year} {2017})}\BibitemShut {NoStop}%
\bibitem [{\citenamefont {Wiersig}(2020{\natexlab{a}})}]{Wiersig:20}%
  \BibitemOpen
  \bibfield  {author} {\bibinfo {author} {\bibfnamefont {J.}~\bibnamefont {Wiersig}},\ }\bibfield  {title} {\bibinfo {title} {Review of exceptional point-based sensors},\ }\href {https://doi.org/10.1364/PRJ.396115} {\bibfield  {journal} {\bibinfo  {journal} {Photon. Res.}\ }\textbf {\bibinfo {volume} {8}},\ \bibinfo {pages} {1457} (\bibinfo {year} {2020}{\natexlab{a}})}\BibitemShut {NoStop}%
\bibitem [{\citenamefont {Wiersig}(2016)}]{PhysRevA.93.033809}%
  \BibitemOpen
  \bibfield  {author} {\bibinfo {author} {\bibfnamefont {J.}~\bibnamefont {Wiersig}},\ }\bibfield  {title} {\bibinfo {title} {Sensors operating at exceptional points: General theory},\ }\href {https://doi.org/10.1103/PhysRevA.93.033809} {\bibfield  {journal} {\bibinfo  {journal} {Phys. Rev. A}\ }\textbf {\bibinfo {volume} {93}},\ \bibinfo {pages} {033809} (\bibinfo {year} {2016})}\BibitemShut {NoStop}%
\bibitem [{\citenamefont {Hodaei}\ \emph {et~al.}(2017)\citenamefont {Hodaei}, \citenamefont {Hassan}, \citenamefont {Wittek}, \citenamefont {Garcia-Gracia}, \citenamefont {El-Ganainy}, \citenamefont {Christodoulides},\ and\ \citenamefont {Khajavikhan}}]{hodaei2017enhanced}%
  \BibitemOpen
  \bibfield  {author} {\bibinfo {author} {\bibfnamefont {H.}~\bibnamefont {Hodaei}}, \bibinfo {author} {\bibfnamefont {A.~U.}\ \bibnamefont {Hassan}}, \bibinfo {author} {\bibfnamefont {S.}~\bibnamefont {Wittek}}, \bibinfo {author} {\bibfnamefont {H.}~\bibnamefont {Garcia-Gracia}}, \bibinfo {author} {\bibfnamefont {R.}~\bibnamefont {El-Ganainy}}, \bibinfo {author} {\bibfnamefont {D.~N.}\ \bibnamefont {Christodoulides}},\ and\ \bibinfo {author} {\bibfnamefont {M.}~\bibnamefont {Khajavikhan}},\ }\bibfield  {title} {\bibinfo {title} {Enhanced sensitivity at higher-order exceptional points},\ }\href {https://doi.org/10.1038/nature23280} {\bibfield  {journal} {\bibinfo  {journal} {Nature}\ }\textbf {\bibinfo {volume} {548}},\ \bibinfo {pages} {187} (\bibinfo {year} {2017})}\BibitemShut {NoStop}%
\bibitem [{\citenamefont {Wiersig}(2020{\natexlab{b}})}]{Wiersig2020}%
  \BibitemOpen
  \bibfield  {author} {\bibinfo {author} {\bibfnamefont {J.}~\bibnamefont {Wiersig}},\ }\bibfield  {title} {\bibinfo {title} {Prospects and fundamental limits in exceptional point-based sensing},\ }\href {https://doi.org/10.1038/s41467-020-16373-8} {\bibfield  {journal} {\bibinfo  {journal} {Nat. Commun.}\ }\textbf {\bibinfo {volume} {11}},\ \bibinfo {pages} {2454} (\bibinfo {year} {2020}{\natexlab{b}})}\BibitemShut {NoStop}%
\bibitem [{\citenamefont {Langbein}(2018{\natexlab{a}})}]{langbein2018no}%
  \BibitemOpen
  \bibfield  {author} {\bibinfo {author} {\bibfnamefont {W.}~\bibnamefont {Langbein}},\ }\bibfield  {title} {\bibinfo {title} {No exceptional precision of exceptional-point sensors},\ }\href@noop {} {\bibfield  {journal} {\bibinfo  {journal} {Phys. Rev. A}\ }\textbf {\bibinfo {volume} {98}},\ \bibinfo {pages} {023805} (\bibinfo {year} {2018}{\natexlab{a}})}\BibitemShut {NoStop}%
\bibitem [{\citenamefont {Lau}\ and\ \citenamefont {Clerk}(2018)}]{lau2018fundamental}%
  \BibitemOpen
  \bibfield  {author} {\bibinfo {author} {\bibfnamefont {H.-K.}\ \bibnamefont {Lau}}\ and\ \bibinfo {author} {\bibfnamefont {A.~A.}\ \bibnamefont {Clerk}},\ }\bibfield  {title} {\bibinfo {title} {Fundamental limits and non-reciprocal approaches in non-{H}ermitian quantum sensing},\ }\href@noop {} {\bibfield  {journal} {\bibinfo  {journal} {Nat. Commun.}\ }\textbf {\bibinfo {volume} {9}},\ \bibinfo {pages} {4320} (\bibinfo {year} {2018})}\BibitemShut {NoStop}%
\bibitem [{\citenamefont {Mortensen}\ \emph {et~al.}(2018)\citenamefont {Mortensen}, \citenamefont {Gon{\c{c}}alves}, \citenamefont {Khajavikhan}, \citenamefont {Christodoulides}, \citenamefont {Tserkezis},\ and\ \citenamefont {Wolff}}]{mortensen2018fluctuations}%
  \BibitemOpen
  \bibfield  {author} {\bibinfo {author} {\bibfnamefont {N.~A.}\ \bibnamefont {Mortensen}}, \bibinfo {author} {\bibfnamefont {P.}~\bibnamefont {Gon{\c{c}}alves}}, \bibinfo {author} {\bibfnamefont {M.}~\bibnamefont {Khajavikhan}}, \bibinfo {author} {\bibfnamefont {D.~N.}\ \bibnamefont {Christodoulides}}, \bibinfo {author} {\bibfnamefont {C.}~\bibnamefont {Tserkezis}},\ and\ \bibinfo {author} {\bibfnamefont {C.}~\bibnamefont {Wolff}},\ }\bibfield  {title} {\bibinfo {title} {Fluctuations and noise-limited sensing near the exceptional point of parity-time-symmetric resonator systems},\ }\href@noop {} {\bibfield  {journal} {\bibinfo  {journal} {Optica}\ }\textbf {\bibinfo {volume} {5}},\ \bibinfo {pages} {1342} (\bibinfo {year} {2018})}\BibitemShut {NoStop}%
\bibitem [{\citenamefont {Wolff}\ \emph {et~al.}(2019)\citenamefont {Wolff}, \citenamefont {Tserkezis},\ and\ \citenamefont {Mortensen}}]{wolff2019time}%
  \BibitemOpen
  \bibfield  {author} {\bibinfo {author} {\bibfnamefont {C.}~\bibnamefont {Wolff}}, \bibinfo {author} {\bibfnamefont {C.}~\bibnamefont {Tserkezis}},\ and\ \bibinfo {author} {\bibfnamefont {N.~A.}\ \bibnamefont {Mortensen}},\ }\bibfield  {title} {\bibinfo {title} {On the time evolution at a fluctuating exceptional point},\ }\href@noop {} {\bibfield  {journal} {\bibinfo  {journal} {Nanophotonics}\ }\textbf {\bibinfo {volume} {8}},\ \bibinfo {pages} {1319} (\bibinfo {year} {2019})}\BibitemShut {NoStop}%
\bibitem [{\citenamefont {Zhang}\ \emph {et~al.}(2019)\citenamefont {Zhang}, \citenamefont {Sweeney}, \citenamefont {Hsu}, \citenamefont {Yang}, \citenamefont {Stone},\ and\ \citenamefont {Jiang}}]{zhang2019quantum}%
  \BibitemOpen
  \bibfield  {author} {\bibinfo {author} {\bibfnamefont {M.}~\bibnamefont {Zhang}}, \bibinfo {author} {\bibfnamefont {W.}~\bibnamefont {Sweeney}}, \bibinfo {author} {\bibfnamefont {C.~W.}\ \bibnamefont {Hsu}}, \bibinfo {author} {\bibfnamefont {L.}~\bibnamefont {Yang}}, \bibinfo {author} {\bibfnamefont {A.}~\bibnamefont {Stone}},\ and\ \bibinfo {author} {\bibfnamefont {L.}~\bibnamefont {Jiang}},\ }\bibfield  {title} {\bibinfo {title} {Quantum noise theory of exceptional point amplifying sensors},\ }\href@noop {} {\bibfield  {journal} {\bibinfo  {journal} {Phys. Rev. Lett.}\ }\textbf {\bibinfo {volume} {123}},\ \bibinfo {pages} {180501} (\bibinfo {year} {2019})}\BibitemShut {NoStop}%
\bibitem [{\citenamefont {Chen}\ \emph {et~al.}(2019)\citenamefont {Chen}, \citenamefont {Jin},\ and\ \citenamefont {Liu}}]{chen2019sensitivity}%
  \BibitemOpen
  \bibfield  {author} {\bibinfo {author} {\bibfnamefont {C.}~\bibnamefont {Chen}}, \bibinfo {author} {\bibfnamefont {L.}~\bibnamefont {Jin}},\ and\ \bibinfo {author} {\bibfnamefont {R.-B.}\ \bibnamefont {Liu}},\ }\bibfield  {title} {\bibinfo {title} {Sensitivity of parameter estimation near the exceptional point of a non-{H}ermitian system},\ }\href@noop {} {\bibfield  {journal} {\bibinfo  {journal} {New J. Phys.}\ }\textbf {\bibinfo {volume} {21}},\ \bibinfo {pages} {083002} (\bibinfo {year} {2019})}\BibitemShut {NoStop}%
\bibitem [{\citenamefont {Naikoo}\ \emph {et~al.}(2023)\citenamefont {Naikoo}, \citenamefont {Chhajlany},\ and\ \citenamefont {Ko{\l}ody{\'n}ski}}]{naikoo2023multiparameter}%
  \BibitemOpen
  \bibfield  {author} {\bibinfo {author} {\bibfnamefont {J.}~\bibnamefont {Naikoo}}, \bibinfo {author} {\bibfnamefont {R.~W.}\ \bibnamefont {Chhajlany}},\ and\ \bibinfo {author} {\bibfnamefont {J.}~\bibnamefont {Ko{\l}ody{\'n}ski}},\ }\bibfield  {title} {\bibinfo {title} {Multiparameter estimation perspective on non-{H}ermitian singularity-enhanced sensing},\ }\href@noop {} {\bibfield  {journal} {\bibinfo  {journal} {Phys. Rev. Lett.}\ }\textbf {\bibinfo {volume} {131}},\ \bibinfo {pages} {220801} (\bibinfo {year} {2023})}\BibitemShut {NoStop}%
\bibitem [{\citenamefont {Loughlin}\ and\ \citenamefont {Sudhir}(2024)}]{loughlin2024exceptional}%
  \BibitemOpen
  \bibfield  {author} {\bibinfo {author} {\bibfnamefont {H.}~\bibnamefont {Loughlin}}\ and\ \bibinfo {author} {\bibfnamefont {V.}~\bibnamefont {Sudhir}},\ }\bibfield  {title} {\bibinfo {title} {Exceptional-point sensors offer no fundamental signal-to-noise ratio enhancement},\ }\href@noop {} {\bibfield  {journal} {\bibinfo  {journal} {Phys. Rev. Lett.}\ }\textbf {\bibinfo {volume} {132}},\ \bibinfo {pages} {243601} (\bibinfo {year} {2024})}\BibitemShut {NoStop}%
\bibitem [{\citenamefont {Wang}\ \emph {et~al.}(2019{\natexlab{a}})\citenamefont {Wang}, \citenamefont {Assawaworrarit},\ and\ \citenamefont {Fan}}]{wang2019dynamics}%
  \BibitemOpen
  \bibfield  {author} {\bibinfo {author} {\bibfnamefont {H.}~\bibnamefont {Wang}}, \bibinfo {author} {\bibfnamefont {S.}~\bibnamefont {Assawaworrarit}},\ and\ \bibinfo {author} {\bibfnamefont {S.}~\bibnamefont {Fan}},\ }\bibfield  {title} {\bibinfo {title} {Dynamics for encircling an exceptional point in a nonlinear non-{H}ermitian system},\ }\href {https://doi.org/10.1364/OL.44.000638} {\bibfield  {journal} {\bibinfo  {journal} {Opt. Lett.}\ }\textbf {\bibinfo {volume} {44}},\ \bibinfo {pages} {638} (\bibinfo {year} {2019}{\natexlab{a}})}\BibitemShut {NoStop}%
\bibitem [{\citenamefont {Ramezanpour}\ and\ \citenamefont {Bogdanov}(2021)}]{PhysRevA.103.043510}%
  \BibitemOpen
  \bibfield  {author} {\bibinfo {author} {\bibfnamefont {S.}~\bibnamefont {Ramezanpour}}\ and\ \bibinfo {author} {\bibfnamefont {A.}~\bibnamefont {Bogdanov}},\ }\bibfield  {title} {\bibinfo {title} {Tuning exceptional points with {Kerr} nonlinearity},\ }\href {https://doi.org/10.1103/PhysRevA.103.043510} {\bibfield  {journal} {\bibinfo  {journal} {Phys. Rev. A}\ }\textbf {\bibinfo {volume} {103}},\ \bibinfo {pages} {043510} (\bibinfo {year} {2021})}\BibitemShut {NoStop}%
\bibitem [{\citenamefont {Felski}\ and\ \citenamefont {Kunst}(2025)}]{PhysRevResearch.7.013326}%
  \BibitemOpen
  \bibfield  {author} {\bibinfo {author} {\bibfnamefont {A.}~\bibnamefont {Felski}}\ and\ \bibinfo {author} {\bibfnamefont {F.~K.}\ \bibnamefont {Kunst}},\ }\bibfield  {title} {\bibinfo {title} {Exceptional points and stability in nonlinear models of population dynamics having $\mathcal{PT}$ symmetry},\ }\href {https://doi.org/10.1103/PhysRevResearch.7.013326} {\bibfield  {journal} {\bibinfo  {journal} {Phys. Rev. Res.}\ }\textbf {\bibinfo {volume} {7}},\ \bibinfo {pages} {013326} (\bibinfo {year} {2025})}\BibitemShut {NoStop}%
\bibitem [{\citenamefont {Opala}\ \emph {et~al.}(2023)\citenamefont {Opala}, \citenamefont {Furman}, \citenamefont {Kr\'{o}l}, \citenamefont {Mirek}, \citenamefont {Tyszka}, \citenamefont {Seredy\'{n}ski}, \citenamefont {Pacuski}, \citenamefont {Szczytko}, \citenamefont {Matuszewski},\ and\ \citenamefont {Piętka}}]{Opala:23}%
  \BibitemOpen
  \bibfield  {author} {\bibinfo {author} {\bibfnamefont {A.}~\bibnamefont {Opala}}, \bibinfo {author} {\bibfnamefont {M.}~\bibnamefont {Furman}}, \bibinfo {author} {\bibfnamefont {M.}~\bibnamefont {Kr\'{o}l}}, \bibinfo {author} {\bibfnamefont {R.}~\bibnamefont {Mirek}}, \bibinfo {author} {\bibfnamefont {K.}~\bibnamefont {Tyszka}}, \bibinfo {author} {\bibfnamefont {B.}~\bibnamefont {Seredy\'{n}ski}}, \bibinfo {author} {\bibfnamefont {W.}~\bibnamefont {Pacuski}}, \bibinfo {author} {\bibfnamefont {J.}~\bibnamefont {Szczytko}}, \bibinfo {author} {\bibfnamefont {M.}~\bibnamefont {Matuszewski}},\ and\ \bibinfo {author} {\bibfnamefont {B.}~\bibnamefont {Piętka}},\ }\bibfield  {title} {\bibinfo {title} {Natural exceptional points in the excitation spectrum of a light-matter system},\ }\href {https://doi.org/10.1364/OPTICA.497170} {\bibfield  {journal} {\bibinfo  {journal} {Optica}\ }\textbf {\bibinfo {volume} {10}},\ \bibinfo {pages} {1111} (\bibinfo {year} {2023})}\BibitemShut {NoStop}%
\bibitem [{\citenamefont {Rahmani}\ \emph {et~al.}(2024)\citenamefont {Rahmani}, \citenamefont {Opala},\ and\ \citenamefont {Matuszewski}}]{PhysRevB.109.085311}%
  \BibitemOpen
  \bibfield  {author} {\bibinfo {author} {\bibfnamefont {A.}~\bibnamefont {Rahmani}}, \bibinfo {author} {\bibfnamefont {A.}~\bibnamefont {Opala}},\ and\ \bibinfo {author} {\bibfnamefont {M.}~\bibnamefont {Matuszewski}},\ }\bibfield  {title} {\bibinfo {title} {Exceptional points and phase transitions in non-{H}ermitian nonlinear binary systems},\ }\href {https://doi.org/10.1103/PhysRevB.109.085311} {\bibfield  {journal} {\bibinfo  {journal} {Phys. Rev. B}\ }\textbf {\bibinfo {volume} {109}},\ \bibinfo {pages} {085311} (\bibinfo {year} {2024})}\BibitemShut {NoStop}%
\bibitem [{\citenamefont {Assawaworrarit}\ \emph {et~al.}(2017)\citenamefont {Assawaworrarit}, \citenamefont {Yu},\ and\ \citenamefont {Fan}}]{assawaworrarit2017robust}%
  \BibitemOpen
  \bibfield  {author} {\bibinfo {author} {\bibfnamefont {S.}~\bibnamefont {Assawaworrarit}}, \bibinfo {author} {\bibfnamefont {X.}~\bibnamefont {Yu}},\ and\ \bibinfo {author} {\bibfnamefont {S.}~\bibnamefont {Fan}},\ }\bibfield  {title} {\bibinfo {title} {Robust wireless power transfer using a nonlinear parity--time-symmetric circuit},\ }\href {https://doi.org/10.1038/nature22404} {\bibfield  {journal} {\bibinfo  {journal} {Nature}\ }\textbf {\bibinfo {volume} {546}},\ \bibinfo {pages} {387} (\bibinfo {year} {2017})}\BibitemShut {NoStop}%
\bibitem [{\citenamefont {Ramezanpour}(2024)}]{Ramezanpour:24}%
  \BibitemOpen
  \bibfield  {author} {\bibinfo {author} {\bibfnamefont {S.}~\bibnamefont {Ramezanpour}},\ }\bibfield  {title} {\bibinfo {title} {Dynamic of time-independent and time-dependent asymmetric {Gross-Pitaevskii} equation around exceptional point},\ }\href {https://doi.org/10.1364/OPTCON.535433} {\bibfield  {journal} {\bibinfo  {journal} {Opt. Continuum}\ }\textbf {\bibinfo {volume} {3}},\ \bibinfo {pages} {1907} (\bibinfo {year} {2024})}\BibitemShut {NoStop}%
\bibitem [{\citenamefont {Wingenbach}\ \emph {et~al.}(2024)\citenamefont {Wingenbach}, \citenamefont {Schumacher},\ and\ \citenamefont {Ma}}]{PhysRevResearch.6.013148}%
  \BibitemOpen
  \bibfield  {author} {\bibinfo {author} {\bibfnamefont {J.}~\bibnamefont {Wingenbach}}, \bibinfo {author} {\bibfnamefont {S.}~\bibnamefont {Schumacher}},\ and\ \bibinfo {author} {\bibfnamefont {X.}~\bibnamefont {Ma}},\ }\bibfield  {title} {\bibinfo {title} {Manipulating spectral topology and exceptional points by nonlinearity in non-{H}ermitian polariton systems},\ }\href {https://doi.org/10.1103/PhysRevResearch.6.013148} {\bibfield  {journal} {\bibinfo  {journal} {Phys. Rev. Res.}\ }\textbf {\bibinfo {volume} {6}},\ \bibinfo {pages} {013148} (\bibinfo {year} {2024})}\BibitemShut {NoStop}%
\bibitem [{\citenamefont {Fang}\ \emph {et~al.}(2025)\citenamefont {Fang}, \citenamefont {Bai}, \citenamefont {Guo}, \citenamefont {Liu}, \citenamefont {Li},\ and\ \citenamefont {Xiao}}]{PhysRevB.111.L161102}%
  \BibitemOpen
  \bibfield  {author} {\bibinfo {author} {\bibfnamefont {L.}~\bibnamefont {Fang}}, \bibinfo {author} {\bibfnamefont {K.}~\bibnamefont {Bai}}, \bibinfo {author} {\bibfnamefont {C.}~\bibnamefont {Guo}}, \bibinfo {author} {\bibfnamefont {T.-R.}\ \bibnamefont {Liu}}, \bibinfo {author} {\bibfnamefont {J.-Z.}\ \bibnamefont {Li}},\ and\ \bibinfo {author} {\bibfnamefont {M.}~\bibnamefont {Xiao}},\ }\bibfield  {title} {\bibinfo {title} {Exceptional features in nonlinear {H}ermitian systems},\ }\href {https://doi.org/10.1103/PhysRevB.111.L161102} {\bibfield  {journal} {\bibinfo  {journal} {Phys. Rev. B}\ }\textbf {\bibinfo {volume} {111}},\ \bibinfo {pages} {L161102} (\bibinfo {year} {2025})}\BibitemShut {NoStop}%
\bibitem [{\citenamefont {Zhen}\ \emph {et~al.}(2015)\citenamefont {Zhen}, \citenamefont {Hsu}, \citenamefont {Igarashi}, \citenamefont {Lu}, \citenamefont {Kaminer}, \citenamefont {Pick}, \citenamefont {Chua}, \citenamefont {Joannopoulos},\ and\ \citenamefont {Solja{\v{c}}i{\'c}}}]{zhen2015spawning}%
  \BibitemOpen
  \bibfield  {author} {\bibinfo {author} {\bibfnamefont {B.}~\bibnamefont {Zhen}}, \bibinfo {author} {\bibfnamefont {C.~W.}\ \bibnamefont {Hsu}}, \bibinfo {author} {\bibfnamefont {Y.}~\bibnamefont {Igarashi}}, \bibinfo {author} {\bibfnamefont {L.}~\bibnamefont {Lu}}, \bibinfo {author} {\bibfnamefont {I.}~\bibnamefont {Kaminer}}, \bibinfo {author} {\bibfnamefont {A.}~\bibnamefont {Pick}}, \bibinfo {author} {\bibfnamefont {S.-L.}\ \bibnamefont {Chua}}, \bibinfo {author} {\bibfnamefont {J.~D.}\ \bibnamefont {Joannopoulos}},\ and\ \bibinfo {author} {\bibfnamefont {M.}~\bibnamefont {Solja{\v{c}}i{\'c}}},\ }\bibfield  {title} {\bibinfo {title} {Spawning rings of exceptional points out of {D}irac cones},\ }\href {https://doi.org/10.1038/nature14889} {\bibfield  {journal} {\bibinfo  {journal} {Nature}\ }\textbf {\bibinfo {volume} {525}},\ \bibinfo {pages} {354} (\bibinfo {year} {2015})}\BibitemShut {NoStop}%
\bibitem [{\citenamefont {Cerjan}\ \emph {et~al.}(2019)\citenamefont {Cerjan}, \citenamefont {Huang}, \citenamefont {Wang}, \citenamefont {Chen}, \citenamefont {Chong},\ and\ \citenamefont {Rechtsman}}]{cerjan2019experimental}%
  \BibitemOpen
  \bibfield  {author} {\bibinfo {author} {\bibfnamefont {A.}~\bibnamefont {Cerjan}}, \bibinfo {author} {\bibfnamefont {S.}~\bibnamefont {Huang}}, \bibinfo {author} {\bibfnamefont {M.}~\bibnamefont {Wang}}, \bibinfo {author} {\bibfnamefont {K.~P.}\ \bibnamefont {Chen}}, \bibinfo {author} {\bibfnamefont {Y.}~\bibnamefont {Chong}},\ and\ \bibinfo {author} {\bibfnamefont {M.~C.}\ \bibnamefont {Rechtsman}},\ }\bibfield  {title} {\bibinfo {title} {Experimental realization of a {W}eyl exceptional ring},\ }\href {https://doi.org/10.1038/s41566-019-0453-z} {\bibfield  {journal} {\bibinfo  {journal} {Nat. Photonics}\ }\textbf {\bibinfo {volume} {13}},\ \bibinfo {pages} {623} (\bibinfo {year} {2019})}\BibitemShut {NoStop}%
\bibitem [{\citenamefont {Peng-Zhen}\ \emph {et~al.}(2025)\citenamefont {Peng-Zhen}, \citenamefont {Zhou-Tao},\ and\ \citenamefont {Yuan-Gang}}]{10.1088/1674-1056/addcbe}%
  \BibitemOpen
  \bibfield  {author} {\bibinfo {author} {\bibfnamefont {S.}~\bibnamefont {Peng-Zhen}}, \bibinfo {author} {\bibfnamefont {L.}~\bibnamefont {Zhou-Tao}},\ and\ \bibinfo {author} {\bibfnamefont {D.}~\bibnamefont {Yuan-Gang}},\ }\bibfield  {title} {\bibinfo {title} {Exceptional rings and non-{A}belian topology in non-{H}ermitian high-spin systems},\ }\href@noop {} {\bibfield  {journal} {\bibinfo  {journal} {Chinese Physics B}\ } (\bibinfo {year} {2025})}\BibitemShut {NoStop}%
\bibitem [{\citenamefont {Yoshida}\ and\ \citenamefont {Hatsugai}(2019)}]{yoshida2019exceptional}%
  \BibitemOpen
  \bibfield  {author} {\bibinfo {author} {\bibfnamefont {T.}~\bibnamefont {Yoshida}}\ and\ \bibinfo {author} {\bibfnamefont {Y.}~\bibnamefont {Hatsugai}},\ }\bibfield  {title} {\bibinfo {title} {Exceptional rings protected by emergent symmetry for mechanical systems},\ }\href {https://doi.org/10.1103/PhysRevB.100.054109} {\bibfield  {journal} {\bibinfo  {journal} {Phys. Rev. B}\ }\textbf {\bibinfo {volume} {100}},\ \bibinfo {pages} {054109} (\bibinfo {year} {2019})}\BibitemShut {NoStop}%
\bibitem [{\citenamefont {He}\ and\ \citenamefont {Huang}(2020)}]{he2020floquet}%
  \BibitemOpen
  \bibfield  {author} {\bibinfo {author} {\bibfnamefont {P.}~\bibnamefont {He}}\ and\ \bibinfo {author} {\bibfnamefont {Z.-H.}\ \bibnamefont {Huang}},\ }\bibfield  {title} {\bibinfo {title} {Floquet engineering and simulating exceptional rings with a quantum spin system},\ }\href {https://doi.org/10.1103/PhysRevA.102.062201} {\bibfield  {journal} {\bibinfo  {journal} {Phys. Rev. A}\ }\textbf {\bibinfo {volume} {102}},\ \bibinfo {pages} {062201} (\bibinfo {year} {2020})}\BibitemShut {NoStop}%
\bibitem [{\citenamefont {Xu}\ \emph {et~al.}(2022)\citenamefont {Xu}, \citenamefont {Li}, \citenamefont {Zhou}, \citenamefont {Li}, \citenamefont {Li}, \citenamefont {Fan}, \citenamefont {Zhang}, \citenamefont {Christodoulides},\ and\ \citenamefont {Qiu}}]{xu2022observation}%
  \BibitemOpen
  \bibfield  {author} {\bibinfo {author} {\bibfnamefont {G.}~\bibnamefont {Xu}}, \bibinfo {author} {\bibfnamefont {W.}~\bibnamefont {Li}}, \bibinfo {author} {\bibfnamefont {X.}~\bibnamefont {Zhou}}, \bibinfo {author} {\bibfnamefont {H.}~\bibnamefont {Li}}, \bibinfo {author} {\bibfnamefont {Y.}~\bibnamefont {Li}}, \bibinfo {author} {\bibfnamefont {S.}~\bibnamefont {Fan}}, \bibinfo {author} {\bibfnamefont {S.}~\bibnamefont {Zhang}}, \bibinfo {author} {\bibfnamefont {D.~N.}\ \bibnamefont {Christodoulides}},\ and\ \bibinfo {author} {\bibfnamefont {C.-W.}\ \bibnamefont {Qiu}},\ }\bibfield  {title} {\bibinfo {title} {Observation of {W}eyl exceptional rings in thermal diffusion},\ }\href {https://doi.org/10.1073/pnas.2110018119} {\bibfield  {journal} {\bibinfo  {journal} {Proc. Natl. Acad. Sci.}\ }\textbf {\bibinfo {volume} {119}},\ \bibinfo {pages} {e2110018119} (\bibinfo {year} {2022})}\BibitemShut {NoStop}%
\bibitem [{\citenamefont {Yoshida}\ \emph {et~al.}(2019)\citenamefont {Yoshida}, \citenamefont {Peters}, \citenamefont {Kawakami},\ and\ \citenamefont {Hatsugai}}]{yoshida2019symmetry}%
  \BibitemOpen
  \bibfield  {author} {\bibinfo {author} {\bibfnamefont {T.}~\bibnamefont {Yoshida}}, \bibinfo {author} {\bibfnamefont {R.}~\bibnamefont {Peters}}, \bibinfo {author} {\bibfnamefont {N.}~\bibnamefont {Kawakami}},\ and\ \bibinfo {author} {\bibfnamefont {Y.}~\bibnamefont {Hatsugai}},\ }\bibfield  {title} {\bibinfo {title} {Symmetry-protected exceptional rings in two-dimensional correlated systems with chiral symmetry},\ }\href {https://doi.org/10.1103/PhysRevB.99.121101} {\bibfield  {journal} {\bibinfo  {journal} {Phys. Rev. B}\ }\textbf {\bibinfo {volume} {99}},\ \bibinfo {pages} {121101} (\bibinfo {year} {2019})}\BibitemShut {NoStop}%
\bibitem [{\citenamefont {Cao}\ \emph {et~al.}(2020)\citenamefont {Cao}, \citenamefont {Yi},\ and\ \citenamefont {Wang}}]{cao2020band}%
  \BibitemOpen
  \bibfield  {author} {\bibinfo {author} {\bibfnamefont {J.}~\bibnamefont {Cao}}, \bibinfo {author} {\bibfnamefont {X.}~\bibnamefont {Yi}},\ and\ \bibinfo {author} {\bibfnamefont {H.-F.}\ \bibnamefont {Wang}},\ }\bibfield  {title} {\bibinfo {title} {Band structure and the exceptional ring in a two-dimensional superconducting circuit lattice},\ }\href {https://doi.org/10.1103/PhysRevA.102.032619} {\bibfield  {journal} {\bibinfo  {journal} {Phys. Rev. A}\ }\textbf {\bibinfo {volume} {102}},\ \bibinfo {pages} {032619} (\bibinfo {year} {2020})}\BibitemShut {NoStop}%
\bibitem [{\citenamefont {Wang}\ \emph {et~al.}(2019{\natexlab{b}})\citenamefont {Wang}, \citenamefont {Xie}, \citenamefont {Gupta}, \citenamefont {Zhu}, \citenamefont {Liu}, \citenamefont {Liu}, \citenamefont {Lu},\ and\ \citenamefont {Chen}}]{PhysRevB.100.165134}%
  \BibitemOpen
  \bibfield  {author} {\bibinfo {author} {\bibfnamefont {H.}~\bibnamefont {Wang}}, \bibinfo {author} {\bibfnamefont {B.}~\bibnamefont {Xie}}, \bibinfo {author} {\bibfnamefont {S.~K.}\ \bibnamefont {Gupta}}, \bibinfo {author} {\bibfnamefont {X.}~\bibnamefont {Zhu}}, \bibinfo {author} {\bibfnamefont {L.}~\bibnamefont {Liu}}, \bibinfo {author} {\bibfnamefont {X.}~\bibnamefont {Liu}}, \bibinfo {author} {\bibfnamefont {M.}~\bibnamefont {Lu}},\ and\ \bibinfo {author} {\bibfnamefont {Y.}~\bibnamefont {Chen}},\ }\bibfield  {title} {\bibinfo {title} {Exceptional concentric rings in a non-hermitian bilayer photonic system},\ }\href {https://doi.org/10.1103/PhysRevB.100.165134} {\bibfield  {journal} {\bibinfo  {journal} {Phys. Rev. B}\ }\textbf {\bibinfo {volume} {100}},\ \bibinfo {pages} {165134} (\bibinfo {year} {2019}{\natexlab{b}})}\BibitemShut {NoStop}%
\bibitem [{\citenamefont {Liu}\ \emph {et~al.}(2022)\citenamefont {Liu}, \citenamefont {Li}, \citenamefont {Chen}, \citenamefont {Tang}, \citenamefont {Chen}, \citenamefont {Liang}, \citenamefont {Ma},\ and\ \citenamefont {Cheng}}]{liu2022experimental}%
  \BibitemOpen
  \bibfield  {author} {\bibinfo {author} {\bibfnamefont {J.-j.}\ \bibnamefont {Liu}}, \bibinfo {author} {\bibfnamefont {Z.-w.}\ \bibnamefont {Li}}, \bibinfo {author} {\bibfnamefont {Z.-G.}\ \bibnamefont {Chen}}, \bibinfo {author} {\bibfnamefont {W.}~\bibnamefont {Tang}}, \bibinfo {author} {\bibfnamefont {A.}~\bibnamefont {Chen}}, \bibinfo {author} {\bibfnamefont {B.}~\bibnamefont {Liang}}, \bibinfo {author} {\bibfnamefont {G.}~\bibnamefont {Ma}},\ and\ \bibinfo {author} {\bibfnamefont {J.-C.}\ \bibnamefont {Cheng}},\ }\bibfield  {title} {\bibinfo {title} {Experimental realization of {W}eyl exceptional rings in a synthetic three-dimensional non-{H}ermitian phononic crystal},\ }\href {https://doi.org/10.1103/PhysRevLett.129.084301} {\bibfield  {journal} {\bibinfo  {journal} {Phys Rev Lett.}\ }\textbf {\bibinfo {volume} {129}},\ \bibinfo {pages} {084301} (\bibinfo {year} {2022})}\BibitemShut {NoStop}%
\bibitem [{\citenamefont {Isobe}\ \emph {et~al.}(2023)\citenamefont {Isobe}, \citenamefont {Yoshida},\ and\ \citenamefont {Hatsugai}}]{isobe2023symmetry}%
  \BibitemOpen
  \bibfield  {author} {\bibinfo {author} {\bibfnamefont {T.}~\bibnamefont {Isobe}}, \bibinfo {author} {\bibfnamefont {T.}~\bibnamefont {Yoshida}},\ and\ \bibinfo {author} {\bibfnamefont {Y.}~\bibnamefont {Hatsugai}},\ }\bibfield  {title} {\bibinfo {title} {A symmetry-protected exceptional ring in a photonic crystal with negative index media},\ }\href {https://doi.org/10.1515/nanoph-2022-0747} {\bibfield  {journal} {\bibinfo  {journal} {Nanophotonics}\ }\textbf {\bibinfo {volume} {12}},\ \bibinfo {pages} {2335} (\bibinfo {year} {2023})}\BibitemShut {NoStop}%
\bibitem [{\citenamefont {Wang}\ \emph {et~al.}(2023)\citenamefont {Wang}, \citenamefont {Ge}, \citenamefont {Yuan}, \citenamefont {Jia},\ and\ \citenamefont {Sun}}]{wang2023exceptional}%
  \BibitemOpen
  \bibfield  {author} {\bibinfo {author} {\bibfnamefont {B.-B.}\ \bibnamefont {Wang}}, \bibinfo {author} {\bibfnamefont {Y.}~\bibnamefont {Ge}}, \bibinfo {author} {\bibfnamefont {S.-Q.}\ \bibnamefont {Yuan}}, \bibinfo {author} {\bibfnamefont {D.}~\bibnamefont {Jia}},\ and\ \bibinfo {author} {\bibfnamefont {H.-X.}\ \bibnamefont {Sun}},\ }\bibfield  {title} {\bibinfo {title} {Exceptional ring by non-{H}ermitian sonic crystals},\ }\href@noop {} {\bibfield  {journal} {\bibinfo  {journal} {Prog. Electromagn. Res.}\ }\textbf {\bibinfo {volume} {176}},\ \bibinfo {pages} {1} (\bibinfo {year} {2023})}\BibitemShut {NoStop}%
\bibitem [{\citenamefont {Isobe}\ \emph {et~al.}(2025)\citenamefont {Isobe}, \citenamefont {Yoshida},\ and\ \citenamefont {Hatsugai}}]{isobe2025topological}%
  \BibitemOpen
  \bibfield  {author} {\bibinfo {author} {\bibfnamefont {T.}~\bibnamefont {Isobe}}, \bibinfo {author} {\bibfnamefont {T.}~\bibnamefont {Yoshida}},\ and\ \bibinfo {author} {\bibfnamefont {Y.}~\bibnamefont {Hatsugai}},\ }\bibfield  {title} {\bibinfo {title} {Topological photonics of generalized and nonlinear eigenvalue equations},\ }\href@noop {} {\bibfield  {journal} {\bibinfo  {journal} {J. Phys. Soc. Jpn.}\ }\textbf {\bibinfo {volume} {94}},\ \bibinfo {pages} {101002} (\bibinfo {year} {2025})}\BibitemShut {NoStop}%
\bibitem [{\citenamefont {Zhao}\ \emph {et~al.}(2025)\citenamefont {Zhao}, \citenamefont {Wang}, \citenamefont {Liu}, \citenamefont {Shi},\ and\ \citenamefont {Zi}}]{zhao2025magnetically}%
  \BibitemOpen
  \bibfield  {author} {\bibinfo {author} {\bibfnamefont {X.}~\bibnamefont {Zhao}}, \bibinfo {author} {\bibfnamefont {J.}~\bibnamefont {Wang}}, \bibinfo {author} {\bibfnamefont {W.}~\bibnamefont {Liu}}, \bibinfo {author} {\bibfnamefont {L.}~\bibnamefont {Shi}},\ and\ \bibinfo {author} {\bibfnamefont {J.}~\bibnamefont {Zi}},\ }\bibfield  {title} {\bibinfo {title} {Magnetically induced topological evolutions of exceptional points in photonic bands},\ }\href@noop {} {\bibfield  {journal} {\bibinfo  {journal} {Phys. Rev. Lett.}\ }\textbf {\bibinfo {volume} {135}},\ \bibinfo {pages} {046203} (\bibinfo {year} {2025})}\BibitemShut {NoStop}%
\bibitem [{\citenamefont {Na}\ \emph {et~al.}(2025)\citenamefont {Na}, \citenamefont {Zhang},\ and\ \citenamefont {Feng}}]{clbx-mh6y}%
  \BibitemOpen
  \bibfield  {author} {\bibinfo {author} {\bibfnamefont {Z.-H.}\ \bibnamefont {Na}}, \bibinfo {author} {\bibfnamefont {X.-L.}\ \bibnamefont {Zhang}},\ and\ \bibinfo {author} {\bibfnamefont {J.}~\bibnamefont {Feng}},\ }\bibfield  {title} {\bibinfo {title} {Multiple exceptional rings in one-dimensional perovskite photonic crystals},\ }\href {https://doi.org/10.1103/clbx-mh6y} {\bibfield  {journal} {\bibinfo  {journal} {Phys. Rev. Lett.}\ }\textbf {\bibinfo {volume} {135}},\ \bibinfo {pages} {083802} (\bibinfo {year} {2025})}\BibitemShut {NoStop}%
\bibitem [{\citenamefont {Kolkowski}\ \emph {et~al.}(2021)\citenamefont {Kolkowski}, \citenamefont {Kovaios},\ and\ \citenamefont {Koenderink}}]{kolkowski2021pseudochirality}%
  \BibitemOpen
  \bibfield  {author} {\bibinfo {author} {\bibfnamefont {R.}~\bibnamefont {Kolkowski}}, \bibinfo {author} {\bibfnamefont {S.}~\bibnamefont {Kovaios}},\ and\ \bibinfo {author} {\bibfnamefont {A.~F.}\ \bibnamefont {Koenderink}},\ }\bibfield  {title} {\bibinfo {title} {Pseudochirality at exceptional rings of optical metasurfaces},\ }\href {https://doi.org/10.1103/PhysRevResearch.3.023185} {\bibfield  {journal} {\bibinfo  {journal} {Phys. Rev. Res.}\ }\textbf {\bibinfo {volume} {3}},\ \bibinfo {pages} {023185} (\bibinfo {year} {2021})}\BibitemShut {NoStop}%
\bibitem [{\citenamefont {Cerjan}\ \emph {et~al.}(2018)\citenamefont {Cerjan}, \citenamefont {Xiao}, \citenamefont {Yuan},\ and\ \citenamefont {Fan}}]{cerjan2018effects}%
  \BibitemOpen
  \bibfield  {author} {\bibinfo {author} {\bibfnamefont {A.}~\bibnamefont {Cerjan}}, \bibinfo {author} {\bibfnamefont {M.}~\bibnamefont {Xiao}}, \bibinfo {author} {\bibfnamefont {L.}~\bibnamefont {Yuan}},\ and\ \bibinfo {author} {\bibfnamefont {S.}~\bibnamefont {Fan}},\ }\bibfield  {title} {\bibinfo {title} {Effects of non-{H}ermitian perturbations on {W}eyl {H}amiltonians with arbitrary topological charges},\ }\href {https://doi.org/10.1103/PhysRevB.97.075128} {\bibfield  {journal} {\bibinfo  {journal} {Phys. Rev. B}\ }\textbf {\bibinfo {volume} {97}},\ \bibinfo {pages} {075128} (\bibinfo {year} {2018})}\BibitemShut {NoStop}%
\bibitem [{\citenamefont {Liu}\ \emph {et~al.}(2021)\citenamefont {Liu}, \citenamefont {He}, \citenamefont {Yang},\ and\ \citenamefont {Nori}}]{liu2021higher}%
  \BibitemOpen
  \bibfield  {author} {\bibinfo {author} {\bibfnamefont {T.}~\bibnamefont {Liu}}, \bibinfo {author} {\bibfnamefont {J.~J.}\ \bibnamefont {He}}, \bibinfo {author} {\bibfnamefont {Z.}~\bibnamefont {Yang}},\ and\ \bibinfo {author} {\bibfnamefont {F.}~\bibnamefont {Nori}},\ }\bibfield  {title} {\bibinfo {title} {Higher-order {W}eyl-exceptional-ring semimetals},\ }\href {https://doi.org/10.1103/PhysRevLett.127.196801} {\bibfield  {journal} {\bibinfo  {journal} {Phys Rev Lett.}\ }\textbf {\bibinfo {volume} {127}},\ \bibinfo {pages} {196801} (\bibinfo {year} {2021})}\BibitemShut {NoStop}%
\bibitem [{\citenamefont {Qin}\ \emph {et~al.}(2025)\citenamefont {Qin}, \citenamefont {Shen},\ and\ \citenamefont {Lee}}]{8g3q-qrpg}%
  \BibitemOpen
  \bibfield  {author} {\bibinfo {author} {\bibfnamefont {F.}~\bibnamefont {Qin}}, \bibinfo {author} {\bibfnamefont {R.}~\bibnamefont {Shen}},\ and\ \bibinfo {author} {\bibfnamefont {C.~H.}\ \bibnamefont {Lee}},\ }\bibfield  {title} {\bibinfo {title} {Nonlinear {H}all effects with an exceptional ring},\ }\href@noop {} {\bibfield  {journal} {\bibinfo  {journal} {Phys. Rev. B}\ }\textbf {\bibinfo {volume} {111}},\ \bibinfo {pages} {245413} (\bibinfo {year} {2025})}\BibitemShut {NoStop}%
\bibitem [{\citenamefont {Kwong}\ \emph {et~al.}(2025)\citenamefont {Kwong}, \citenamefont {Wingenbach}, \citenamefont {Ares}, \citenamefont {Sperling}, \citenamefont {Ma}, \citenamefont {Schumacher},\ and\ \citenamefont {Binder}}]{kwong2025universal}%
  \BibitemOpen
  \bibfield  {author} {\bibinfo {author} {\bibfnamefont {N.~H.}\ \bibnamefont {Kwong}}, \bibinfo {author} {\bibfnamefont {J.}~\bibnamefont {Wingenbach}}, \bibinfo {author} {\bibfnamefont {L.}~\bibnamefont {Ares}}, \bibinfo {author} {\bibfnamefont {J.}~\bibnamefont {Sperling}}, \bibinfo {author} {\bibfnamefont {X.}~\bibnamefont {Ma}}, \bibinfo {author} {\bibfnamefont {S.}~\bibnamefont {Schumacher}},\ and\ \bibinfo {author} {\bibfnamefont {R.}~\bibnamefont {Binder}},\ }\bibfield  {title} {\bibinfo {title} {Universal topology of exceptional points in nonlinear non-{H}ermitian systems},\ }\href@noop {} {\bibfield  {journal} {\bibinfo  {journal} {arXiv:2502.19236 [physics.optics]}\ } (\bibinfo {year} {2025})}\BibitemShut {NoStop}%
\bibitem [{\citenamefont {Kr{\'o}l}\ \emph {et~al.}(2022)\citenamefont {Kr{\'o}l}, \citenamefont {Septembre}, \citenamefont {Oliwa}, \citenamefont {K{\k{e}}dziora}, \citenamefont {{\L}empicka-Mirek}, \citenamefont {Muszy{\'n}ski}, \citenamefont {Mazur}, \citenamefont {Morawiak}, \citenamefont {Piecek}, \citenamefont {Kula}, \citenamefont {Bardyszewski}, \citenamefont {Lagoudakis}, \citenamefont {Solnyshkov}, \citenamefont {Malpuech}, \citenamefont {Pi{\k{e}}tka},\ and\ \citenamefont {Szczytko}}]{Krol2022}%
  \BibitemOpen
  \bibfield  {author} {\bibinfo {author} {\bibfnamefont {M.}~\bibnamefont {Kr{\'o}l}}, \bibinfo {author} {\bibfnamefont {I.}~\bibnamefont {Septembre}}, \bibinfo {author} {\bibfnamefont {P.}~\bibnamefont {Oliwa}}, \bibinfo {author} {\bibfnamefont {M.}~\bibnamefont {K{\k{e}}dziora}}, \bibinfo {author} {\bibfnamefont {K.}~\bibnamefont {{\L}empicka-Mirek}}, \bibinfo {author} {\bibfnamefont {M.}~\bibnamefont {Muszy{\'n}ski}}, \bibinfo {author} {\bibfnamefont {R.}~\bibnamefont {Mazur}}, \bibinfo {author} {\bibfnamefont {P.}~\bibnamefont {Morawiak}}, \bibinfo {author} {\bibfnamefont {W.}~\bibnamefont {Piecek}}, \bibinfo {author} {\bibfnamefont {P.}~\bibnamefont {Kula}}, \bibinfo {author} {\bibfnamefont {W.}~\bibnamefont {Bardyszewski}}, \bibinfo {author} {\bibfnamefont {P.~G.}\ \bibnamefont {Lagoudakis}}, \bibinfo {author} {\bibfnamefont {D.~D.}\ \bibnamefont {Solnyshkov}}, \bibinfo {author} {\bibfnamefont {G.}~\bibnamefont {Malpuech}}, \bibinfo {author} {\bibfnamefont {B.}~\bibnamefont {Pi{\k{e}}tka}},\ and\
  \bibinfo {author} {\bibfnamefont {J.}~\bibnamefont {Szczytko}},\ }\bibfield  {title} {\bibinfo {title} {Annihilation of exceptional points from different {D}irac valleys in a 2{D} photonic system},\ }\href {https://doi.org/10.1038/s41467-022-33001-9} {\bibfield  {journal} {\bibinfo  {journal} {Nat. Commun.}\ }\textbf {\bibinfo {volume} {13}},\ \bibinfo {pages} {5340} (\bibinfo {year} {2022})}\BibitemShut {NoStop}%
\bibitem [{\citenamefont {Oliwa}\ \emph {et~al.}(2024)\citenamefont {Oliwa}, \citenamefont {Bardyszewski},\ and\ \citenamefont {Szczytko}}]{PhysRevResearch.6.013324}%
  \BibitemOpen
  \bibfield  {author} {\bibinfo {author} {\bibfnamefont {P.}~\bibnamefont {Oliwa}}, \bibinfo {author} {\bibfnamefont {W.}~\bibnamefont {Bardyszewski}},\ and\ \bibinfo {author} {\bibfnamefont {J.}~\bibnamefont {Szczytko}},\ }\bibfield  {title} {\bibinfo {title} {Quantum mechanical-like approach with non-{H}ermitian effective {H}amiltonians in spin-orbit coupled optical cavities},\ }\href {https://doi.org/10.1103/PhysRevResearch.6.013324} {\bibfield  {journal} {\bibinfo  {journal} {Phys. Rev. Res.}\ }\textbf {\bibinfo {volume} {6}},\ \bibinfo {pages} {013324} (\bibinfo {year} {2024})}\BibitemShut {NoStop}%
\bibitem [{\citenamefont {Su}\ \emph {et~al.}(2021)\citenamefont {Su}, \citenamefont {Estrecho}, \citenamefont {Biegańska}, \citenamefont {Huang}, \citenamefont {Wurdack}, \citenamefont {Pieczarka}, \citenamefont {Truscott}, \citenamefont {Liew}, \citenamefont {Ostrovskaya},\ and\ \citenamefont {Xiong}}]{doi:10.1126/sciadv.abj8905}%
  \BibitemOpen
  \bibfield  {author} {\bibinfo {author} {\bibfnamefont {R.}~\bibnamefont {Su}}, \bibinfo {author} {\bibfnamefont {E.}~\bibnamefont {Estrecho}}, \bibinfo {author} {\bibfnamefont {D.}~\bibnamefont {Biegańska}}, \bibinfo {author} {\bibfnamefont {Y.}~\bibnamefont {Huang}}, \bibinfo {author} {\bibfnamefont {M.}~\bibnamefont {Wurdack}}, \bibinfo {author} {\bibfnamefont {M.}~\bibnamefont {Pieczarka}}, \bibinfo {author} {\bibfnamefont {A.~G.}\ \bibnamefont {Truscott}}, \bibinfo {author} {\bibfnamefont {T.~C.~H.}\ \bibnamefont {Liew}}, \bibinfo {author} {\bibfnamefont {E.~A.}\ \bibnamefont {Ostrovskaya}},\ and\ \bibinfo {author} {\bibfnamefont {Q.}~\bibnamefont {Xiong}},\ }\bibfield  {title} {\bibinfo {title} {Direct measurement of a non-{H}ermitian topological invariant in a hybrid light-matter system},\ }\href {https://doi.org/10.1126/sciadv.abj8905} {\bibfield  {journal} {\bibinfo  {journal} {Science Advances}\ }\textbf {\bibinfo {volume} {7}},\ \bibinfo {pages} {eabj8905} (\bibinfo {year} {2021})}\BibitemShut
  {NoStop}%
\bibitem [{\citenamefont {Hu}\ \emph {et~al.}(2023)\citenamefont {Hu}, \citenamefont {Ostrovskaya},\ and\ \citenamefont {Estrecho}}]{PhysRevB.108.115404}%
  \BibitemOpen
  \bibfield  {author} {\bibinfo {author} {\bibfnamefont {Y.-M.~R.}\ \bibnamefont {Hu}}, \bibinfo {author} {\bibfnamefont {E.~A.}\ \bibnamefont {Ostrovskaya}},\ and\ \bibinfo {author} {\bibfnamefont {E.}~\bibnamefont {Estrecho}},\ }\bibfield  {title} {\bibinfo {title} {Wave-packet dynamics in a non-{H}ermitian exciton-polariton system},\ }\href {https://doi.org/10.1103/PhysRevB.108.115404} {\bibfield  {journal} {\bibinfo  {journal} {Phys. Rev. B}\ }\textbf {\bibinfo {volume} {108}},\ \bibinfo {pages} {115404} (\bibinfo {year} {2023})}\BibitemShut {NoStop}%
\bibitem [{\citenamefont {Rajaei}\ \emph {et~al.}(2019)\citenamefont {Rajaei}, \citenamefont {Zeng}, \citenamefont {Albooyeh}, \citenamefont {Kamandi}, \citenamefont {Hanifeh}, \citenamefont {Capolino},\ and\ \citenamefont {Wickramasinghe}}]{Rajaei2019}%
  \BibitemOpen
  \bibfield  {author} {\bibinfo {author} {\bibfnamefont {M.}~\bibnamefont {Rajaei}}, \bibinfo {author} {\bibfnamefont {J.}~\bibnamefont {Zeng}}, \bibinfo {author} {\bibfnamefont {M.}~\bibnamefont {Albooyeh}}, \bibinfo {author} {\bibfnamefont {M.}~\bibnamefont {Kamandi}}, \bibinfo {author} {\bibfnamefont {M.}~\bibnamefont {Hanifeh}}, \bibinfo {author} {\bibfnamefont {F.}~\bibnamefont {Capolino}},\ and\ \bibinfo {author} {\bibfnamefont {H.~K.}\ \bibnamefont {Wickramasinghe}},\ }\bibfield  {title} {\bibinfo {title} {Giant circular dichroism at visible frequencies enabled by plasmonic ramp-shaped nanostructures},\ }\href {https://doi.org/10.1021/acsphotonics.8b01584} {\bibfield  {journal} {\bibinfo  {journal} {ACS Photonics}\ }\textbf {\bibinfo {volume} {6}},\ \bibinfo {pages} {924} (\bibinfo {year} {2019})}\BibitemShut {NoStop}%
\bibitem [{\citenamefont {Qian}\ \emph {et~al.}(2024)\citenamefont {Qian}, \citenamefont {Wang}, \citenamefont {Yang}, \citenamefont {Wang},\ and\ \citenamefont {Yan}}]{PhysRevApplied.22.054016}%
  \BibitemOpen
  \bibfield  {author} {\bibinfo {author} {\bibfnamefont {Y.-J.}\ \bibnamefont {Qian}}, \bibinfo {author} {\bibfnamefont {Y.-B.}\ \bibnamefont {Wang}}, \bibinfo {author} {\bibfnamefont {X.-D.}\ \bibnamefont {Yang}}, \bibinfo {author} {\bibfnamefont {J.}~\bibnamefont {Wang}},\ and\ \bibinfo {author} {\bibfnamefont {Y.-Z.}\ \bibnamefont {Yan}},\ }\bibfield  {title} {\bibinfo {title} {Manipulation of circular polarization and mechanical waves in gyroscopic metamaterials},\ }\href {https://doi.org/10.1103/PhysRevApplied.22.054016} {\bibfield  {journal} {\bibinfo  {journal} {Phys. Rev. Appl.}\ }\textbf {\bibinfo {volume} {22}},\ \bibinfo {pages} {054016} (\bibinfo {year} {2024})}\BibitemShut {NoStop}%
\bibitem [{\citenamefont {Rahimi-Iman}\ \emph {et~al.}(2011)\citenamefont {Rahimi-Iman}, \citenamefont {Schneider}, \citenamefont {Fischer}, \citenamefont {Holzinger}, \citenamefont {Amthor}, \citenamefont {H\"ofling}, \citenamefont {Reitzenstein}, \citenamefont {Worschech}, \citenamefont {Kamp},\ and\ \citenamefont {Forchel}}]{PhysRevB.84.165325}%
  \BibitemOpen
  \bibfield  {author} {\bibinfo {author} {\bibfnamefont {A.}~\bibnamefont {Rahimi-Iman}}, \bibinfo {author} {\bibfnamefont {C.}~\bibnamefont {Schneider}}, \bibinfo {author} {\bibfnamefont {J.}~\bibnamefont {Fischer}}, \bibinfo {author} {\bibfnamefont {S.}~\bibnamefont {Holzinger}}, \bibinfo {author} {\bibfnamefont {M.}~\bibnamefont {Amthor}}, \bibinfo {author} {\bibfnamefont {S.}~\bibnamefont {H\"ofling}}, \bibinfo {author} {\bibfnamefont {S.}~\bibnamefont {Reitzenstein}}, \bibinfo {author} {\bibfnamefont {L.}~\bibnamefont {Worschech}}, \bibinfo {author} {\bibfnamefont {M.}~\bibnamefont {Kamp}},\ and\ \bibinfo {author} {\bibfnamefont {A.}~\bibnamefont {Forchel}},\ }\bibfield  {title} {\bibinfo {title} {Zeeman splitting and diamagnetic shift of spatially confined quantum-well exciton polaritons in an external magnetic field},\ }\href {https://doi.org/10.1103/PhysRevB.84.165325} {\bibfield  {journal} {\bibinfo  {journal} {Phys. Rev. B}\ }\textbf {\bibinfo {volume} {84}},\ \bibinfo {pages} {165325} (\bibinfo
  {year} {2011})}\BibitemShut {NoStop}%
\bibitem [{\citenamefont {Polimeno}\ \emph {et~al.}(2021)\citenamefont {Polimeno}, \citenamefont {Lerario}, \citenamefont {De~Giorgi}, \citenamefont {De~Marco}, \citenamefont {Dominici}, \citenamefont {Todisco}, \citenamefont {Coriolano}, \citenamefont {Ardizzone}, \citenamefont {Pugliese}, \citenamefont {Prontera}, \citenamefont {Maiorano}, \citenamefont {Moliterni}, \citenamefont {Giannini}, \citenamefont {Olieric}, \citenamefont {Gigli}, \citenamefont {Ballarini}, \citenamefont {Xiong}, \citenamefont {Fieramosca}, \citenamefont {Solnyshkov}, \citenamefont {Malpuech},\ and\ \citenamefont {Sanvitto}}]{Polimeno2021}%
  \BibitemOpen
  \bibfield  {author} {\bibinfo {author} {\bibfnamefont {L.}~\bibnamefont {Polimeno}}, \bibinfo {author} {\bibfnamefont {G.}~\bibnamefont {Lerario}}, \bibinfo {author} {\bibfnamefont {M.}~\bibnamefont {De~Giorgi}}, \bibinfo {author} {\bibfnamefont {L.}~\bibnamefont {De~Marco}}, \bibinfo {author} {\bibfnamefont {L.}~\bibnamefont {Dominici}}, \bibinfo {author} {\bibfnamefont {F.}~\bibnamefont {Todisco}}, \bibinfo {author} {\bibfnamefont {A.}~\bibnamefont {Coriolano}}, \bibinfo {author} {\bibfnamefont {V.}~\bibnamefont {Ardizzone}}, \bibinfo {author} {\bibfnamefont {M.}~\bibnamefont {Pugliese}}, \bibinfo {author} {\bibfnamefont {C.~T.}\ \bibnamefont {Prontera}}, \bibinfo {author} {\bibfnamefont {V.}~\bibnamefont {Maiorano}}, \bibinfo {author} {\bibfnamefont {A.}~\bibnamefont {Moliterni}}, \bibinfo {author} {\bibfnamefont {C.}~\bibnamefont {Giannini}}, \bibinfo {author} {\bibfnamefont {V.}~\bibnamefont {Olieric}}, \bibinfo {author} {\bibfnamefont {G.}~\bibnamefont {Gigli}}, \bibinfo {author} {\bibfnamefont
  {D.}~\bibnamefont {Ballarini}}, \bibinfo {author} {\bibfnamefont {Q.}~\bibnamefont {Xiong}}, \bibinfo {author} {\bibfnamefont {A.}~\bibnamefont {Fieramosca}}, \bibinfo {author} {\bibfnamefont {D.~D.}\ \bibnamefont {Solnyshkov}}, \bibinfo {author} {\bibfnamefont {G.}~\bibnamefont {Malpuech}},\ and\ \bibinfo {author} {\bibfnamefont {D.}~\bibnamefont {Sanvitto}},\ }\bibfield  {title} {\bibinfo {title} {Tuning of the {Berry curvature in 2D} perovskite polaritons},\ }\href {https://doi.org/10.1038/s41565-021-00977-2} {\bibfield  {journal} {\bibinfo  {journal} {Nat. Nanotechnol}\ }\textbf {\bibinfo {volume} {16}},\ \bibinfo {pages} {1349} (\bibinfo {year} {2021})}\BibitemShut {NoStop}%
\bibitem [{\citenamefont {Kammann}\ \emph {et~al.}(2012)\citenamefont {Kammann}, \citenamefont {Liew}, \citenamefont {Ohadi}, \citenamefont {Cilibrizzi}, \citenamefont {Tsotsis}, \citenamefont {Hatzopoulos}, \citenamefont {Savvidis}, \citenamefont {Kavokin},\ and\ \citenamefont {Lagoudakis}}]{PhysRevLett.109.036404}%
  \BibitemOpen
  \bibfield  {author} {\bibinfo {author} {\bibfnamefont {E.}~\bibnamefont {Kammann}}, \bibinfo {author} {\bibfnamefont {T.~C.~H.}\ \bibnamefont {Liew}}, \bibinfo {author} {\bibfnamefont {H.}~\bibnamefont {Ohadi}}, \bibinfo {author} {\bibfnamefont {P.}~\bibnamefont {Cilibrizzi}}, \bibinfo {author} {\bibfnamefont {P.}~\bibnamefont {Tsotsis}}, \bibinfo {author} {\bibfnamefont {Z.}~\bibnamefont {Hatzopoulos}}, \bibinfo {author} {\bibfnamefont {P.~G.}\ \bibnamefont {Savvidis}}, \bibinfo {author} {\bibfnamefont {A.~V.}\ \bibnamefont {Kavokin}},\ and\ \bibinfo {author} {\bibfnamefont {P.~G.}\ \bibnamefont {Lagoudakis}},\ }\bibfield  {title} {\bibinfo {title} {Nonlinear optical spin {H}all effect and long-range spin transport in polariton lasers},\ }\href {https://doi.org/10.1103/PhysRevLett.109.036404} {\bibfield  {journal} {\bibinfo  {journal} {Phys. Rev. Lett.}\ }\textbf {\bibinfo {volume} {109}},\ \bibinfo {pages} {036404} (\bibinfo {year} {2012})}\BibitemShut {NoStop}%
\bibitem [{\citenamefont {Waldherr}\ \emph {et~al.}(2018)\citenamefont {Waldherr}, \citenamefont {Lundt}, \citenamefont {Klaas}, \citenamefont {Betzold}, \citenamefont {Wurdack}, \citenamefont {Baumann}, \citenamefont {Estrecho}, \citenamefont {Nalitov}, \citenamefont {Cherotchenko}, \citenamefont {Cai}, \citenamefont {Ostrovskaya}, \citenamefont {Kavokin}, \citenamefont {Tongay}, \citenamefont {Klembt}, \citenamefont {Höfling},\ and\ \citenamefont {Schneider}}]{waldherr2018observation}%
  \BibitemOpen
  \bibfield  {author} {\bibinfo {author} {\bibfnamefont {M.}~\bibnamefont {Waldherr}}, \bibinfo {author} {\bibfnamefont {N.}~\bibnamefont {Lundt}}, \bibinfo {author} {\bibfnamefont {M.}~\bibnamefont {Klaas}}, \bibinfo {author} {\bibfnamefont {S.}~\bibnamefont {Betzold}}, \bibinfo {author} {\bibfnamefont {M.}~\bibnamefont {Wurdack}}, \bibinfo {author} {\bibfnamefont {V.}~\bibnamefont {Baumann}}, \bibinfo {author} {\bibfnamefont {E.}~\bibnamefont {Estrecho}}, \bibinfo {author} {\bibfnamefont {A.}~\bibnamefont {Nalitov}}, \bibinfo {author} {\bibfnamefont {E.}~\bibnamefont {Cherotchenko}}, \bibinfo {author} {\bibfnamefont {H.}~\bibnamefont {Cai}}, \bibinfo {author} {\bibfnamefont {E.~A.}\ \bibnamefont {Ostrovskaya}}, \bibinfo {author} {\bibfnamefont {A.~V.}\ \bibnamefont {Kavokin}}, \bibinfo {author} {\bibfnamefont {S.}~\bibnamefont {Tongay}}, \bibinfo {author} {\bibfnamefont {S.}~\bibnamefont {Klembt}}, \bibinfo {author} {\bibfnamefont {S.}~\bibnamefont {Höfling}},\ and\ \bibinfo {author} {\bibfnamefont
  {C.}~\bibnamefont {Schneider}},\ }\bibfield  {title} {\bibinfo {title} {Observation of bosonic condensation in a hybrid monolayer {MoSe2-GaAs} microcavity},\ }\href {https://doi.org/10.1038/s41467-018-05532-7} {\bibfield  {journal} {\bibinfo  {journal} {Nat. Commun.}\ }\textbf {\bibinfo {volume} {9}},\ \bibinfo {pages} {3286} (\bibinfo {year} {2018})}\BibitemShut {NoStop}%
\bibitem [{\citenamefont {Panzarini}\ \emph {et~al.}(1999{\natexlab{a}})\citenamefont {Panzarini}, \citenamefont {Andreani}, \citenamefont {Armitage}, \citenamefont {Baxter}, \citenamefont {Skolnick}, \citenamefont {Astratov}, \citenamefont {Roberts}, \citenamefont {Kavokin}, \citenamefont {Vladimirova},\ and\ \citenamefont {Kaliteevski}}]{panzarini1999exciton}%
  \BibitemOpen
  \bibfield  {author} {\bibinfo {author} {\bibfnamefont {G.}~\bibnamefont {Panzarini}}, \bibinfo {author} {\bibfnamefont {L.~C.}\ \bibnamefont {Andreani}}, \bibinfo {author} {\bibfnamefont {A.}~\bibnamefont {Armitage}}, \bibinfo {author} {\bibfnamefont {D.}~\bibnamefont {Baxter}}, \bibinfo {author} {\bibfnamefont {M.}~\bibnamefont {Skolnick}}, \bibinfo {author} {\bibfnamefont {V.}~\bibnamefont {Astratov}}, \bibinfo {author} {\bibfnamefont {J.}~\bibnamefont {Roberts}}, \bibinfo {author} {\bibfnamefont {A.~V.}\ \bibnamefont {Kavokin}}, \bibinfo {author} {\bibfnamefont {M.~R.}\ \bibnamefont {Vladimirova}},\ and\ \bibinfo {author} {\bibfnamefont {M.}~\bibnamefont {Kaliteevski}},\ }\bibfield  {title} {\bibinfo {title} {Exciton-light coupling in single and coupled semiconductor microcavities: Polariton dispersion and polarization splitting},\ }\href@noop {} {\bibfield  {journal} {\bibinfo  {journal} {Phys. Rev. B}\ }\textbf {\bibinfo {volume} {59}},\ \bibinfo {pages} {5082} (\bibinfo {year}
  {1999}{\natexlab{a}})}\BibitemShut {NoStop}%
\bibitem [{\citenamefont {Kavokin}\ \emph {et~al.}(2004)\citenamefont {Kavokin}, \citenamefont {Shelykh}, \citenamefont {Kavokin}, \citenamefont {Malpuech},\ and\ \citenamefont {Bigenwald}}]{kavokin2004quantum}%
  \BibitemOpen
  \bibfield  {author} {\bibinfo {author} {\bibfnamefont {K.}~\bibnamefont {Kavokin}}, \bibinfo {author} {\bibfnamefont {I.}~\bibnamefont {Shelykh}}, \bibinfo {author} {\bibfnamefont {A.}~\bibnamefont {Kavokin}}, \bibinfo {author} {\bibfnamefont {G.}~\bibnamefont {Malpuech}},\ and\ \bibinfo {author} {\bibfnamefont {P.}~\bibnamefont {Bigenwald}},\ }\bibfield  {title} {\bibinfo {title} {Quantum theory of spin dynamics of exciton-polaritons in microcavities},\ }\href@noop {} {\bibfield  {journal} {\bibinfo  {journal} {Phys. Rev. Lett.}\ }\textbf {\bibinfo {volume} {92}},\ \bibinfo {pages} {017401} (\bibinfo {year} {2004})}\BibitemShut {NoStop}%
\bibitem [{\citenamefont {Shelykh}\ \emph {et~al.}(2005)\citenamefont {Shelykh}, \citenamefont {Kavokin},\ and\ \citenamefont {Malpuech}}]{shelykh-etal.2005}%
  \BibitemOpen
  \bibfield  {author} {\bibinfo {author} {\bibfnamefont {I.~A.}\ \bibnamefont {Shelykh}}, \bibinfo {author} {\bibfnamefont {A.~V.}\ \bibnamefont {Kavokin}},\ and\ \bibinfo {author} {\bibfnamefont {G.}~\bibnamefont {Malpuech}},\ }\bibfield  {title} {\bibinfo {title} {Spin dynamics of exciton polaritons in microcavities},\ }\href@noop {} {\bibfield  {journal} {\bibinfo  {journal} {physica status solidi (b)}\ }\textbf {\bibinfo {volume} {242}},\ \bibinfo {pages} {2271} (\bibinfo {year} {2005})}\BibitemShut {NoStop}%
\bibitem [{\citenamefont {Schumacher}\ \emph {et~al.}(2007)\citenamefont {Schumacher}, \citenamefont {Kwong},\ and\ \citenamefont {Binder}}]{PhysRevB.76.245324}%
  \BibitemOpen
  \bibfield  {author} {\bibinfo {author} {\bibfnamefont {S.}~\bibnamefont {Schumacher}}, \bibinfo {author} {\bibfnamefont {N.~H.}\ \bibnamefont {Kwong}},\ and\ \bibinfo {author} {\bibfnamefont {R.}~\bibnamefont {Binder}},\ }\bibfield  {title} {\bibinfo {title} {Influence of exciton-exciton correlations on the polarization characteristics of polariton amplification in semiconductor microcavities},\ }\href {https://doi.org/10.1103/PhysRevB.76.245324} {\bibfield  {journal} {\bibinfo  {journal} {Phys. Rev. B}\ }\textbf {\bibinfo {volume} {76}},\ \bibinfo {pages} {245324} (\bibinfo {year} {2007})}\BibitemShut {NoStop}%
\bibitem [{\citenamefont {Panzarini}\ \emph {et~al.}(1999{\natexlab{b}})\citenamefont {Panzarini}, \citenamefont {Andreani}, \citenamefont {Armitage}, \citenamefont {Baxter}, \citenamefont {Skolnick}, \citenamefont {Astratov}, \citenamefont {Roberts}, \citenamefont {Kavokin}, \citenamefont {Vladimirova},\ and\ \citenamefont {Kaliteevski}}]{PhysRevB.59.5082}%
  \BibitemOpen
  \bibfield  {author} {\bibinfo {author} {\bibfnamefont {G.}~\bibnamefont {Panzarini}}, \bibinfo {author} {\bibfnamefont {L.~C.}\ \bibnamefont {Andreani}}, \bibinfo {author} {\bibfnamefont {A.}~\bibnamefont {Armitage}}, \bibinfo {author} {\bibfnamefont {D.}~\bibnamefont {Baxter}}, \bibinfo {author} {\bibfnamefont {M.~S.}\ \bibnamefont {Skolnick}}, \bibinfo {author} {\bibfnamefont {V.~N.}\ \bibnamefont {Astratov}}, \bibinfo {author} {\bibfnamefont {J.~S.}\ \bibnamefont {Roberts}}, \bibinfo {author} {\bibfnamefont {A.~V.}\ \bibnamefont {Kavokin}}, \bibinfo {author} {\bibfnamefont {M.~R.}\ \bibnamefont {Vladimirova}},\ and\ \bibinfo {author} {\bibfnamefont {M.~A.}\ \bibnamefont {Kaliteevski}},\ }\bibfield  {title} {\bibinfo {title} {Exciton-light coupling in single and coupled semiconductor microcavities: Polariton dispersion and polarization splitting},\ }\href {https://doi.org/10.1103/PhysRevB.59.5082} {\bibfield  {journal} {\bibinfo  {journal} {Phys. Rev. B}\ }\textbf {\bibinfo {volume} {59}},\ \bibinfo
  {pages} {5082} (\bibinfo {year} {1999}{\natexlab{b}})}\BibitemShut {NoStop}%
\bibitem [{\citenamefont {Spencer}\ \emph {et~al.}(2021)\citenamefont {Spencer}, \citenamefont {Fu}, \citenamefont {Schlaus}, \citenamefont {Hwang}, \citenamefont {Dai}, \citenamefont {Smith}, \citenamefont {Gamelin},\ and\ \citenamefont {Zhu}}]{doi:10.1126/sciadv.abj7667}%
  \BibitemOpen
  \bibfield  {author} {\bibinfo {author} {\bibfnamefont {M.~S.}\ \bibnamefont {Spencer}}, \bibinfo {author} {\bibfnamefont {Y.}~\bibnamefont {Fu}}, \bibinfo {author} {\bibfnamefont {A.~P.}\ \bibnamefont {Schlaus}}, \bibinfo {author} {\bibfnamefont {D.}~\bibnamefont {Hwang}}, \bibinfo {author} {\bibfnamefont {Y.}~\bibnamefont {Dai}}, \bibinfo {author} {\bibfnamefont {M.~D.}\ \bibnamefont {Smith}}, \bibinfo {author} {\bibfnamefont {D.~R.}\ \bibnamefont {Gamelin}},\ and\ \bibinfo {author} {\bibfnamefont {X.-Y.}\ \bibnamefont {Zhu}},\ }\bibfield  {title} {\bibinfo {title} {Spin-orbit–coupled exciton-polariton condensates in lead halide perovskites},\ }\href {https://doi.org/10.1126/sciadv.abj7667} {\bibfield  {journal} {\bibinfo  {journal} {Science Advances}\ }\textbf {\bibinfo {volume} {7}},\ \bibinfo {pages} {eabj7667} (\bibinfo {year} {2021})}\BibitemShut {NoStop}%
\bibitem [{\citenamefont {Sun}\ \emph {et~al.}(2017)\citenamefont {Sun}, \citenamefont {Wen}, \citenamefont {Yoon}, \citenamefont {Liu}, \citenamefont {Steger}, \citenamefont {Pfeiffer}, \citenamefont {West}, \citenamefont {Snoke},\ and\ \citenamefont {Nelson}}]{PhysRevLett.118.016602}%
  \BibitemOpen
  \bibfield  {author} {\bibinfo {author} {\bibfnamefont {Y.}~\bibnamefont {Sun}}, \bibinfo {author} {\bibfnamefont {P.}~\bibnamefont {Wen}}, \bibinfo {author} {\bibfnamefont {Y.}~\bibnamefont {Yoon}}, \bibinfo {author} {\bibfnamefont {G.}~\bibnamefont {Liu}}, \bibinfo {author} {\bibfnamefont {M.}~\bibnamefont {Steger}}, \bibinfo {author} {\bibfnamefont {L.~N.}\ \bibnamefont {Pfeiffer}}, \bibinfo {author} {\bibfnamefont {K.}~\bibnamefont {West}}, \bibinfo {author} {\bibfnamefont {D.~W.}\ \bibnamefont {Snoke}},\ and\ \bibinfo {author} {\bibfnamefont {K.~A.}\ \bibnamefont {Nelson}},\ }\bibfield  {title} {\bibinfo {title} {{B}ose-{E}instein condensation of long-lifetime polaritons in thermal equilibrium},\ }\href {https://doi.org/10.1103/PhysRevLett.118.016602} {\bibfield  {journal} {\bibinfo  {journal} {Phys. Rev. Lett.}\ }\textbf {\bibinfo {volume} {118}},\ \bibinfo {pages} {016602} (\bibinfo {year} {2017})}\BibitemShut {NoStop}%
\bibitem [{\citenamefont {Su}\ \emph {et~al.}(2018)\citenamefont {Su}, \citenamefont {Wang}, \citenamefont {Zhao}, \citenamefont {Xing}, \citenamefont {Zhao}, \citenamefont {Diederichs}, \citenamefont {Liew},\ and\ \citenamefont {Xiong}}]{su2018room}%
  \BibitemOpen
  \bibfield  {author} {\bibinfo {author} {\bibfnamefont {R.}~\bibnamefont {Su}}, \bibinfo {author} {\bibfnamefont {J.}~\bibnamefont {Wang}}, \bibinfo {author} {\bibfnamefont {J.}~\bibnamefont {Zhao}}, \bibinfo {author} {\bibfnamefont {J.}~\bibnamefont {Xing}}, \bibinfo {author} {\bibfnamefont {W.}~\bibnamefont {Zhao}}, \bibinfo {author} {\bibfnamefont {C.}~\bibnamefont {Diederichs}}, \bibinfo {author} {\bibfnamefont {T.~C.}\ \bibnamefont {Liew}},\ and\ \bibinfo {author} {\bibfnamefont {Q.}~\bibnamefont {Xiong}},\ }\bibfield  {title} {\bibinfo {title} {Room temperature long-range coherent exciton polariton condensate flow in lead halide perovskites},\ }\href@noop {} {\bibfield  {journal} {\bibinfo  {journal} {Science advances}\ }\textbf {\bibinfo {volume} {4}},\ \bibinfo {pages} {eaau0244} (\bibinfo {year} {2018})}\BibitemShut {NoStop}%
\bibitem [{\citenamefont {Wingenbach}\ \emph {et~al.}(2025)\citenamefont {Wingenbach}, \citenamefont {Bauch}, \citenamefont {Ma}, \citenamefont {Schade}, \citenamefont {Plessl},\ and\ \citenamefont {Schumacher}}]{wingenbach2024phoenixpaderbornhighly}%
  \BibitemOpen
  \bibfield  {author} {\bibinfo {author} {\bibfnamefont {J.}~\bibnamefont {Wingenbach}}, \bibinfo {author} {\bibfnamefont {D.}~\bibnamefont {Bauch}}, \bibinfo {author} {\bibfnamefont {X.}~\bibnamefont {Ma}}, \bibinfo {author} {\bibfnamefont {R.}~\bibnamefont {Schade}}, \bibinfo {author} {\bibfnamefont {C.}~\bibnamefont {Plessl}},\ and\ \bibinfo {author} {\bibfnamefont {S.}~\bibnamefont {Schumacher}},\ }\bibfield  {title} {\bibinfo {title} {{PHOENIX -- Paderborn highly optimized and energy efficient solver for two-dimensional nonlinear Schrödinger equations with integrated extensions}},\ }\href@noop {} {\bibfield  {journal} {\bibinfo  {journal} {Computer Physics Communications}\ }\textbf {\bibinfo {volume} {315}},\ \bibinfo {pages} {109689} (\bibinfo {year} {2025})}\BibitemShut {NoStop}%
\bibitem [{\citenamefont {Wingenbach}\ \emph {et~al.}()\citenamefont {Wingenbach}, \citenamefont {Bauch}, \citenamefont {Ma}, \citenamefont {Schade}, \citenamefont {Plessl},\ and\ \citenamefont {Schumacher}}]{phoenix_example}%
  \BibitemOpen
  \bibfield  {author} {\bibinfo {author} {\bibfnamefont {J.}~\bibnamefont {Wingenbach}}, \bibinfo {author} {\bibfnamefont {D.}~\bibnamefont {Bauch}}, \bibinfo {author} {\bibfnamefont {X.}~\bibnamefont {Ma}}, \bibinfo {author} {\bibfnamefont {R.}~\bibnamefont {Schade}}, \bibinfo {author} {\bibfnamefont {C.}~\bibnamefont {Plessl}},\ and\ \bibinfo {author} {\bibfnamefont {S.}~\bibnamefont {Schumacher}},\ }\href@noop {} {\bibinfo {title} {Phoenix examples}},\ \bibinfo {howpublished} {\url{https://github.com/Schumacher-Group-UPB/PHOENIX/tree/master/examples}},\ \bibinfo {note} {accessed: 2025-05-21}\BibitemShut {NoStop}%
\bibitem [{\citenamefont {Montag}\ and\ \citenamefont {Kunst}(2024)}]{PhysRevResearch.6.023205}%
  \BibitemOpen
  \bibfield  {author} {\bibinfo {author} {\bibfnamefont {A.}~\bibnamefont {Montag}}\ and\ \bibinfo {author} {\bibfnamefont {F.~K.}\ \bibnamefont {Kunst}},\ }\bibfield  {title} {\bibinfo {title} {Symmetry-induced higher-order exceptional points in two dimensions},\ }\href {https://doi.org/10.1103/PhysRevResearch.6.023205} {\bibfield  {journal} {\bibinfo  {journal} {Phys. Rev. Res.}\ }\textbf {\bibinfo {volume} {6}},\ \bibinfo {pages} {023205} (\bibinfo {year} {2024})}\BibitemShut {NoStop}%
\bibitem [{\citenamefont {Bai}\ \emph {et~al.}(2023{\natexlab{a}})\citenamefont {Bai}, \citenamefont {Fang}, \citenamefont {Liu}, \citenamefont {Li}, \citenamefont {Wan},\ and\ \citenamefont {Xiao}}]{bai2023nonlinearity}%
  \BibitemOpen
  \bibfield  {author} {\bibinfo {author} {\bibfnamefont {K.}~\bibnamefont {Bai}}, \bibinfo {author} {\bibfnamefont {L.}~\bibnamefont {Fang}}, \bibinfo {author} {\bibfnamefont {T.-R.}\ \bibnamefont {Liu}}, \bibinfo {author} {\bibfnamefont {J.-Z.}\ \bibnamefont {Li}}, \bibinfo {author} {\bibfnamefont {D.}~\bibnamefont {Wan}},\ and\ \bibinfo {author} {\bibfnamefont {M.}~\bibnamefont {Xiao}},\ }\bibfield  {title} {\bibinfo {title} {Nonlinearity-enabled higher-order exceptional singularities with ultra-enhanced signal-to-noise ratio},\ }\href {https://doi.org/10.1093/nsr/nwac259} {\bibfield  {journal} {\bibinfo  {journal} {Natl. Sci. Rev.}\ }\textbf {\bibinfo {volume} {10}},\ \bibinfo {pages} {nwac259} (\bibinfo {year} {2023}{\natexlab{a}})}\BibitemShut {NoStop}%
\bibitem [{\citenamefont {Bai}\ \emph {et~al.}(2023{\natexlab{b}})\citenamefont {Bai}, \citenamefont {Li}, \citenamefont {Liu}, \citenamefont {Fang}, \citenamefont {Wan},\ and\ \citenamefont {Xiao}}]{PhysRevLett.130.266901}%
  \BibitemOpen
  \bibfield  {author} {\bibinfo {author} {\bibfnamefont {K.}~\bibnamefont {Bai}}, \bibinfo {author} {\bibfnamefont {J.-Z.}\ \bibnamefont {Li}}, \bibinfo {author} {\bibfnamefont {T.-R.}\ \bibnamefont {Liu}}, \bibinfo {author} {\bibfnamefont {L.}~\bibnamefont {Fang}}, \bibinfo {author} {\bibfnamefont {D.}~\bibnamefont {Wan}},\ and\ \bibinfo {author} {\bibfnamefont {M.}~\bibnamefont {Xiao}},\ }\bibfield  {title} {\bibinfo {title} {Nonlinear exceptional points with a complete basis in dynamics},\ }\href {https://doi.org/10.1103/PhysRevLett.130.266901} {\bibfield  {journal} {\bibinfo  {journal} {Phys. Rev. Lett.}\ }\textbf {\bibinfo {volume} {130}},\ \bibinfo {pages} {266901} (\bibinfo {year} {2023}{\natexlab{b}})}\BibitemShut {NoStop}%
\bibitem [{\citenamefont {Tang}\ \emph {et~al.}(2020)\citenamefont {Tang}, \citenamefont {Jiang}, \citenamefont {Ding}, \citenamefont {Xiao}, \citenamefont {Zhang}, \citenamefont {Chan},\ and\ \citenamefont {Ma}}]{tang2020exceptional}%
  \BibitemOpen
  \bibfield  {author} {\bibinfo {author} {\bibfnamefont {W.}~\bibnamefont {Tang}}, \bibinfo {author} {\bibfnamefont {X.}~\bibnamefont {Jiang}}, \bibinfo {author} {\bibfnamefont {K.}~\bibnamefont {Ding}}, \bibinfo {author} {\bibfnamefont {Y.-X.}\ \bibnamefont {Xiao}}, \bibinfo {author} {\bibfnamefont {Z.-Q.}\ \bibnamefont {Zhang}}, \bibinfo {author} {\bibfnamefont {C.~T.}\ \bibnamefont {Chan}},\ and\ \bibinfo {author} {\bibfnamefont {G.}~\bibnamefont {Ma}},\ }\bibfield  {title} {\bibinfo {title} {Exceptional nexus with a hybrid topological invariant},\ }\href@noop {} {\bibfield  {journal} {\bibinfo  {journal} {Science}\ }\textbf {\bibinfo {volume} {370}},\ \bibinfo {pages} {1077} (\bibinfo {year} {2020})}\BibitemShut {NoStop}%
\bibitem [{\citenamefont {Savvidis}\ \emph {et~al.}(2000)\citenamefont {Savvidis}, \citenamefont {Baumberg}, \citenamefont {Stevenson}, \citenamefont {Skolnick}, \citenamefont {Whittaker},\ and\ \citenamefont {Roberts}}]{PhysRevLett.84.1547}%
  \BibitemOpen
  \bibfield  {author} {\bibinfo {author} {\bibfnamefont {P.~G.}\ \bibnamefont {Savvidis}}, \bibinfo {author} {\bibfnamefont {J.~J.}\ \bibnamefont {Baumberg}}, \bibinfo {author} {\bibfnamefont {R.~M.}\ \bibnamefont {Stevenson}}, \bibinfo {author} {\bibfnamefont {M.~S.}\ \bibnamefont {Skolnick}}, \bibinfo {author} {\bibfnamefont {D.~M.}\ \bibnamefont {Whittaker}},\ and\ \bibinfo {author} {\bibfnamefont {J.~S.}\ \bibnamefont {Roberts}},\ }\bibfield  {title} {\bibinfo {title} {Angle-resonant stimulated polariton amplifier},\ }\href {https://doi.org/10.1103/PhysRevLett.84.1547} {\bibfield  {journal} {\bibinfo  {journal} {Phys. Rev. Lett.}\ }\textbf {\bibinfo {volume} {84}},\ \bibinfo {pages} {1547} (\bibinfo {year} {2000})}\BibitemShut {NoStop}%
\bibitem [{\citenamefont {Hashemi}\ \emph {et~al.}(2009)\citenamefont {Hashemi}, \citenamefont {Rodriguez}, \citenamefont {Joannopoulos}, \citenamefont {Solja\ifmmode \check{c}\else \v{c}\fi{}i\ifmmode~\acute{c}\else \'{c}\fi{}},\ and\ \citenamefont {Johnson}}]{PhysRevA.79.013812}%
  \BibitemOpen
  \bibfield  {author} {\bibinfo {author} {\bibfnamefont {H.}~\bibnamefont {Hashemi}}, \bibinfo {author} {\bibfnamefont {A.~W.}\ \bibnamefont {Rodriguez}}, \bibinfo {author} {\bibfnamefont {J.~D.}\ \bibnamefont {Joannopoulos}}, \bibinfo {author} {\bibfnamefont {M.}~\bibnamefont {Solja\ifmmode \check{c}\else \v{c}\fi{}i\ifmmode~\acute{c}\else \'{c}\fi{}}},\ and\ \bibinfo {author} {\bibfnamefont {S.~G.}\ \bibnamefont {Johnson}},\ }\bibfield  {title} {\bibinfo {title} {Nonlinear harmonic generation and devices in doubly resonant {Kerr} cavities},\ }\href {https://doi.org/10.1103/PhysRevA.79.013812} {\bibfield  {journal} {\bibinfo  {journal} {Phys. Rev. A}\ }\textbf {\bibinfo {volume} {79}},\ \bibinfo {pages} {013812} (\bibinfo {year} {2009})}\BibitemShut {NoStop}%
\bibitem [{\citenamefont {Li}\ and\ \citenamefont {Kevrekidis}(2011)}]{PhysRevE.83.066608}%
  \BibitemOpen
  \bibfield  {author} {\bibinfo {author} {\bibfnamefont {K.}~\bibnamefont {Li}}\ and\ \bibinfo {author} {\bibfnamefont {P.~G.}\ \bibnamefont {Kevrekidis}},\ }\bibfield  {title} {\bibinfo {title} {$\mathcal{PT}$-symmetric oligomers: Analytical solutions, linear stability, and nonlinear dynamics},\ }\href {https://doi.org/10.1103/PhysRevE.83.066608} {\bibfield  {journal} {\bibinfo  {journal} {Phys. Rev. E}\ }\textbf {\bibinfo {volume} {83}},\ \bibinfo {pages} {066608} (\bibinfo {year} {2011})}\BibitemShut {NoStop}%
\bibitem [{\citenamefont {Graefe}\ \emph {et~al.}(2008)\citenamefont {Graefe}, \citenamefont {Günther}, \citenamefont {Korsch},\ and\ \citenamefont {Niederle}}]{Graefe_2008}%
  \BibitemOpen
  \bibfield  {author} {\bibinfo {author} {\bibfnamefont {E.~M.}\ \bibnamefont {Graefe}}, \bibinfo {author} {\bibfnamefont {U.}~\bibnamefont {Günther}}, \bibinfo {author} {\bibfnamefont {H.~J.}\ \bibnamefont {Korsch}},\ and\ \bibinfo {author} {\bibfnamefont {A.~E.}\ \bibnamefont {Niederle}},\ }\bibfield  {title} {\bibinfo {title} {A non-{H}ermitian symmetric {B}ose–{H}ubbard model: eigenvalue rings from unfolding higher-order exceptional points},\ }\href {https://doi.org/10.1088/1751-8113/41/25/255206} {\bibfield  {journal} {\bibinfo  {journal} {J. Phys. A: Math. Theor.}\ }\textbf {\bibinfo {volume} {41}},\ \bibinfo {pages} {255206} (\bibinfo {year} {2008})}\BibitemShut {NoStop}%
\bibitem [{\citenamefont {Graefe}\ \emph {et~al.}(2010)\citenamefont {Graefe}, \citenamefont {Korsch},\ and\ \citenamefont {Niederle}}]{PhysRevA.82.013629}%
  \BibitemOpen
  \bibfield  {author} {\bibinfo {author} {\bibfnamefont {E.-M.}\ \bibnamefont {Graefe}}, \bibinfo {author} {\bibfnamefont {H.~J.}\ \bibnamefont {Korsch}},\ and\ \bibinfo {author} {\bibfnamefont {A.~E.}\ \bibnamefont {Niederle}},\ }\bibfield  {title} {\bibinfo {title} {Quantum-classical correspondence for a non-{H}ermitian {B}ose-{H}ubbard dimer},\ }\href {https://doi.org/10.1103/PhysRevA.82.013629} {\bibfield  {journal} {\bibinfo  {journal} {Phys. Rev. A}\ }\textbf {\bibinfo {volume} {82}},\ \bibinfo {pages} {013629} (\bibinfo {year} {2010})}\BibitemShut {NoStop}%
\bibitem [{\citenamefont {Ramezani}\ \emph {et~al.}(2010)\citenamefont {Ramezani}, \citenamefont {Kottos}, \citenamefont {El-Ganainy},\ and\ \citenamefont {Christodoulides}}]{PhysRevA.82.043803}%
  \BibitemOpen
  \bibfield  {author} {\bibinfo {author} {\bibfnamefont {H.}~\bibnamefont {Ramezani}}, \bibinfo {author} {\bibfnamefont {T.}~\bibnamefont {Kottos}}, \bibinfo {author} {\bibfnamefont {R.}~\bibnamefont {El-Ganainy}},\ and\ \bibinfo {author} {\bibfnamefont {D.~N.}\ \bibnamefont {Christodoulides}},\ }\bibfield  {title} {\bibinfo {title} {Unidirectional nonlinear $\mathcal{PT}$-symmetric optical structures},\ }\href {https://doi.org/10.1103/PhysRevA.82.043803} {\bibfield  {journal} {\bibinfo  {journal} {Phys. Rev. A}\ }\textbf {\bibinfo {volume} {82}},\ \bibinfo {pages} {043803} (\bibinfo {year} {2010})}\BibitemShut {NoStop}%
\bibitem [{\citenamefont {Graefe}(2012)}]{Graefe_2012}%
  \BibitemOpen
  \bibfield  {author} {\bibinfo {author} {\bibfnamefont {E.-M.}\ \bibnamefont {Graefe}},\ }\bibfield  {title} {\bibinfo {title} {Stationary states of a {PT} symmetric two-mode {B}ose–{E}instein condensate},\ }\href {https://doi.org/10.1088/1751-8113/45/44/444015} {\bibfield  {journal} {\bibinfo  {journal} {J. Phys. A: Math. Theor.}\ }\textbf {\bibinfo {volume} {45}},\ \bibinfo {pages} {444015} (\bibinfo {year} {2012})}\BibitemShut {NoStop}%
\bibitem [{\citenamefont {Gu}\ \emph {et~al.}(2024)\citenamefont {Gu}, \citenamefont {Qu},\ and\ \citenamefont {Zhang}}]{GU2024107736}%
  \BibitemOpen
  \bibfield  {author} {\bibinfo {author} {\bibfnamefont {Q.}~\bibnamefont {Gu}}, \bibinfo {author} {\bibfnamefont {C.}~\bibnamefont {Qu}},\ and\ \bibinfo {author} {\bibfnamefont {Y.}~\bibnamefont {Zhang}},\ }\bibfield  {title} {\bibinfo {title} {Adjusting exceptional points using saturable nonlinearities},\ }\href {https://doi.org/https://doi.org/10.1016/j.rinp.2024.107736} {\bibfield  {journal} {\bibinfo  {journal} {Results Phys.}\ }\textbf {\bibinfo {volume} {61}},\ \bibinfo {pages} {107736} (\bibinfo {year} {2024})}\BibitemShut {NoStop}%
\bibitem [{\citenamefont {Sperling}\ and\ \citenamefont {Vogel}(2013)}]{PhysRevLett.111.110503}%
  \BibitemOpen
  \bibfield  {author} {\bibinfo {author} {\bibfnamefont {J.}~\bibnamefont {Sperling}}\ and\ \bibinfo {author} {\bibfnamefont {W.}~\bibnamefont {Vogel}},\ }\bibfield  {title} {\bibinfo {title} {Multipartite entanglement witnesses},\ }\href {https://doi.org/10.1103/PhysRevLett.111.110503} {\bibfield  {journal} {\bibinfo  {journal} {Phys. Rev. Lett.}\ }\textbf {\bibinfo {volume} {111}},\ \bibinfo {pages} {110503} (\bibinfo {year} {2013})}\BibitemShut {NoStop}%
\bibitem [{\citenamefont {Estrecho}\ \emph {et~al.}(2019)\citenamefont {Estrecho}, \citenamefont {Gao}, \citenamefont {Bobrovska}, \citenamefont {Comber-Todd}, \citenamefont {Fraser}, \citenamefont {Steger}, \citenamefont {West}, \citenamefont {Pfeiffer}, \citenamefont {Levinsen}, \citenamefont {Parish}, \citenamefont {Liew}, \citenamefont {Matuszewski}, \citenamefont {Snoke}, \citenamefont {Truscott},\ and\ \citenamefont {Ostrovskaya}}]{PhysRevB.100.035306}%
  \BibitemOpen
  \bibfield  {author} {\bibinfo {author} {\bibfnamefont {E.}~\bibnamefont {Estrecho}}, \bibinfo {author} {\bibfnamefont {T.}~\bibnamefont {Gao}}, \bibinfo {author} {\bibfnamefont {N.}~\bibnamefont {Bobrovska}}, \bibinfo {author} {\bibfnamefont {D.}~\bibnamefont {Comber-Todd}}, \bibinfo {author} {\bibfnamefont {M.~D.}\ \bibnamefont {Fraser}}, \bibinfo {author} {\bibfnamefont {M.}~\bibnamefont {Steger}}, \bibinfo {author} {\bibfnamefont {K.}~\bibnamefont {West}}, \bibinfo {author} {\bibfnamefont {L.~N.}\ \bibnamefont {Pfeiffer}}, \bibinfo {author} {\bibfnamefont {J.}~\bibnamefont {Levinsen}}, \bibinfo {author} {\bibfnamefont {M.~M.}\ \bibnamefont {Parish}}, \bibinfo {author} {\bibfnamefont {T.~C.~H.}\ \bibnamefont {Liew}}, \bibinfo {author} {\bibfnamefont {M.}~\bibnamefont {Matuszewski}}, \bibinfo {author} {\bibfnamefont {D.~W.}\ \bibnamefont {Snoke}}, \bibinfo {author} {\bibfnamefont {A.~G.}\ \bibnamefont {Truscott}},\ and\ \bibinfo {author} {\bibfnamefont {E.~A.}\ \bibnamefont {Ostrovskaya}},\ }\bibfield
  {title} {\bibinfo {title} {Direct measurement of polariton-polariton interaction strength in the {T}homas-{F}ermi regime of exciton-polariton condensation},\ }\href {https://doi.org/10.1103/PhysRevB.100.035306} {\bibfield  {journal} {\bibinfo  {journal} {Phys. Rev. B}\ }\textbf {\bibinfo {volume} {100}},\ \bibinfo {pages} {035306} (\bibinfo {year} {2019})}\BibitemShut {NoStop}%
\bibitem [{\citenamefont {Polimeno}\ \emph {et~al.}(2024)\citenamefont {Polimeno}, \citenamefont {Coriolano}, \citenamefont {Mastria}, \citenamefont {Todisco}, \citenamefont {De~Giorgi}, \citenamefont {Fieramosca}, \citenamefont {Pugliese}, \citenamefont {Prontera}, \citenamefont {Rizzo}, \citenamefont {De~Marco} \emph {et~al.}}]{polimeno2024room}%
  \BibitemOpen
  \bibfield  {author} {\bibinfo {author} {\bibfnamefont {L.}~\bibnamefont {Polimeno}}, \bibinfo {author} {\bibfnamefont {A.}~\bibnamefont {Coriolano}}, \bibinfo {author} {\bibfnamefont {R.}~\bibnamefont {Mastria}}, \bibinfo {author} {\bibfnamefont {F.}~\bibnamefont {Todisco}}, \bibinfo {author} {\bibfnamefont {M.}~\bibnamefont {De~Giorgi}}, \bibinfo {author} {\bibfnamefont {A.}~\bibnamefont {Fieramosca}}, \bibinfo {author} {\bibfnamefont {M.}~\bibnamefont {Pugliese}}, \bibinfo {author} {\bibfnamefont {C.~T.}\ \bibnamefont {Prontera}}, \bibinfo {author} {\bibfnamefont {A.}~\bibnamefont {Rizzo}}, \bibinfo {author} {\bibfnamefont {L.}~\bibnamefont {De~Marco}}, \emph {et~al.},\ }\bibfield  {title} {\bibinfo {title} {Room temperature polariton condensation from whispering gallery modes in {CsPbBr3} microplatelets},\ }\href@noop {} {\bibfield  {journal} {\bibinfo  {journal} {Adv. Mater.}\ }\textbf {\bibinfo {volume} {36}},\ \bibinfo {pages} {2312131} (\bibinfo {year} {2024})}\BibitemShut {NoStop}%
\bibitem [{\citenamefont {Wiersig}(2023)}]{wiersig2023petermann}%
  \BibitemOpen
  \bibfield  {author} {\bibinfo {author} {\bibfnamefont {J.}~\bibnamefont {Wiersig}},\ }\bibfield  {title} {\bibinfo {title} {Petermann factors and phase rigidities near exceptional points},\ }\href {https://doi.org/10.1103/PhysRevResearch.5.033042} {\bibfield  {journal} {\bibinfo  {journal} {Phys. Rev. Res.}\ }\textbf {\bibinfo {volume} {5}},\ \bibinfo {pages} {033042} (\bibinfo {year} {2023})}\BibitemShut {NoStop}%
\bibitem [{\citenamefont {Jin}\ \emph {et~al.}(2018)\citenamefont {Jin}, \citenamefont {Tan}, \citenamefont {Zhang}, \citenamefont {Wu}, \citenamefont {Chen}, \citenamefont {Zhang},\ and\ \citenamefont {Wu}}]{jin2018high}%
  \BibitemOpen
  \bibfield  {author} {\bibinfo {author} {\bibfnamefont {B.}~\bibnamefont {Jin}}, \bibinfo {author} {\bibfnamefont {W.}~\bibnamefont {Tan}}, \bibinfo {author} {\bibfnamefont {C.}~\bibnamefont {Zhang}}, \bibinfo {author} {\bibfnamefont {J.}~\bibnamefont {Wu}}, \bibinfo {author} {\bibfnamefont {J.}~\bibnamefont {Chen}}, \bibinfo {author} {\bibfnamefont {S.}~\bibnamefont {Zhang}},\ and\ \bibinfo {author} {\bibfnamefont {P.}~\bibnamefont {Wu}},\ }\bibfield  {title} {\bibinfo {title} {High-performance {T}erahertz sensing at exceptional points in a bilayer structure},\ }\href@noop {} {\bibfield  {journal} {\bibinfo  {journal} {Advanced Theory and Simulations}\ }\textbf {\bibinfo {volume} {1}},\ \bibinfo {pages} {1800070} (\bibinfo {year} {2018})}\BibitemShut {NoStop}%
\bibitem [{\citenamefont {Niu}\ \emph {et~al.}(2025)\citenamefont {Niu}, \citenamefont {Wang}, \citenamefont {Zhang},\ and\ \citenamefont {Wang}}]{niu2025enhancing}%
  \BibitemOpen
  \bibfield  {author} {\bibinfo {author} {\bibfnamefont {W.}~\bibnamefont {Niu}}, \bibinfo {author} {\bibfnamefont {T.}~\bibnamefont {Wang}}, \bibinfo {author} {\bibfnamefont {S.}~\bibnamefont {Zhang}},\ and\ \bibinfo {author} {\bibfnamefont {H.-F.}\ \bibnamefont {Wang}},\ }\bibfield  {title} {\bibinfo {title} {Enhancing exceptional-point-based sensing via pump gain in reversed-dissipation cavity optomechanics},\ }\href {https://doi.org/10.1002/qute.202400665} {\bibfield  {journal} {\bibinfo  {journal} {Adv. Quantum Technol.}\ ,\ \bibinfo {pages} {2400665}} (\bibinfo {year} {2025})}\BibitemShut {NoStop}%
\bibitem [{\citenamefont {Langbein}(2018{\natexlab{b}})}]{PhysRevA.98.023805}%
  \BibitemOpen
  \bibfield  {author} {\bibinfo {author} {\bibfnamefont {W.}~\bibnamefont {Langbein}},\ }\bibfield  {title} {\bibinfo {title} {No exceptional precision of exceptional-point sensors},\ }\href {https://doi.org/10.1103/PhysRevA.98.023805} {\bibfield  {journal} {\bibinfo  {journal} {Phys. Rev. A}\ }\textbf {\bibinfo {volume} {98}},\ \bibinfo {pages} {023805} (\bibinfo {year} {2018}{\natexlab{b}})}\BibitemShut {NoStop}%
\bibitem [{\citenamefont {Wang}\ \emph {et~al.}(2020{\natexlab{a}})\citenamefont {Wang}, \citenamefont {Lai}, \citenamefont {Yuan}, \citenamefont {Suh},\ and\ \citenamefont {Vahala}}]{wang2020petermann}%
  \BibitemOpen
  \bibfield  {author} {\bibinfo {author} {\bibfnamefont {H.}~\bibnamefont {Wang}}, \bibinfo {author} {\bibfnamefont {Y.-H.}\ \bibnamefont {Lai}}, \bibinfo {author} {\bibfnamefont {Z.}~\bibnamefont {Yuan}}, \bibinfo {author} {\bibfnamefont {M.-G.}\ \bibnamefont {Suh}},\ and\ \bibinfo {author} {\bibfnamefont {K.}~\bibnamefont {Vahala}},\ }\bibfield  {title} {\bibinfo {title} {Petermann-factor sensitivity limit near an exceptional point in a {B}rillouin ring laser gyroscope},\ }\href {https://doi.org/10.1038/s41467-020-15341-6} {\bibfield  {journal} {\bibinfo  {journal} {Nat. Commun.}\ }\textbf {\bibinfo {volume} {11}},\ \bibinfo {pages} {1610} (\bibinfo {year} {2020}{\natexlab{a}})}\BibitemShut {NoStop}%
\bibitem [{\citenamefont {Li}\ \emph {et~al.}(2024)\citenamefont {Li}, \citenamefont {Chen}, \citenamefont {Wu}, \citenamefont {Wang}, \citenamefont {Wang}, \citenamefont {Zhong}, \citenamefont {Huang}, \citenamefont {Liu}, \citenamefont {Chen}, \citenamefont {Luo},\ and\ \citenamefont {Chen}}]{Li2024}%
  \BibitemOpen
  \bibfield  {author} {\bibinfo {author} {\bibfnamefont {H.}~\bibnamefont {Li}}, \bibinfo {author} {\bibfnamefont {L.}~\bibnamefont {Chen}}, \bibinfo {author} {\bibfnamefont {W.}~\bibnamefont {Wu}}, \bibinfo {author} {\bibfnamefont {H.}~\bibnamefont {Wang}}, \bibinfo {author} {\bibfnamefont {T.}~\bibnamefont {Wang}}, \bibinfo {author} {\bibfnamefont {Y.}~\bibnamefont {Zhong}}, \bibinfo {author} {\bibfnamefont {F.}~\bibnamefont {Huang}}, \bibinfo {author} {\bibfnamefont {G.-S.}\ \bibnamefont {Liu}}, \bibinfo {author} {\bibfnamefont {Y.}~\bibnamefont {Chen}}, \bibinfo {author} {\bibfnamefont {Y.}~\bibnamefont {Luo}},\ and\ \bibinfo {author} {\bibfnamefont {Z.}~\bibnamefont {Chen}},\ }\bibfield  {title} {\bibinfo {title} {Enhanced sensitivity with nonlinearity-induced exceptional points degeneracy lifting},\ }\href {https://doi.org/10.1038/s42005-024-01609-6} {\bibfield  {journal} {\bibinfo  {journal} {Commun. Phys.}\ }\textbf {\bibinfo {volume} {7}},\ \bibinfo {pages} {117} (\bibinfo {year} {2024})}\BibitemShut
  {NoStop}%
\bibitem [{\citenamefont {Zheng}\ and\ \citenamefont {Chong}(2025)}]{PhysRevLett.134.133801}%
  \BibitemOpen
  \bibfield  {author} {\bibinfo {author} {\bibfnamefont {X.}~\bibnamefont {Zheng}}\ and\ \bibinfo {author} {\bibfnamefont {Y.~D.}\ \bibnamefont {Chong}},\ }\bibfield  {title} {\bibinfo {title} {Noise constraints for nonlinear exceptional point sensing},\ }\href {https://doi.org/10.1103/PhysRevLett.134.133801} {\bibfield  {journal} {\bibinfo  {journal} {Phys. Rev. Lett.}\ }\textbf {\bibinfo {volume} {134}},\ \bibinfo {pages} {133801} (\bibinfo {year} {2025})}\BibitemShut {NoStop}%
\bibitem [{\citenamefont {Kullig}\ \emph {et~al.}(2025)\citenamefont {Kullig}, \citenamefont {Wiersig},\ and\ \citenamefont {Schomerus}}]{kullig2025}%
  \BibitemOpen
  \bibfield  {author} {\bibinfo {author} {\bibfnamefont {J.}~\bibnamefont {Kullig}}, \bibinfo {author} {\bibfnamefont {J.}~\bibnamefont {Wiersig}},\ and\ \bibinfo {author} {\bibfnamefont {H.}~\bibnamefont {Schomerus}},\ }\bibfield  {title} {\bibinfo {title} {Generalized {P}etermann factor of non-{H}ermitian systems at exceptional points},\ }\href@noop {} {\bibfield  {journal} {\bibinfo  {journal} {arXiv:2506.15807 [physics.optics]}\ } (\bibinfo {year} {2025})}\BibitemShut {NoStop}%
\bibitem [{\citenamefont {Xu}\ \emph {et~al.}(2017)\citenamefont {Xu}, \citenamefont {Wang},\ and\ \citenamefont {Duan}}]{xu2017weyl}%
  \BibitemOpen
  \bibfield  {author} {\bibinfo {author} {\bibfnamefont {Y.}~\bibnamefont {Xu}}, \bibinfo {author} {\bibfnamefont {S.-T.}\ \bibnamefont {Wang}},\ and\ \bibinfo {author} {\bibfnamefont {L.-M.}\ \bibnamefont {Duan}},\ }\bibfield  {title} {\bibinfo {title} {{W}eyl exceptional rings in a three-dimensional dissipative cold atomic gas},\ }\href {https://doi.org/10.1103/PhysRevLett.118.045701} {\bibfield  {journal} {\bibinfo  {journal} {Phys Rev Lett.}\ }\textbf {\bibinfo {volume} {118}},\ \bibinfo {pages} {045701} (\bibinfo {year} {2017})}\BibitemShut {NoStop}%
\bibitem [{\citenamefont {Shen}\ \emph {et~al.}(2018)\citenamefont {Shen}, \citenamefont {Zhen},\ and\ \citenamefont {Fu}}]{shen-etal.2018}%
  \BibitemOpen
  \bibfield  {author} {\bibinfo {author} {\bibfnamefont {H.}~\bibnamefont {Shen}}, \bibinfo {author} {\bibfnamefont {B.}~\bibnamefont {Zhen}},\ and\ \bibinfo {author} {\bibfnamefont {L.}~\bibnamefont {Fu}},\ }\bibfield  {title} {\bibinfo {title} {Topological band theory for non-{H}ermitian {H}amiltonians},\ }\href {https://doi.org/10.1103/PhysRevLett.120.146402} {\bibfield  {journal} {\bibinfo  {journal} {Phys. Rev. Lett.}\ }\textbf {\bibinfo {volume} {120}},\ \bibinfo {pages} {146402} (\bibinfo {year} {2018})}\BibitemShut {NoStop}%
\bibitem [{\citenamefont {Mailybaev}\ \emph {et~al.}(2005)\citenamefont {Mailybaev}, \citenamefont {Kirillov},\ and\ \citenamefont {Seyranian}}]{mailybaev-etal.2005}%
  \BibitemOpen
  \bibfield  {author} {\bibinfo {author} {\bibfnamefont {A.~A.}\ \bibnamefont {Mailybaev}}, \bibinfo {author} {\bibfnamefont {O.~N.}\ \bibnamefont {Kirillov}},\ and\ \bibinfo {author} {\bibfnamefont {A.~P.}\ \bibnamefont {Seyranian}},\ }\bibfield  {title} {\bibinfo {title} {Geometric phase around exceptional points},\ }\href {https://doi.org/10.1103/PhysRevA.72.014104} {\bibfield  {journal} {\bibinfo  {journal} {Phys. Rev. A}\ }\textbf {\bibinfo {volume} {72}},\ \bibinfo {pages} {014104} (\bibinfo {year} {2005})}\BibitemShut {NoStop}%
\bibitem [{\citenamefont {Wang}\ \emph {et~al.}(2020{\natexlab{b}})\citenamefont {Wang}, \citenamefont {Zhang}, \citenamefont {Yuan}, \citenamefont {Zhong},\ and\ \citenamefont {Lu}}]{Wang2020}%
  \BibitemOpen
  \bibfield  {author} {\bibinfo {author} {\bibfnamefont {C.}~\bibnamefont {Wang}}, \bibinfo {author} {\bibfnamefont {H.}~\bibnamefont {Zhang}}, \bibinfo {author} {\bibfnamefont {H.}~\bibnamefont {Yuan}}, \bibinfo {author} {\bibfnamefont {J.}~\bibnamefont {Zhong}},\ and\ \bibinfo {author} {\bibfnamefont {C.}~\bibnamefont {Lu}},\ }\bibfield  {title} {\bibinfo {title} {Universal numerical calculation method for the {B}erry curvature and {C}hern numbers of typical topological photonic crystals},\ }\href {https://doi.org/10.1007/s12200-019-0963-9} {\bibfield  {journal} {\bibinfo  {journal} {Front. Optoelectron.}\ }\textbf {\bibinfo {volume} {13}},\ \bibinfo {pages} {73} (\bibinfo {year} {2020}{\natexlab{b}})}\BibitemShut {NoStop}%
\bibitem [{\citenamefont {Zhang}\ and\ \citenamefont {Chen}(2023)}]{PhysRevB.107.224306}%
  \BibitemOpen
  \bibfield  {author} {\bibinfo {author} {\bibfnamefont {Y.}~\bibnamefont {Zhang}}\ and\ \bibinfo {author} {\bibfnamefont {S.}~\bibnamefont {Chen}},\ }\bibfield  {title} {\bibinfo {title} {Engineering an imaginary {S}tark ladder in a dissipative lattice: Passive $\mathcal{PT}$ symmetry, $k$ symmetry, and localized damping},\ }\href {https://doi.org/10.1103/PhysRevB.107.224306} {\bibfield  {journal} {\bibinfo  {journal} {Phys. Rev. B}\ }\textbf {\bibinfo {volume} {107}},\ \bibinfo {pages} {224306} (\bibinfo {year} {2023})}\BibitemShut {NoStop}%
\bibitem [{\citenamefont {Montag}\ and\ \citenamefont {Kunst}(2025)}]{privateKunst}%
  \BibitemOpen
  \bibfield  {author} {\bibinfo {author} {\bibfnamefont {A.}~\bibnamefont {Montag}}\ and\ \bibinfo {author} {\bibfnamefont {F.}~\bibnamefont {Kunst}},\ }\href@noop {} {\bibinfo {title} {Private communication}} (\bibinfo {year} {2025})\BibitemShut {NoStop}%
\bibitem [{\citenamefont {Rechci{\'n}ska}\ \emph {et~al.}(2019)\citenamefont {Rechci{\'n}ska}, \citenamefont {Kr{\'o}l}, \citenamefont {Mazur}, \citenamefont {Morawiak}, \citenamefont {Mirek}, \citenamefont {{\L}empicka}, \citenamefont {Bardyszewski}, \citenamefont {Matuszewski}, \citenamefont {Kula}, \citenamefont {Piecek} \emph {et~al.}}]{rechcinska2019engineering}%
  \BibitemOpen
  \bibfield  {author} {\bibinfo {author} {\bibfnamefont {K.}~\bibnamefont {Rechci{\'n}ska}}, \bibinfo {author} {\bibfnamefont {M.}~\bibnamefont {Kr{\'o}l}}, \bibinfo {author} {\bibfnamefont {R.}~\bibnamefont {Mazur}}, \bibinfo {author} {\bibfnamefont {P.}~\bibnamefont {Morawiak}}, \bibinfo {author} {\bibfnamefont {R.}~\bibnamefont {Mirek}}, \bibinfo {author} {\bibfnamefont {K.}~\bibnamefont {{\L}empicka}}, \bibinfo {author} {\bibfnamefont {W.}~\bibnamefont {Bardyszewski}}, \bibinfo {author} {\bibfnamefont {M.}~\bibnamefont {Matuszewski}}, \bibinfo {author} {\bibfnamefont {P.}~\bibnamefont {Kula}}, \bibinfo {author} {\bibfnamefont {W.}~\bibnamefont {Piecek}}, \emph {et~al.},\ }\bibfield  {title} {\bibinfo {title} {Engineering spin-orbit synthetic {H}amiltonians in liquid-crystal optical cavities},\ }\href@noop {} {\bibfield  {journal} {\bibinfo  {journal} {Science}\ }\textbf {\bibinfo {volume} {366}},\ \bibinfo {pages} {727} (\bibinfo {year} {2019})}\BibitemShut {NoStop}%
\bibitem [{\citenamefont {Muszy\ifmmode~\acute{n}\else \'{n}\fi{}ski}\ \emph {et~al.}(2022)\citenamefont {Muszy\ifmmode~\acute{n}\else \'{n}\fi{}ski}, \citenamefont {Kr\'ol}, \citenamefont {Rechci\ifmmode~\acute{n}\else \'{n}\fi{}ska}, \citenamefont {Oliwa}, \citenamefont {K\ifmmode~\mbox{\k{e}}\else \k{e}\fi{}dziora}, \citenamefont {\L{}empicka-Mirek}, \citenamefont {Mazur}, \citenamefont {Morawiak}, \citenamefont {Piecek}, \citenamefont {Kula}, \citenamefont {Lagoudakis}, \citenamefont {Pi\ifmmode~\mbox{\k{e}}\else \k{e}\fi{}tka},\ and\ \citenamefont {Szczytko}}]{PhysRevApplied.17.014041}%
  \BibitemOpen
  \bibfield  {author} {\bibinfo {author} {\bibfnamefont {M.}~\bibnamefont {Muszy\ifmmode~\acute{n}\else \'{n}\fi{}ski}}, \bibinfo {author} {\bibfnamefont {M.}~\bibnamefont {Kr\'ol}}, \bibinfo {author} {\bibfnamefont {K.}~\bibnamefont {Rechci\ifmmode~\acute{n}\else \'{n}\fi{}ska}}, \bibinfo {author} {\bibfnamefont {P.}~\bibnamefont {Oliwa}}, \bibinfo {author} {\bibfnamefont {M.}~\bibnamefont {K\ifmmode~\mbox{\k{e}}\else \k{e}\fi{}dziora}}, \bibinfo {author} {\bibfnamefont {K.}~\bibnamefont {\L{}empicka-Mirek}}, \bibinfo {author} {\bibfnamefont {R.}~\bibnamefont {Mazur}}, \bibinfo {author} {\bibfnamefont {P.}~\bibnamefont {Morawiak}}, \bibinfo {author} {\bibfnamefont {W.}~\bibnamefont {Piecek}}, \bibinfo {author} {\bibfnamefont {P.}~\bibnamefont {Kula}}, \bibinfo {author} {\bibfnamefont {P.~G.}\ \bibnamefont {Lagoudakis}}, \bibinfo {author} {\bibfnamefont {B.}~\bibnamefont {Pi\ifmmode~\mbox{\k{e}}\else \k{e}\fi{}tka}},\ and\ \bibinfo {author} {\bibfnamefont {J.}~\bibnamefont {Szczytko}},\ }\bibfield  {title}
  {\bibinfo {title} {Realizing persistent-spin-helix lasing in the regime of {Rashba-Dresselhaus} spin-orbit coupling in a dye-filled liquid-crystal optical microcavity},\ }\href@noop {} {\bibfield  {journal} {\bibinfo  {journal} {Phys. Rev. Appl.}\ }\textbf {\bibinfo {volume} {17}},\ \bibinfo {pages} {014041} (\bibinfo {year} {2022})}\BibitemShut {NoStop}%
\bibitem [{\citenamefont {Li}\ \emph {et~al.}(2022{\natexlab{b}})\citenamefont {Li}, \citenamefont {Ma}, \citenamefont {Zhai}, \citenamefont {Gao}, \citenamefont {Dai}, \citenamefont {Schumacher},\ and\ \citenamefont {Gao}}]{Li2022}%
  \BibitemOpen
  \bibfield  {author} {\bibinfo {author} {\bibfnamefont {Y.}~\bibnamefont {Li}}, \bibinfo {author} {\bibfnamefont {X.}~\bibnamefont {Ma}}, \bibinfo {author} {\bibfnamefont {X.}~\bibnamefont {Zhai}}, \bibinfo {author} {\bibfnamefont {M.}~\bibnamefont {Gao}}, \bibinfo {author} {\bibfnamefont {H.}~\bibnamefont {Dai}}, \bibinfo {author} {\bibfnamefont {S.}~\bibnamefont {Schumacher}},\ and\ \bibinfo {author} {\bibfnamefont {T.}~\bibnamefont {Gao}},\ }\bibfield  {title} {\bibinfo {title} {Manipulating polariton condensates by {Rashba-Dresselhaus} coupling at room temperature},\ }\href {https://doi.org/10.1038/s41467-022-31529-4} {\bibfield  {journal} {\bibinfo  {journal} {Nat. Commun.}\ }\textbf {\bibinfo {volume} {13}},\ \bibinfo {pages} {3785} (\bibinfo {year} {2022}{\natexlab{b}})}\BibitemShut {NoStop}%
\bibitem [{\citenamefont {Sedov}\ \emph {et~al.}(2024)\citenamefont {Sedov}, \citenamefont {Glazov}, \citenamefont {Lagoudakis},\ and\ \citenamefont {Kavokin}}]{Sedov2024}%
  \BibitemOpen
  \bibfield  {author} {\bibinfo {author} {\bibfnamefont {E.~S.}\ \bibnamefont {Sedov}}, \bibinfo {author} {\bibfnamefont {M.~M.}\ \bibnamefont {Glazov}}, \bibinfo {author} {\bibfnamefont {P.~G.}\ \bibnamefont {Lagoudakis}},\ and\ \bibinfo {author} {\bibfnamefont {A.~V.}\ \bibnamefont {Kavokin}},\ }\bibfield  {title} {\bibinfo {title} {Light-matter coupling and spin-orbit interaction of polariton modes in liquid crystal optical microcavities},\ }\href@noop {} {\bibfield  {journal} {\bibinfo  {journal} {Phys. Rev. Res.}\ }\textbf {\bibinfo {volume} {6}},\ \bibinfo {pages} {023220} (\bibinfo {year} {2024})}\BibitemShut {NoStop}%
\bibitem [{\citenamefont {Yakovlev}\ \emph {et~al.}(2024)\citenamefont {Yakovlev}, \citenamefont {Crooker}, \citenamefont {Semina}, \citenamefont {Rautert}, \citenamefont {Mund}, \citenamefont {Dirin}, \citenamefont {Kovalenko},\ and\ \citenamefont {Bayer}}]{yakovlev2024exciton}%
  \BibitemOpen
  \bibfield  {author} {\bibinfo {author} {\bibfnamefont {D.~R.}\ \bibnamefont {Yakovlev}}, \bibinfo {author} {\bibfnamefont {S.~A.}\ \bibnamefont {Crooker}}, \bibinfo {author} {\bibfnamefont {M.~A.}\ \bibnamefont {Semina}}, \bibinfo {author} {\bibfnamefont {J.}~\bibnamefont {Rautert}}, \bibinfo {author} {\bibfnamefont {J.}~\bibnamefont {Mund}}, \bibinfo {author} {\bibfnamefont {D.~N.}\ \bibnamefont {Dirin}}, \bibinfo {author} {\bibfnamefont {M.~V.}\ \bibnamefont {Kovalenko}},\ and\ \bibinfo {author} {\bibfnamefont {M.}~\bibnamefont {Bayer}},\ }\bibfield  {title} {\bibinfo {title} {Exciton--polaritons in {CsPbBr3} crystals revealed by optical reflectivity in high magnetic fields and two-photon spectroscopy},\ }\href {https://doi.org/10.1002/pssr.202300407} {\bibfield  {journal} {\bibinfo  {journal} {Phys. Status Solidi Rapid Res. Lett.}\ }\textbf {\bibinfo {volume} {18}},\ \bibinfo {pages} {2300407} (\bibinfo {year} {2024})}\BibitemShut {NoStop}%
\bibitem [{\citenamefont {Sperling}\ and\ \citenamefont {Vogel}(2009)}]{PhysRevA.79.022318}%
  \BibitemOpen
  \bibfield  {author} {\bibinfo {author} {\bibfnamefont {J.}~\bibnamefont {Sperling}}\ and\ \bibinfo {author} {\bibfnamefont {W.}~\bibnamefont {Vogel}},\ }\bibfield  {title} {\bibinfo {title} {Necessary and sufficient conditions for bipartite entanglement},\ }\href {https://doi.org/10.1103/PhysRevA.79.022318} {\bibfield  {journal} {\bibinfo  {journal} {Phys. Rev. A}\ }\textbf {\bibinfo {volume} {79}},\ \bibinfo {pages} {022318} (\bibinfo {year} {2009})}\BibitemShut {NoStop}%
\bibitem [{\citenamefont {Lang}(2012)}]{lang2012algebra}%
  \BibitemOpen
  \bibfield  {author} {\bibinfo {author} {\bibfnamefont {S.}~\bibnamefont {Lang}},\ }\href@noop {} {\emph {\bibinfo {title} {Algebra}}},\ Vol.\ \bibinfo {volume} {211}\ (\bibinfo  {publisher} {Springer Science \& Business Media},\ \bibinfo {year} {2012})\BibitemShut {NoStop}%
\bibitem [{\citenamefont {Ara\'ujo}\ \emph {et~al.}(2019)\citenamefont {Ara\'ujo}, \citenamefont {Maciel}, \citenamefont {Dornelas}, \citenamefont {Varjas},\ and\ \citenamefont {Ferreira}}]{araujo-etal.2019}%
  \BibitemOpen
  \bibfield  {author} {\bibinfo {author} {\bibfnamefont {A.~L.}\ \bibnamefont {Ara\'ujo}}, \bibinfo {author} {\bibfnamefont {R.~P.}\ \bibnamefont {Maciel}}, \bibinfo {author} {\bibfnamefont {R.~G.~F.}\ \bibnamefont {Dornelas}}, \bibinfo {author} {\bibfnamefont {D.}~\bibnamefont {Varjas}},\ and\ \bibinfo {author} {\bibfnamefont {G.~J.}\ \bibnamefont {Ferreira}},\ }\bibfield  {title} {\bibinfo {title} {Interplay between boundary conditions and {W}ilson's mass in {D}irac-like {H}amiltonians},\ }\href {https://doi.org/10.1103/PhysRevB.100.205111} {\bibfield  {journal} {\bibinfo  {journal} {Phys. Rev. B}\ }\textbf {\bibinfo {volume} {100}},\ \bibinfo {pages} {205111} (\bibinfo {year} {2019})}\BibitemShut {NoStop}%
\end{thebibliography}

%

\end{document}